\documentclass[a4paper,11pt]{article}
\pdfoutput=1
\usepackage{jheppub}
\usepackage{amsmath}
\usepackage{amssymb}
\usepackage{amsfonts}
\usepackage{booktabs}
\usepackage{multirow}
\usepackage{graphicx}
\usepackage{longtable}
\usepackage{tikz}

\graphicspath{{jhep/figs/}}

\providecommand{\openone}{\leavevmode\hbox{\small1\normalsize\kern-.33em1}}

\newcommand{\SItfortyYzero}{23/17} %
\newcommand{\SItfortyDots}{4} %
\newcommand{\SItfortySisp}{4} %
\newcommand{\SIduhrNint}{8} %

\title{Landau singularity of the K3 maximal cut in black-hole scattering at fifth post-Minkowskian order}

\author{Matthew D. Schwartz}
\affiliation{Department of Physics, Harvard University, Cambridge, MA 02138, U.S.A.}
\emailAdd{schwartz@g.harvard.edu}

\abstract{\noindent Modeling the gravitational waves from merging black holes requires precise
calculations of the two-body problem. The post-Minkowskian expansion of black-hole
scattering, in powers of Newton's constant, now stands at fifth order, four loops. At this
order one part of the deflection diverges when the black holes pass at about 94 percent of
the speed of light. The divergence is known to come from periods of a K3 surface, which has
a node at this speed. We show that the node lies on the integration contour of the K3
maximal cut, which makes this speed a Landau singularity of the cut. Requiring the
divergence to cancel, without introducing new infinities, has been shown to fix the memory
contribution, named for the permanent displacement that the waves leave in test masses, up
to one constant. We derive the leading coefficients of its boundary integrals in closed
form. We obtain both results from maximal cuts and boundary limits, without an
integration-by-parts reduction of the full four-loop integrand. We also continue the K3 and
Calabi--Yau periods of these integrals to the energies of bound orbits, as in the inspiral
before a merger.
}

\begin{document}
\maketitle
\flushbottom

\section{Introduction}
\label{sec:intro}

The observation of gravitational waves from merging black
holes~\cite{LIGOScientific:2016aoc}, and the accuracy that next-generation
detectors~\cite{ET:2019dnz,Evans:2021gyd} will require of waveform
models~\cite{Purrer:2019jcp}, have motivated precise analytic calculations of the two-body
problem~\cite{Buonanno:2022pgc}.
One of these is the post-Minkowskian (PM) expansion of black-hole scattering, in powers of
Newton's constant $G$ and exact in the velocity, whose results are connected to bound
orbits~\cite{Kalin:2019rwq,Kalin:2019inp} and enter effective-one-body waveform
models~\cite{Damour:2016gwp,Buonanno:2024byg}.
Since its revival a decade ago~\cite{Damour:2016gwp} it has been developed with scattering
amplitudes~\cite{Cheung:2018wkq,Kosower:2018adc}, worldline effective field
theory~\cite{Kalin:2020mvi}, and worldline quantum field
theory~\cite{Mogull:2020sak}, and it now stands at fifth order (5PM), four
loops~\cite{Driesse:2024xad,Driesse:2026qiz}.

As the techniques for computing this expansion have developed, similarities to scattering
amplitudes have emerged.
Each order in $G$ adds a loop, and the loop integrals are reduced to master integrals and
solved through differential equations, as in collider physics.
There are differences: the massive lines are linearized worldline propagators, the
classical limit separates potential and radiation regions, and dissipative observables
require retarded graviton propagators in place of Feynman ones.
One can nevertheless ask which tools, techniques, and properties of the functions transfer.
Many Feynman integrals evaluate to multiple polylogarithms, while others involve elliptic
curves, K3 surfaces, or Calabi--Yau
manifolds~\cite{Bourjaily:2022bwx,Brown:2010bw,Bourjaily:2018yfy,Bonisch:2020qmm}, and the
same kinds of geometry have recently been found in the PM expansion.
K3 surfaces, which govern the three-loop photon self-energy of
QED~\cite{Bloch:2014qca,Bonisch:2020qmm,Forner:2024ojj}, also appear in black-hole scattering
from 4PM on, although as different families~\cite{Klemm:2024wtd,Duhr:2025lbz}, and the
four-loop integrals involve two K3 surfaces and two Calabi--Yau
threefolds~\cite{Frellesvig:2023bbf,Klemm:2024wtd,Frellesvig:2024zph,Brammer:2025rqo}.
The space of these functions, however, and whether techniques such as the Landau
bootstrap~\cite{Hannesdottir:2021kpd,Hannesdottir:2024hke}, which constrains an integral through the
singularities that the Landau equations~\cite{Landau:1959fi,Eden:1966dnq} allow, can be
applied fruitfully, have only begun to be explored.

This paper is an exploration of these connections.
We use methods developed for Feynman integrals, in particular ideas from the BootLoops
toolkit~\cite{Schwartz:2026afx,BootLoops:v1}, and test how they generalize.
For example, we generalize \textsc{AMFlow}~\cite{Liu:2022chg}, which computes Feynman
integrals numerically, to \textsc{PMflow} for the integrals of the PM expansion, and use it
to obtain the numerical boundary data and linear relations that a bootstrap requires.
More generally, the idea of characterizing the analytic structure of an integral from its
maximal cut appears to transfer as well.
Our main example is the divergence found at 5PM and second order in the mass ratio $\nu$
(5PM-2SF), where the potential-region contribution diverges at relative Lorentz factor
$\gamma=3$, about 94 percent of the speed of light, through the integrals of a K3
sector~\cite{Driesse:2026qiz}.
The Picard--Fuchs differential equation of a period determines where the period can be
singular, but not whether the integral over the physical contour is singular there or what
kind of singularity it is.
Our main result is that $\gamma=3$ is a Landau singularity of the maximal cut of the K3
sector: the surface is known to have a node at this velocity, and we show that the node lies
on the integration contour of the cut.
The singularity is of the second type~\cite{Fairlie:1962jmp}, a pinch at infinite loop
momentum that no real on-shell configuration produces~\cite{Coleman:1965xm}.

The apparent divergence at $\gamma=3$ is not physical: there is nothing special at that speed.
Similar unphysical divergences that must cancel appear in many places in scattering amplitudes, such as UV divergences in a CFT or gauge-dependent poles at unphysical kinematic points, for example the poles at $n\cdot k=0$ of individual diagrams in axial gauge.
Here the cancellation must be between regions of the same integrals: Driesse et al.\ found that the divergence arises in the potential region, through the K3 sector, and that only the memory region can cancel it~\cite{Driesse:2026qiz}.
The memory region is named for the memory effect, the permanent displacement a burst of radiation leaves in freely falling test masses, with a nonlinear part sourced by the radiated energy flux~\cite{ZeldovichPolnarev1974,BraginskyThorne,Christodoulou1991,Blanchet:1992br}.
In this region two of the four gravitons are radiative, and the impulse depends on the $i0^+$ prescription of their propagators, which we call the routing.
Driesse et al.\ impose this cancellation, together with that of the poles in the dimensional regulator $\epsilon$, and find that it fixes the two boundary integrals of the memory region that survive in the static limit up to a single constant $c_M$~\cite{Driesse:2026qiz}.
The cancellation then holds for any $c_M$ and does not explain the origin of the divergence.

The constant $c_M$ enters the conservative scattering angle at fifth post-Newtonian order~\cite{Driesse:2026qiz}.
Fixing it requires a rule for the routing, and no such rule has yet been derived from first principles.
Driesse et al.\ fix $c_M$ by a routing prescription:
retarded propagators directed toward the three-graviton vertex, averaged
under time reversal.
With this routing they evaluate the memory-region integrals, obtain $c_M=1$ and a finite impulse, and note that the physical motivation for the prescription is not yet clear~\cite{Driesse:2026qiz}.
Porto and Riva, analyzing the corresponding sector at fifth post-Newtonian
order, reproduce the result of this routing where the two computations
overlap, but advocate instead a local-in-time (Poincar\'e--Bertrand)
version of Feynman's prescription, under which the memory-like
contribution at $O(G^5\nu^2)$ has the opposite sign~\cite{Porto:2026fsd}.
They also observe that the analogous retarded rerouting at $O(G^6\nu^2)$ would be
incompatible with the conjecture~\cite{Bini:2025vuk} that all $\pi^2$ terms arise from the
potential region~\cite{Porto:2026fsd}.
Dlapa, K\"alin, Liu, and Porto note that the 5PM-2SF results of Driesse
et al.\ are fixed only up to the two boundary
integrals~\cite{Dlapa:2026oyq}.

Our goal is not to demonstrate the cancellation, which the routing of Driesse et al.\ achieves, but to understand where the singularity comes from in the analytic structure of the K3 integrals and whether it singles out a routing.
We also begin a study of the 2SF dissipative integrals, which contain the same families with retarded propagators.
We work from maximal cuts and boundary limits, without reducing the full four-loop integrand, so none of our results is a 5PM observable.

The geometry of the PM integrals is known in detail through four loops.
The conservative functions are polylogarithmic at 3PM~\cite{Bern:2019nnu,Bern:2019crd,Kalin:2020fhe}, and at 4PM, including radiation-reaction tail terms~\cite{Bern:2021dqo,Bern:2021yeh,Dlapa:2021npj,Dlapa:2021vgp}, they contain products of complete elliptic integrals, the periods of another K3 surface~\cite{Bern:2021dqo,Dlapa:2021npj,Dlapa:2022wdu,Dlapa:2023hsl,Klemm:2024wtd,Frellesvig:2024zph}.
At 5PM the conservative sector is known at first~\cite{Driesse:2024xad} and second~\cite{Driesse:2026qiz} order in $\nu$, the orders 1SF and 2SF of the self-force expansion~\cite{Barack:2018yvs,Pound:2019lzj,Wardell:2021fyy}, and Bern et al.\ gave its potential-graviton part as a velocity expansion~\cite{Bern:2025wyd}.
Frellesvig, Morales, and Wilhelm identified the first Calabi--Yau threefold of its four-loop integrals in the conservative sector at 5PM-2SF~\cite{Frellesvig:2023bbf}, see also Refs.~\cite{Frellesvig:2024rea,Duhr:2025lbz}, and the periods of a second govern the 5PM-1SF dissipative integrals and enter the radiated energy and recoil~\cite{Driesse:2024feo}.
Klemm et al.\ found the second threefold and, at 2SF, the K3 surface of the divergence, governed by the Ap\'ery operator, and gave their Picard--Fuchs operators~\cite{Klemm:2024wtd}.
Brammer et al.\ completed the loop-by-loop Baikov classification of Ref.~\cite{Frellesvig:2024zph} and showed that no other geometry occurs in the four-loop families, conservative or dissipative, at any self-force order~\cite{Brammer:2025rqo}.
In the conservative 5PM-2SF results~\cite{Bern:2025wyd,Driesse:2026qiz} the Ap\'ery K3 period remains while the threefolds drop out, and Bern et al.\ derive their absence from the ultraviolet structure of the integrand~\cite{Bern:2026oqp}.
The divergence at $\gamma=3$ arises in the only geometry that survives in the conservative result.

To find where the singularity comes from, we use the Baikov polynomial of Brammer et al.\ to map the surface of the maximal cut of the planar K3 sector, topology~40 of Ref.~\cite{Brammer:2025rqo}, onto the Ap\'ery--Fermi K3~\cite{PetersStienstra:1989}, obtain the torus period of the cut as a series, and prove that the Ap\'ery operator annihilates it.
Klemm et al.\ found this operator from differential equations~\cite{Klemm:2024wtd}, and Duhr et al.\ from the maximal cut in other variables~\cite{Duhr:2025lbz}.
We then show that the known node~\cite{BeukersPeters,PetersStienstra:1989,Festi:2018qip} of the Ap\'ery K3 surface lies on the integration torus in the scattering region, where the torus period is pinched.
The divergence is therefore a second-type Landau singularity of the maximal cut.
The routing of the graviton propagators does not follow from the pinch, and solving the Landau equations in Feynman parameters~\cite{Mizera:2021icv,Fevola:2023kaw}, we find nothing that singles out $\gamma=3$ on the subtopologies and polytope faces that we could resolve.

The dissipative sector at 2SF is much less developed.
The radiated energy and the radiation-reacted impulse are known at 3PM~\cite{Damour:2020tta,DiVecchia:2021bdo,Herrmann:2021tct,Jakobsen:2022psy,Herrmann:2021lqe,Riva:2021vnj,Kalin:2022hph,Bini:2021gat} and 4PM~\cite{Dlapa:2022lmu,Damgaard:2023ttc,Jakobsen:2023hig}, where nonlinear memory terms in the radiation-reaction force contribute to the impulse~\cite{Porto:2024cwd}, but at 5PM only at 1SF~\cite{Driesse:2024feo}; see also Refs.~\cite{Geralico:2026efi,Jakobsen:2026ayw,Heissenberg:2025ocy}.
In the post-Newtonian expansion two gauge-invariant observables at 5PM-2SF have been predicted~\cite{Bini:2022enm,Bini:2025vuk}, and the even-in-velocity relative impulse with nonlinear radiation-reaction and memory effects at $O(G^5\nu^2)$ is known to fifth post-Newtonian order~\cite{Porto:2026fsd}.
To our knowledge no 5PM-2SF results for radiation reaction, radiated energy, or recoil with full velocity dependence have appeared, and the 2SF dissipative function space, its boundary data, and its continuation to bound orbits have not been determined.

We determine the constants relating the maximal cut to the holomorphic period for all four geometries (Table~\ref{tab:four-geometry}), recovering that of the 2SF threefold~\cite{Klemm:2024wtd}, and obtain a closed formula for the Frobenius coefficients of its period, and a similar formula follows from tabulated Calabi--Yau operators~\cite{Klemm:2024wtd,Almkvist:2005qoo,Almkvist:2023yja}.
We continue the Frobenius bases of the four Picard--Fuchs operators to bound-orbit energies~\cite{Kalin:2019rwq,Kalin:2019inp} and find the expected monodromies.
For the four dissipative families of Ref.~\cite{Driesse:2026qiz}, with both worldline routings retained, we give first-stage master-integral counts.
From the response matrices of the auxiliary-mass flow~\cite{Liu:2017jxz,Liu:2022chg}, which we compute with \textsc{PMflow}, we find ten linear relations among master integrals on three of the four leading top sectors of PX2, the smallest family, and all ten are exact reduction identities.
As boundary data, we evaluate twelve PX2 master integrals at one velocity with Feynman graviton propagators.

For the memory region we derive in closed form the Laurent coefficients of the boundary integral $I_1^{(M)}$ and of an auxiliary integral $I_3^{(M)}$ through order $\epsilon$, and, for Feynman graviton propagators, the three leading poles of $I_2^{(M)}$.
Ref.~\cite{Driesse:2026qiz} reports the $\epsilon^{-4}$ pole without its coefficient.
We evaluate the memory-region integrals in the frequency representation of Driesse et al., with their retarded routing as input.
We confirm $c_M=1$, in agreement with Ref.~\cite{Driesse:2026qiz} and with Porto and Riva's fifth-post-Newtonian implementation of the same routing~\cite{Porto:2026fsd}.

Section~\ref{sec:geometry-fifth} sets up the integral families, the four geometries and the memory-region integrals. Section~\ref{sec:bootstrap-validation} contains the analysis of the maximal cut of topology~40 and the constants of the other three geometries, Section~\ref{sec:functionspace} the dissipative master integrals and boundary constants of PX2, Section~\ref{sec:boundorbits} the continuation to bound kinematics, and Section~\ref{sec:memory} the memory-region integrals and $c_M$. We conclude in Section~\ref{sec:outlook}. The appendices cover the auxiliary-mass flow (\ref{app:flow}), the reductions and master integrals (\ref{app:masters}), the boundary integrals (\ref{app:boundary}), the Ap\'ery proofs and the position-space integral (\ref{app:apery}), and the 4PM K3 sector beyond the maximal cut (\ref{app:fourpm}).

\section{Post-Minkowskian integrals at fifth order}
\label{sec:geometry-fifth}
\label{sec:setup}

In this section we fix the single-variable kinematics and the four integral families of 5PM scattering at second self-force order. We then list the four non-polylogarithmic geometries that these integrals and the 4PM integrals involve, and we recall the two boundary integrals to which the memory-region contribution of Driesse et al.\ reduces at the static point of zero relative velocity.

\subsection{Kinematics and integral families}
\label{sec:kinematics}
\label{sec:families}

We use the single-scale kinematics of
Refs.~\cite{Parra-Martinez:2020dzs,Dlapa:2021npj,Dlapa:2022wdu,Brammer:2025rqo}, in which the two
bodies scatter with four-velocities $u_1,u_2$ ($u_i^2=1$) and
momentum transfer $q$, with $u_i\cdot q=0$. The masses factor out, and after we set $q^2=-1$
the integrals depend on a single variable,
\begin{equation}
  \gamma \;=\; u_1\cdot u_2 \;=\; \frac{1+x^2}{2x}\,,
\label{eq:yx}
\end{equation}
the relative Lorentz factor, where we use $x=\gamma-\sqrt{\gamma^2-1}$
to rationalize $\sqrt{\gamma^2-1}$ and the scattering region is
$0<x<1$.

Besides quadratic graviton propagators $1/k^2$, the integrals have
linearized worldline propagators $1/(2u_i\cdot k)$. In the worldline
formalisms~\cite{Kalin:2020mvi,Mogull:2020sak,Jakobsen:2022psy},
$L$ of them are replaced at $L$ loops by delta functions
$\delta(2k\cdot u_i)$.
We assign the self-force order $\min(n_1,n_2)$SF to an integral with $n_i$ of these delta functions on worldline $i$.
The integrals have definite parity under $u_i\to-u_i$, and
the even and odd sectors decouple~\cite{Frellesvig:2024zph,Klemm:2024wtd}.

At 5PM-2SF the integrals form four families, which differ in
how the four graviton lines cross and which we denote P, PX1, PX2, and NP
(P, NP1, NP2, and NPM in the notation of Driesse et al.~\cite{Driesse:2026qiz}).
For the crossed family PX2 we write the
integrals as
\begingroup\small
\begin{equation}
\begin{gathered}
I_{n_5\cdots n_{22}}
 \;=\; \int_{k_1\cdots k_4}
 \frac{\delta(D_1)\,\delta(D_2)\,\delta(D_3)\,\delta(D_4)}
      {D_5^{\,n_5}\,D_6^{\,n_6}\cdots D_{22}^{\,n_{22}}}\,,\\[2pt]
\begin{aligned}
D_1&=2k_1\cdot u_1, & D_2&=2k_2\cdot u_1, & D_3&=2k_3\cdot u_2, & D_4&=2k_4\cdot u_2,\\
D_5&=2k_1\cdot u_2, & D_6&=2k_1\cdot u_2-2k_2\cdot u_2, & D_7&=2k_3\cdot u_1, & D_8&=2k_3\cdot u_1-2k_4\cdot u_1,\\
D_9&=(k_1-k_2)^2, & D_{10}&=(k_1-k_3)^2, & D_{11}&=(k_1-k_4)^2, & D_{12}&=(k_2-k_3)^2,\\
D_{13}&=(k_2-k_4)^2, & D_{14}&=(k_3-k_4)^2, & D_{14+j}&=(k_j+q)^2, & D_{18+j}&=k_j^2,
\end{aligned}
\end{gathered}
\label{eq:px2props}
\end{equation}
\endgroup
with loop momenta $k_1,\dots,k_4$ and $j=1,\dots,4$.
Here the four cut propagators $D_1,\dots,D_4$ appear as delta functions, two on each
worldline (2SF), as in the conservative families of Driesse et al. We take the uncut linearized
propagators $D_5,\dots,D_8$ as $1/(D_i+i\sigma 0)$ with a routing
sign $\sigma=\pm1$, and we give the graviton propagators $D_9,\dots,D_{22}$ no $i0$ prescription in the family definition.

A top sector of these families has thirteen of the $D_i$ as denominators, including the four cut
ones, and the remaining $D_i$ as numerators. We call a top sector leading if we include it in its family's first-stage reduction
(Section~\ref{sec:functionspace}).
PX2 has four leading top sectors, whose denominators are
$D_1,\dots,D_9$ together with $\{D_{13},D_{14},D_{15},D_{21}\}$ in sector $S_1$, $\{D_{11},D_{14},D_{16},D_{21}\}$ in $S_2$,
$\{D_{11},D_{14},D_{17},D_{20}\}$ in $S_3$, and $\{D_{13},D_{14},D_{17},D_{19}\}$ in the fourth.
The families P and PX1 differ from PX2 only in uncut linearized
propagators and NP also in its
graviton propagators, and we give all four
definitions in the ancillary file \texttt{families\_dissipative.json}.

For the dissipative problem we retain both routings
$\sigma=\pm1$~\cite{Jakobsen:2022psy,Kalin:2022hph}. Since the
integration-by-parts (IBP) system is the same for either sign, $\sigma$
enters only through the boundary
constants, the starting values at a reference point.
We compute the boundary constants
with Feynman graviton propagators, from which retarded propagators differ by a cut term on
the negative-frequency mass shell~\cite{Jakobsen:2022psy,Duhr:2025lbz}.

In the first-stage reductions we use the automatic sector
symmetries of \textsc{Kira}~\cite{Maierhofer:2017gsa,Klappert:2020nbg,Lange:2025fba}, 52 for PX2,
each of which maps
the linearized propagators $D_1,\dots,D_8$ into one another without sign
changes and is therefore valid for a common routing sign $\sigma$.
However, thirty of the 52 symmetries also reverse the momentum of a graviton line, which
leaves Feynman graviton propagators unchanged but turns a
retarded propagator into an advanced one~\cite{Duhr:2025lbz}.
We do not impose relations that hold
only for time-symmetric worldline propagators, such as those under
$u_i\to-u_i$, which are not among the 52.

\subsection{The four geometries}
\label{sec:geometry-map}

On the maximal cut of a top sector in the loop-by-loop Baikov
representation~\cite{Frellesvig:2017aai,Frellesvig:2024zph,Brammer:2025rqo}, a contour
integral over the remaining $n$ Baikov variables survives, which we write as
\begin{equation}
  \varpi(x)\;=\;\frac{1}{(2\pi i)^n}\oint_{T^n}\frac{d^nt}{\sqrt{Q(t;x)}}\,,
  \label{eq:maxcutperiod}
\end{equation}
and call the maximal-cut leading singularity. It is the holomorphic period of the cut surface of the sector, which is the double cover
$y^2=Q(t;x)$ of the $t$-torus, and it normalizes the canonical
basis~\cite{Dlapa:2022wdu,Klemm:2024wtd}.

The periods of the cut surface solve the sector's homogeneous differential
equation in $x$ and are annihilated by a Picard--Fuchs operator, of order three
for K3 surfaces and four for
Calabi--Yau threefolds. Each of the operators that we list in Table~\ref{tab:four-geometry}, in the variable $z$ given there, has a
point of maximal unipotent monodromy (MUM point) at $z=0$. There we
denote the Frobenius basis by $\varpi_0,\dots,\varpi_n$: one power series $\varpi_0(z)=1+O(z)$, up
to an overall power of $z$, and companions $\varpi_1,\dots,\varpi_{n}$,
with $\varpi_k$ of degree $k$ in $\ln z$. The torus period~\eqref{eq:maxcutperiod} is then proportional to $\varpi_0$.

Through 3PM the scattering observables are multiple polylogarithms with
the alphabet $\{x,\,1\pm x\}$~\cite{Parra-Martinez:2020dzs} (Section~\ref{sec:linear-landau}).
At 4PM the cut surface is a K3 surface of $\mathrm{Sym}^2$-Legendre
type~\cite{Dlapa:2022wdu,Dlapa:2023hsl,Klemm:2024wtd,Frellesvig:2024zph}, and
at 5PM three more non-polylogarithmic geometries
appear~\cite{Frellesvig:2023bbf,Klemm:2024wtd}, none of banana type. By the classification of
Brammer et al.~\cite{Brammer:2025rqo} these four are all that occur through 5PM.
With $\theta=z\,d/dz$ we write the four Picard--Fuchs operators, in
the normalization of Klemm et al., as
\begin{align}
  L_{\rm K3} &= (2\theta-1)^3+z^2(2\theta+1)^3-4z\theta\,(4\theta^2+1)\,,
  & z&=x^2, \label{eq:PFlegendre}\\
  L_{{\rm CY}_3} &= \theta^4-2^8z\bigl(\theta+\tfrac12\bigr)^4\,,
  & z&=2^{-8}x^4, \label{eq:PFhyper}\\
  L_A &= \theta^3-z(2\theta+1)(17\theta^2+17\theta+5)+z^2(\theta+1)^3\,,
  & z&=x^2, \label{eq:PFapery}\\
  L_{{\rm CY}_3'} &= \theta^4
  -2^4z\,(192\theta^4+128\theta^3+112\theta^2+48\theta+7) \nonumber\\
  &\quad+2^{14}z^2(192\theta^4+256\theta^3+208\theta^2+64\theta+7)
  -2^{30}z^3\bigl(\theta+\tfrac12\bigr)^4\,,
  & z&=\frac{1-\gamma^2}{2^{10}}. \label{eq:PFhadamard}
\end{align}

\begin{table}[t]
\centering
\renewcommand{\arraystretch}{1.25}
\small
\begin{tabular}{@{}lllll@{}}
\toprule
geometry & order, sector & operator & variable & finite singular points \\
\midrule
K3 ($\mathrm{Sym}^2$ Legendre) & 4PM & $L_{\rm K3}$~\eqref{eq:PFlegendre} & $z=x^2$ & $z=1$ \\
K3$'$ (Ap\'ery) & 5PM-2SF, top.~40 & $L_A$~\eqref{eq:PFapery} & $z=x^2$ & roots of $z^2-34z+1$ \\
CY$_3$ (${}_4F_3$) & 5PM-1SF, top.~3 & $L_{{\rm CY}_3}$~\eqref{eq:PFhyper} & $z=2^{-8}x^4$ & $z=2^{-8}$ \\
CY$_3'$ (Hadamard, $\chi=80$) & 5PM-2SF, top.~37 & $L_{{\rm CY}_3'}$~\eqref{eq:PFhadamard} & $z=\frac{1-\gamma^2}{2^{10}}$ & $z=2^{-10}$ \\
\bottomrule
\end{tabular}\\[8pt]
\begin{tabular}{lll}
\toprule
geometry & cut surface, cycle & maximal cut \\
\midrule
K3 & Gram determinant, Eq.~\eqref{eq:euler1d} &
  $\mathrm{LS}=8\,x\,K(x^2)^2=2\pi^2 x\,\varpi_0(x^2)$ \\
K3$'$ & $Q_6=P_+P_-$, $T^2$ &
  $\varpi(x^2)=\varpi_0(x^2)=\sum_n A_n x^{2n}$ \\
CY$_3$ & $P_8$ (odd parity), $T^3$ &
  $-\tfrac{i}{2}(1-x^2)^{-1/2}\,{}_4F_3\bigl(\tfrac12,\tfrac12,\tfrac12,\tfrac12;1,1,1;x^4\bigr)$ \\
CY$_3'$ & $Q_8$ (even parity), $T^3$ &
  $x\,\varpi_{\rm maxcut}(x)=\tfrac18\,\varpi_0(z)$ \\
\bottomrule
\end{tabular}
\caption{The four non-polylogarithmic geometries of post-Minkowskian
two-body scattering through 5PM~\cite{Frellesvig:2023bbf,Klemm:2024wtd,Brammer:2025rqo}, with the topology labels of
Ref.~\cite{Brammer:2025rqo}.
The singular points listed are those at finite $z\neq0$, $Q$ is the polynomial of Eq.~\eqref{eq:maxcutperiod}, with $Q_6$, $P_8$ and $Q_8$ those of Ref.~\cite{Brammer:2025rqo}, and $(2\pi i)^{-n}$ is absorbed into the torus cycle $T^n$.
The maximal-cut leading singularity $\mathrm{LS}$ of each top sector is proportional to the holomorphic period
$\varpi_0$ of its Picard--Fuchs operator, in the
normalization of Ref.~\cite{Klemm:2024wtd} (Section~\ref{sec:bootstrap-validation}). The parities in
parentheses are those of the sector under $u_i\to-u_i$. Here $\varpi_0(z)=(2/\pi)^2K(z)^2$ for the 4PM K3, with $K$ the complete
elliptic integral of the first kind, $A_n$ are the Ap\'ery
numbers~\eqref{eq:aperynumbers}, and $\varpi_{\rm maxcut}$ is the torus
period~\eqref{eq:maxcutperiod} of the CY$_3'$ sector.}
\label{tab:four-geometry}
\end{table}

The 4PM operator $L_{\rm K3}$ is the
symmetric square of the Legendre operator
$\theta^2-z(\theta+\tfrac12)^2$, up to a factor $\sqrt z$ absorbed in the
master integral. Similarly, the operator $L_{{\rm CY}_3}$ is the
Hadamard
square of the Legendre operator in a rescaled variable, with
holomorphic solution
\begin{equation}
  \varpi_0={}_4F_3\!\bigl(\tfrac12,\tfrac12,\tfrac12,\tfrac12;1,1,1;x^4\bigr),
  \label{eq:period4F3}
\end{equation}
a threefold period~\cite{Klemm:2024wtd}.
The Ap\'ery operator $L_A$ has holomorphic solution $\varpi_0=\sum_{n\ge0}A_nz^n$, with $A_n$ the
Ap\'ery numbers
\begin{equation}
  A_n=\sum_{k=0}^{n}\binom{n}{k}^{\!2}\binom{n+k}{k}^{\!2}
  =1,\ 5,\ 73,\ 1445,\ \dots\,,
  \label{eq:aperynumbers}
\end{equation}
and the associated surface is the K3
of Beukers and Peters~\cite{BeukersPeters,Klemm:2024wtd}.
Finally, the threefold of $L_{{\rm CY}_3'}$, of Euler characteristic $\chi=80$, is related to a Hadamard construction~\cite{Klemm:2024wtd,Almkvist:2023yja}. For the coefficients $a_n=1,\,112,\,47376,\dots$ of the
holomorphic solution $\varpi_0=\sum_n a_nz^n$ of $L_{{\rm CY}_3'}$ we derive
the closed formula~\eqref{eq:cy3p-identity} in Section~\ref{sec:four-geometry-sweep}.

The Ap\'ery operator is singular at a velocity inside the scattering
region~\cite{Driesse:2026qiz}, since the smaller
root of $z^2-34z+1$ is
\begin{equation}
z_+ \;=\; 17 - 12\sqrt{2} \;=\; x^2\big|_{\gamma=3}\,,
\label{eq:zplus}
\end{equation}
that is, $x=3-2\sqrt2$ and $\gamma=3$, or $v/c=\sqrt8/3$, about 94 percent of the speed of light.

\subsection{The memory-region integrals}
\label{sec:memregion}\label{sec:memloop}

Driesse et al.\ identified in the four-loop radiative integrals a distinct integration region, the memory region, besides the potential and tail regions~\cite{Driesse:2026qiz}.
In it two of the four gravitons are
radiative (RPPR and PRRP in their notation; R radiative, P potential), and at 5PM-2SF it gives rise to a memory back-reaction on the impulse.
They named the region for gravitational memory~\cite{Favata:2010zu,BraginskyThorne}, the linear memory of Zel'do\-vich
and Polnarev and the nonlinear memory of Christodoulou and
of Blanchet and Damour~\cite{ZeldovichPolnarev1974,Christodoulou1991,Blanchet:1992br}. The nonlinear memory of the emitted radiation is known at leading PM order~\cite{Georgoudis:2025vkk} and in a multipole expansion through $O(G^5)$~\cite{Bini:2026vaq}.

We call the two radiative gravitons of the memory region the memory gravitons. They can have
retarded, advanced, or Feynman propagators, and the impulse depends on this choice even in dimensional
regularization~\cite{Driesse:2026qiz}.
In the conservative calculation of Driesse et
al.\ the memory-region contribution
reduces, at the static point $\gamma=1$, to two boundary master
integrals,
\begin{equation}
I_1^{(M)}
 = \int_{\ell_{1\ldots4}}
 \frac{\hat\delta(\ell_1\!\cdot\! v_1)\,\hat\delta(\ell_2\!\cdot\! v_1)\,
       \hat\delta(\ell_3\!\cdot\! v_2)\,\hat\delta(\ell_4\!\cdot\! v_2)}
      {D_{q3}\,D_{34}\,D_{04}\,D_{31}\,D_{24}}\,,
\label{memmasters}
\end{equation}
and $I_2^{(M)}$, the same integral with the memory-graviton propagators $D_{31}$ and $D_{24}$ squared.
Here we keep their notation, in which $v_{1,2}$ are the
worldline velocities, $\ell_i$ the loop momenta,
$\int_{\ell}=\int d^{4-2\epsilon}\ell/(2\pi)^{4-2\epsilon}$,
$\hat\delta=2\pi\,\delta$, $D_{ij}=(\ell_i-\ell_j)^2$,
$D_{q3}=(\ell_3+q)^2$, and $D_{04}=\ell_4^2$.\footnote{Reference~\cite{Driesse:2026qiz} prints $D_4$, which there denotes $\ell_3\cdot v_1$ and makes the integral scaleless. The frequency representation~\eqref{I1rep} requires $D_{04}=\ell_4^2$.}

In the conservative sector of Ref.~\cite{Driesse:2026qiz} the
$\epsilon$ pole of the potential region cancels against that of the tail
region. The memory region must introduce no new pole in the sum over
regions and cancel the spurious $(\gamma-3)$ divergence of
the potential region.
These two conditions fix the leading poles up to one
constant~\cite{Driesse:2026qiz},
\begin{align}
I_1^{(M)} &= \frac{1}{15\,(8\pi)^4\,\epsilon} + \mathcal{O}(\epsilon^0)\,,
\nonumber\\
I_2^{(M)} &= -\frac{5\,c_M}{6\,(8\pi)^4\,\epsilon^2}
            + \mathcal{O}(\epsilon^{-1})\,.
\label{cMdef}
\end{align}
Driesse et al.~\cite{Driesse:2026qiz} fix $c_M$ by placing retarded propagators on $D_{31}$ and $D_{24}$, oriented toward the
three-graviton vertex, and averaging with the time-reversed routing.
This is their $\gamma$-3 prescription, named for the divergence at $\gamma=3$,
Eq.~\eqref{eq:zplus}, that it cancels, and under it $c_M$ equals one~\cite{Driesse:2026qiz}. The constant first enters the conservative scattering
angle~\cite{Driesse:2026qiz} at fifth post-Newtonian order, through
$\tfrac{64}{5}\,c_M-\tfrac{11456}{135}$, and Porto and Riva, implementing
the same routing in their fifth-post-Newtonian computation, reproduce the
value corresponding to $c_M=1$~\cite{Porto:2026fsd}.

\section{Maximal cuts and the Landau singularity}
\label{sec:bootstrap-validation}

For the planar K3 sector we show that the node of the Ap\'ery K3 surface lies on the integration contour of the maximal cut, which makes $\gamma=3$ a Landau singularity of the cut.
The singularity is of the second type, a pinch at infinite loop momentum, and in the Feynman-parameter Landau equations of the sector's subtopologies we find, on the faces that we could resolve, only the letters $\{x,\,1\pm x\}$.
The node and the pinch follow from the sector's loop-by-loop Baikov polynomial, which we map onto the Ap\'ery--Fermi K3 of Peters and Stienstra~\cite{PetersStienstra:1989}.
We also expand the torus period of the cut in a series, continue it beyond its disk of convergence and prove that it satisfies the Ap\'ery equation.
New, to our knowledge, are the location of the node on the integration torus, the Landau analysis of the sector's cut surface and subtopologies, the route from this cut surface to the lattice identity~\eqref{eq:fermi}, and the proof of $L_A\varpi=0$ for the cut integral without integration-by-parts reduction.

For each of the four geometries in Table~\ref{tab:four-geometry} we then compute the
maximal-cut leading singularity of the top sector,
Eq.~\eqref{eq:maxcutperiod}, and the constant relating it to the holomorphic
period $\varpi_0$ of the sector's Picard--Fuchs
operator.
Thus we fix the normalization of the canonical basis in each of the four sectors but do not evaluate the master integrals.

\subsection{The cut surface and the Ap\'ery--Fermi K3}
\label{sec:apery-baikov}

The K3 sector of the planar 5PM-2SF family, topology~40 of Brammer et
al.~\cite{Brammer:2025rqo}, is governed by the Ap\'ery operator $L_A$~\eqref{eq:PFapery}, which Klemm et al.\ established
from IBP relations and differential equations on its nine
master integrals~\cite{Klemm:2024wtd}.
Duhr et al.\ obtained $L_A$ from the maximal cut by expanding the torus period as a power series in $x$ whose coefficients are the Ap\'ery numbers~\cite{Duhr:2025lbz}.

Here we study the cut surface itself, starting from the loop-by-loop Baikov representation of
Brammer et al.
In their Eq.~(3.26) the leading singularity of the
sector is proportional to the torus integral of $1/\sqrt{Q_6}$ with
$Q_6=\bigl(x(1{+}t_1)(t_1{+}t_2^2)-(1{+}x^2)t_1t_2\bigr)^2-4x^2t_1^2t_2^2$.
The cut surface is the double cover $y^2=Q_6$ of the $(t_1,t_2)$
torus, branched along the curve $Q_6=0$.
With $w:=(1+t_1)(t_1+t_2^2)/(t_1t_2)$ and $2\gamma=x+1/x$, we factorize
the difference of squares $Q_6$ using $(1+x^2)\pm 2x=(1\pm x)^2$:
\begin{equation}
  Q_6 = P_+\,P_-\,,\quad
  P_\pm := x(1{+}t_1)(t_1{+}t_2^2)-(1{\pm}x)^2 t_1t_2
        = x\,t_1t_2\,(w-2\gamma\mp 2)\,, \label{eq:Q6factor}
\end{equation}
or $Q_6=x^2t_1^2t_2^2\bigl[(w-2\gamma)^2-4\bigr]$, a
spectral-curve form.
Thus the two components $P_\pm=0$ of the branch curve are the level sets $w=2\gamma\pm2$ of the single function $w$ of $t_1$ and $t_2$.
The loop-by-loop reduction of Duhr et al.\ (Section~7.2.1 of their paper) leads instead to the fully factored polynomial
$P_{4,4}=(1{-}t_1^2)\allowbreak(1{-}t_2^2)\allowbreak[(1{-}x)^2{+}4xt_1t_2]\allowbreak[(1{+}x)^2{+}4xt_1t_2]$.

To map the cut surface to the Ap\'ery--Fermi K3 of Peters and Stienstra, we lift the square root with
a Poisson kernel in a third unit-circle variable $s$.
In the spectral-curve
form the lifted integrand has a pole where $w+s+1/s=2\gamma$.
Under the unimodular monomial change $\xi_1=t_1/t_2$, $\xi_2=t_2$ we have
$w=\xi_1+\xi_1^{-1}+\xi_2+\xi_2^{-1}$, and with $\xi_3=s$ the pole lies on the simple-cubic Fermi
surface $\xi_1+\xi_1^{-1}+\xi_2+\xi_2^{-1}+\xi_3+\xi_3^{-1}=2\gamma$.
The Ap\'ery--Fermi pencil also arises in two-loop Bhabha scattering~\cite{Festi:2018qip}, and other Feynman-integral K3 surfaces include the
two-parameter families of Duhr and Maggio~\cite{Duhr:2025ppd}.

\subsection{The node on the integration torus}
\label{sec:apery-k3-bootstrap}

On the unit torus ($t_j=e^{i\phi_j}$), the integration contour of the maximal cut, we have
$w=4\cos\frac{\phi_1}{2}\cos\frac{\phi_1-2\phi_2}{2}$, which is real with $|w|\le4$, and $w=4$ only at $t_1=t_2=1$.
Since $w=2\gamma\pm2$ on $P_\pm=0$, the branch curve misses the torus for $\gamma>3$, touches it at $\gamma=3$, and meets it along a closed curve for $1<\gamma<3$, where $x\to x\pm i0$ defines the integral.
At $\gamma=3$ the surface acquires an ordinary node~\cite{BeukersPeters,PetersStienstra:1989,Festi:2018qip}, which we locate at $t_1=t_2=1$ on the curve $P_-=0$ of Eq.~\eqref{eq:Q6factor}, a point of the
integration torus.
Hence the torus period, Eq.~\eqref{eq:period} below, is pinched there, as at a conifold point~\cite{Bonisch:2021yfw}.

The finite singular points of $L_A$, the roots of
$z^2-34z+1$, are Landau singularities of the maximal cut.
Since the cut surface is singular where its branch
curve is singular, these singularities in $x$ are the projection of
$\{Q_6=\partial_{t_1}Q_6=\partial_{t_2}Q_6=0\}$, the discriminant locus of the maximal-cut polynomial~\cite{Correia:2025yao}.
By an iterated
resultant we find
\begin{equation}
  \mathrm{disc}_t\,Q_6 \;\propto\; x\,(x^2-1)\,(x^4-34x^2+1)
  \;=\; x\,(x^2-1)\,(x^2-6x+1)(x^2+6x+1)\,,
  \label{eq:disc-q6}
\end{equation}
whose zeros in $z$ are the exceptional values of Beukers and Peters.
The unique root in the scattering region $x\in(0,1)$ is $x=3-2\sqrt2$,
i.e.\ $\gamma=(1+x^2)/2x=3$, as noted by Driesse et al.~\cite{Driesse:2026qiz}.
Up to the sign of the radicand, the square root of the quartic $x^4-34x^2+1$ is the algebraic letter $r(x)$ of their canonical differential equation.

\subsection{The torus period}
\label{sec:apery-period}\label{sec:S8}\label{sec:S8outside}\label{sec:S8inside}

In terms of $a:=x/(1+x)^2$, $b:=x/(1-x)^2$ and $z:=x^2$, the torus
period is
\begin{equation}
  \varpi(z):=\frac{1}{(2\pi i)^2}\oint_{T^2}\frac{\mathrm{d}t_1\,\mathrm{d}t_2}{\sqrt{Q_6}}
  =\frac{1}{1-z}\sum_{m\ge0}\frac{\binom{2m}{m}^{\!2}}{16^{m}}
   \sum_{j+k=2m}\binom{2j}{j}\binom{2k}{k}\,a^jb^k
  =\sum_{n\ge0}A_n z^n . \label{eq:period}
\end{equation}
Here $A_n$ are the Ap\'ery numbers~\eqref{eq:aperynumbers}.
To derive the second equality we use, in the variables $\xi_1$ and $\xi_2$,
the constant-term identity (OEIS A002894)
\begin{equation}
  \mathrm{CT}_{t_1,t_2}\bigl[w^n\bigr]
  =\binom{n}{n/2}^{\!2}\,[\,n\ \text{even}\,]
  \qquad\text{for all }n\ge 0 . \label{eq:CTlemma}
\end{equation}
We apply this identity in Eq.~\eqref{eq:period}, with $(w-2\gamma)^2-4=(w-1/a)(w-1/b)$, after expanding each inverse square root in powers of $w$.
The last equality holds because $\varpi$ is annihilated by $L_A$, Eq.~(14) of Klemm et al., whose holomorphic solution at $z=0$ is unique.
Beukers and Peters proved the corresponding statement for their period integral.
For the cut integral we give two proofs that $L_A\varpi=0$ in Appendix~\ref{app:apery}, one by a Griffiths--Dwork reduction of the integrand at symbolic $x$ and one through the lattice identity~\eqref{eq:fermi} below.
Therefore the Taylor coefficients $c_n$ of $\varpi$ obey
$n^3c_n-\allowbreak(2n{-}1)(17n^2{-}17n{+}5)c_{n-1}\allowbreak+(n{-}1)^3c_{n-2}=0$
with $c_0=1$, the recurrence and initial value of the Ap\'ery numbers.

Closed forms of this period are known: modular~\cite{Beukers1987}, hypergeometric~\cite{Chan2010}, and the square of a second-order solution~\cite{Almkvist2011}.
The full leading singularity of Eq.~(3.26) of Brammer et al.\ is
$\mathrm{LS}\propto -\frac{z}{1-z}\varpi(z)$, in their master-integral
normalization, and we find the same prefactor as in the maximal-cut analysis of Duhr et al.

Beyond the disk in which the series converges we evaluate the period as an integral over $w$.
With $w=4\cos\alpha\cos\beta$ on the torus ($\alpha$, $\beta$ independent angles) and the form of $Q_6$ below Eq.~\eqref{eq:Q6factor}, the torus period~\eqref{eq:period} is
the average
\begin{equation}
\varpi(z)=\Big\langle\frac{1}{x\sqrt{(w-2\gamma)^2-4}}\Big\rangle_{T^2}
=\frac1x\int_{-4}^{4}dw\;\frac{\rho(w)}{\sqrt{(w-2\gamma)^2-4}}\,,\quad
\rho(w)=\frac{1}{2\pi^2}\,K\Big(1-\frac{w^2}{16}\Big)\,,
\label{eq:S8density}
\end{equation}
with the branch of the square root that tends to $2\gamma-w$ as $x\to0$ and $K$ the complete elliptic integral of the first
kind as a function of its parameter.
The average~\eqref{eq:S8density} is analytic in $x$ for $\gamma\notin[-3,3]$, that is in the unit disk cut along
$3-2\sqrt2\le|x|<1$ on the real axis, where it is the analytic continuation of $\sum_nA_nz^n$, $z=x^2$.
On the cut the
integrand has an integrable inverse-square-root singularity at $w=2\gamma-2$, and $x\to x\pm i0$ selects the branch
$\pm i/\sqrt{4-(2\gamma-w)^2}$ beyond it.

We compared the average~\eqref{eq:S8density} with the holomorphic Frobenius solution $\varpi_0$ of $L_A$, which we continued from $z=1/200$ by Taylor
expansion of the solutions along a polygonal path (Table~\ref{tab:S8apery}).
At the four real points of the table, which lie beyond the singular point $\gamma=3$, the polygonal path passes through
$\operatorname{Im}z>0$ for $x+i0$ and through $\operatorname{Im}z<0$ for $x-i0$, and the period has the imaginary parts of opposite sign listed there.
At the seven points of the table, all outside the disk $|z|<(3-2\sqrt2)^2$, the average agrees to 38 digits with the continued solution (ancillary file \texttt{tables/apery\_torus\_period\_values.txt}).

\begin{table}[!htb]
\centering
\caption{The torus period $\varpi$ of topology~40, Eq.~\eqref{eq:S8density}, outside the disk $|z|<(3-2\sqrt2)^2$ in which the Frobenius series converges. It is equal to the holomorphic Frobenius solution of $L_A$ continued to $z=x^2$ to the
digits shown (30 significant digits, truncated). For real $x$ the upper and lower signs are the values at $x+i0$ and $x-i0$.}
\label{tab:S8apery}
\small
\begin{tabular}{ll}
\toprule
$x$ & $\varpi$ \\
\midrule
$\tfrac{3}{10}+\tfrac{1}{5}\,i$ & $0.776983133462095384589441291719\ldots + 0.297765770874528585773167695183\ldots\,i$ \\
$\tfrac12+\tfrac14\,i$ & $0.601951922153713601983201850952\ldots + 0.358044619356407929500424134141\ldots\,i$ \\
$\tfrac35\,i$ & $0.590044791850286383411229981177\ldots$ \\
$\tfrac15$ & $1.31517618518367723058022046913\ldots \pm 0.395934065585605585514622044326\ldots\,i$ \\
$\tfrac13$ & $0.878344091497487592753182971886\ldots \pm 0.603799322912014361019053162443\ldots\,i$ \\
$\tfrac12$ & $0.619445680073979053983647953364\ldots \pm 0.585272129641175117030268029231\ldots\,i$ \\
$\tfrac45$ & $0.400259331029054283013497044548\ldots \pm 0.502066562093489943639627129552\ldots\,i$ \\
\bottomrule
\end{tabular}
\end{table}

With the lift of Section~\ref{sec:apery-baikov} we obtain an identity between the torus period and a lattice sum,
\begin{equation}
  (1+z)\,\varpi(z)=\sum_{N\ge0}B_N\Bigl(\frac{z}{(1+z)^2}\Bigr)^{\!N}, \label{eq:fermi}
\end{equation}
where $B_N$ are the return numbers of the simple cubic lattice~\cite{Guttmann2010} (OEIS
A002896).
This identity relates the holomorphic periods of the Ap\'ery pencil and the Fermi K3, which share a double cover.
It follows from Peters and Stienstra and is given by Bertin~\cite{Bertin2005}, and we obtain it from the cut surface, with the parameter map $1/(4\gamma^2)=z/(1+z)^2$ of Peters and Stienstra for $z=x^2$.
By Eq.~\eqref{eq:fermi}, $(1+z)\,\varpi$ is the generating function of the return numbers, that is, the Green function of the simple cubic lattice at the origin, whose evaluations as a product of complete elliptic integrals are reviewed in Ref.~\cite{Guttmann2010}.

\subsection{Landau equations and first-type letters}
\label{sec:linear-landau}\label{sec:S9}\label{sec:S9census}

In the Landau equations in Feynman parameters we find, on the subtopologies and polytope faces that we could resolve, no letter of topology~40 beyond $\{x,\,1\pm x\}$, and none of these searches singles out $\gamma=3$.
The searches are not exhaustive, and we state below what each of them excludes.
Post-Minkowskian integral families in the worldline formalism~\cite{Kalin:2020mvi,Mogull:2020sak} contain both quadratic
graviton propagators $D_j=\ell_j^2$ and the linearized propagators $D_e = 2u_{a(e)}\!\cdot\ell_e$ of worldline $a(e)$.
With loop momenta $k_l$, line momenta $\ell_i=\sum_{l}c_{il}\,k_l+r_i$ and Feynman parameters $\alpha_i$,
the Landau equations~\cite{Landau:1959fi} are
\begin{equation}
\sum_{j}\alpha_j\,c_{jl}\,\ell_j+\sum_{e}\alpha_e\,c_{el}\,u_{a(e)}=0\quad(l=1,\dots,L)\,,\qquad
\alpha_j\,\ell_j^2=0\,,\qquad \alpha_e\,(2u_{a(e)}\cdot\ell_e)=0\,,
\end{equation}
where $j$ labels graviton lines and $e$ worldline lines.
The first Symanzik polynomial $U$ is that of the graviton subgraph, and the second, $F$, is at most quadratic in each
$\alpha_e$.
The variable $\gamma$ enters $F$ only through the cross-terms $\alpha_e\alpha_{e'}(u_{a(e)}\cdot u_{a(e')})$.
Since $U$ contains no kinematic invariants, the second-type
condition $U=0$ (a pinch at infinite loop momentum) alone does not constrain $\gamma$.

At one and two loops we solve the Landau equations by saturated Gr\"obner-basis elimination of the Feynman parameters~\cite{Mizera:2021icv}.
For the 2PM box and, at 3PM, for the H graph and the graph with three worldline propagators~\cite{Bern:2019crd,Kalin:2020fhe}, we find the first-type Landau locus
\begin{equation}
  \{x,\; 1-x,\; 1+x\},
  \label{eq:alphabet}
\end{equation}
which is the polylogarithmic alphabet at 3PM~\cite{Parra-Martinez:2020dzs,Dlapa:2023hsl}, while the alphabet at 2PM contains only the letter $x$.
On one face of the H topology an unsaturated resultant chain produces a spurious letter $\gamma$, which
saturation removes (the elimination ideal there is $\langle 1\rangle$).

For the three-loop K3 sector at 4PM we carry out the same elimination on its 96 subtopologies with a linearized propagator of each worldline and find only $\gamma=\pm1$.
This agrees with the rational singular points $x=0,\pm1$ of the $\epsilon$-form elliptic block of Refs.~\cite{Dlapa:2022wdu,Dlapa:2023hsl}, and the full three-loop system of the second reference contains in addition the kernel $x/(1+x^2)$.
The letter $1+x^2$, which appears at three loops and at 5PM-1SF~\cite{Driesse:2024feo}, is a first-type component of none of these topologies. It arises on no face, coordinate or mixed, of the Newton polytope of $U+F$ of any of them, and we have not derived it.

For the four-loop topology~40 the second Symanzik polynomial $F$ has degree one in $\gamma$.
We take as faces the 544 subtopologies with at least four propagators and a linearized propagator of each worldline, which we obtain by setting Feynman parameters to zero in $F$.
On 204 of them $F$ vanishes identically and on 121 it does not depend on $\gamma$.
On the other 219 we saturate the Landau ideal by the product of the Feynman parameters, eliminate them at symbolic $\gamma$
over $\mathbb{Q}$, and compute the minimal associated primes of the ideal.
The Landau equations have no solution at any $\gamma$ on 133 faces and solutions only at $\gamma=\pm1$ (the letters $\{x,\,1\pm x\}$) on 65, for one of which the result follows from the symmetry that exchanges the two worldlines.
On 20 faces the equations have solutions at every $\gamma$, with further components only over $\gamma=\pm1$.

On the ten-propagator topology the elimination at symbolic $\gamma$ did not terminate.
There we use the loop equations of the two triangle subloops, each with two graviton propagators and one linearized propagator,
which exclude solutions with all $\alpha_i\neq0$ unless $U=0$.
Hence on this topology the Landau equations have no solution at finite loop momenta
for any $\gamma$.
On $U=0$ we probed only $\gamma_0\in\{0,\pm\tfrac12,\pm3\}$, the zeros in $\gamma$ of $1+x^2$,
$x^2\pm x+1$ and $x^2\pm6x+1$, and found no solution at any of them.
With $U\neq0$ imposed on the 219 faces, 152 have no solution at any $\gamma$, two have solutions at every $\gamma$ and no
other component, and $U$ vanishes identically on the remaining 65.
Thus the letters $\{x,\,1\pm x\}$ of
topology~40 arise only on $U=0$.

The Newton polytope of $U+F$~\cite{Fevola:2023kaw,Fevola:2023fzn} has 23003 faces, coordinate and mixed, of which 9468 have initial forms that depend on $\gamma$.
We tested each of these faces for solvability at $\gamma=3$, $0$ and $\tfrac12$, with fixed time limits, and retested the faces solvable there at the generic value $\gamma=7/11$.
At $\gamma=3$ the equations have no solution on 7858 faces and solutions both there and at the generic value on 1575. We could not resolve 35 faces, 9 that exceeded the limit and 26 that are solvable at $\gamma=3$ but for which the test at the generic value did not finish.
The corresponding counts are 7864, 1544 and 60 at $\gamma=0$ and 7858, 1575 and 35 at $\gamma=\tfrac12$.
On no resolved face are the equations solvable at one of the three values and unsolvable at the generic one.
This test compares solvability only: on the faces solvable at both values we did not decompose the solution set, so a further component over the special value is not excluded there.
Within these limits we find the polylogarithmic letters $\{1+x^2,\ x^2\pm x+1\}$ of the 5PM-2SF alphabet of Driesse et al.\ on no face, and they are not factors of the discriminant~\eqref{eq:disc-q6} (for $1+x^2$ in the non-planar family see Eq.~\eqref{eq:np41-alphabet}).

\subsection{\texorpdfstring{The second-type singularity at $\gamma=3$}{The second-type singularity at gamma = 3}}
\label{sec:S9second}

\begin{sloppypar}
On the maximal cut, though not in the Feynman-parameter equations above, the singularity at $\gamma=3$ is of the second type~\cite{Fairlie:1962jmp}: the pinch lies at infinite loop momentum.
We show this in the loop-by-loop Baikov representation~\cite{Frellesvig:2017aai} of Brammer et al., in which the maximal cut is the integral of
$[G(u_1,k_2)\,G(u_2,k_2)\,B_3\,B_2]^{-1/2}$ over $k_2^2$, $(k_2-q)^2$, $2u_1\cdot k_2$ and $2u_2\cdot k_2$ (their Eq.~(3.23)).
Here $G(u_i,k_2)=k_2^2-(u_i\cdot k_2)^2$ are the Gram determinants of the triangle subloops, $B_2$ is the Gram
determinant of $k_2$, $u_1$, $u_2$ and $q$, and the Baikov polynomial of the box subloop $k_3$ on its cut is
$4B_3=[(\gamma-1)(k_2-q)^2-2(u_1\cdot k_2)(u_2\cdot k_2)]\,[(\gamma+1)(k_2-q)^2-2(u_1\cdot k_2)(u_2\cdot k_2)]$.
We change variables to $\varepsilon=1/k_2^2$, $R=(k_2-q)^2/k_2^2$ and $c_i=2u_i\cdot n$, where $n=k_2/\sqrt{k_2^2}$ is the direction of $k_2$.
With $q^2=-1$ the integrand of the maximal cut becomes
\begin{equation*}
\frac{dk_2^2\;d(k_2-q)^2\;d(2u_1\!\cdot\!k_2)\;d(2u_2\!\cdot\!k_2)}{\sqrt{G(u_1,k_2)\,G(u_2,k_2)\,B_3\,B_2}}
=-\frac{32}{\varepsilon}\,\frac{d\varepsilon\;dR\;dc_1\,dc_2}{\sqrt{(4-c_1^2)(4-c_2^2)\,b_3\,b_2}}\,,
\end{equation*}
with $b_3=[2(\gamma-1)R-c_1c_2]\,[2(\gamma+1)R-c_1c_2]$ and
$b_2=(\gamma^2-1)(1-R)^2+\varepsilon\,[2(\gamma^2-1)(1+R)+c_1^2+c_2^2-2\gamma c_1c_2]+\varepsilon^2(\gamma^2-1)$.
It has a simple pole at $\varepsilon=0$, that is at infinite $k_2^2$, where $b_2=(\gamma^2-1)(1-R)^2$, and the residue there has a simple pole at $R=1$, where $b_3=(c_1c_2-2\gamma)^2-4$.
With $c_i=\tau_i+1/\tau_i$, $t_1=\tau_1^2$ and $t_2=\tau_1\tau_2$, the two residues give, up to a sign,
\begin{equation*}
\frac{32}{\sqrt{\gamma^2-1}}\,
\frac{dc_1\,dc_2}{\sqrt{(4-c_1^2)(4-c_2^2)\,[(c_1c_2-2\gamma)^2-4]}}
=\frac{32\,x^2}{1-x^2}\,\frac{dt_1\,dt_2}{\sqrt{Q_6}}\,,
\end{equation*}
the integrand of Eq.~(3.26) of Brammer et al.
The first residue is the one that they take at $t_{11}=0$, and we checked the steps against their ancillary notebook.

Hence the whole integration torus lies at infinite $k_2^2$ with $(k_2-q)^2/k_2^2=1$.
The other three loop momenta are infinite there as well: on the cut $2k_1\cdot k_2=k_2^2=-2k_4\cdot k_2$ and $2k_3\cdot(k_2-q)=(k_2-q)^2$, while in four dimensions the components of $k_2$ grow no faster than $\sqrt{k_2^2}$ and $(1-R)\,k_2^2$, so that the components of $k_1$, $k_3$ and $k_4$ are unbounded as $\varepsilon\to0$ and $R\to1$.
The torus variables depend only on the direction of $k_2$: $(u_1\cdot n)^2=(1+t_1)^2/(4t_1)$, $(u_2\cdot n)^2=(t_1+t_2^2)^2/(4t_1t_2^2)$ and $w=4(u_1\cdot n)(u_2\cdot n)=c_1c_2$.

In the variables $c_i$ the unit torus is the square $|c_i|\le2$, with $c_i=2\cos\theta_i$, $t_1=e^{2i\theta_1}$ and $t_2=e^{i(\theta_1+\theta_2)}$, and the residue is singular on its edges $c_i=\pm2$, where $G(u_i,k_2)/k_2^2$ vanishes, and on the curves $c_1c_2=2\gamma\pm2$, where $B_3/(k_2^2)^2$ vanishes.
At $\gamma=3$ the curve $c_1c_2=2\gamma-2$ meets the square only at its corners $c_1=c_2=\pm2$, both of which are the node $t_1=t_2=1$.
There $G(u_1,k_2)$, $G(u_2,k_2)$ and $B_3$ vanish together relative to $k_2^2$, and the torus is pinched at infinite loop momentum.
The momenta are infinite because $k_2^2$ is, not because the Gram determinants vanish: for complex $k_2$ with $G(u_1,k_2)=0$ the cut conditions of the triangle subloop have finite solutions.
The Gram determinants that vanish at the node are those of the subloops, whereas the Gram determinant of the external vectors $u_1$, $u_2$ and $q$ vanishes only at $\gamma=\pm1$.
\end{sloppypar}

\subsection{The other three geometries}
\label{sec:four-geometry-sweep}
\label{sec:route-beta}\label{sec:baikov-drain}\label{sec:worked-4pm}\label{sec:fourpm-validation}

At 4PM we derive the cut surface of the K3 sector from the Gram
determinant, whereas Ref.~\cite{Frellesvig:2024zph} obtained it loop by loop.
On the maximal cut the worldline propagators
$\delta(2u_i\cdot k)$ are linear in the Baikov variables and fix scalar products inside the Gram
determinant, so no quartic extension of the Baikov polynomial is needed.
We integrate out the remaining irreducible scalar products with Dodgson/Jacobi--Sylvester determinant identities~\cite{Jiang:2023qnl}, and each integration produces an Euler Beta function and lowers the order of the determinant by one.
After four such integrations we reduce the $d=4$ maximal cut to the
two-fold integral
\begin{equation}
  \mathrm{LS}_{\rm shape}(x)
  =\int_{-\infty}^{-1/2}\!\frac{dc_2}{\sqrt{-2c_2-1}}
  \oint\frac{db_2}{\sqrt{\bigl(b_2^2+c_2^2(\gamma^2-1)\bigr)\bigl(b_2^2+2c_2+1\bigr)}},
  \qquad \gamma=\frac{1+x^2}{2x}.
  \label{eq:euler1d}
\end{equation}
For $c_2<-\tfrac{1}{2}$ the radicand has real branch points
$b_2=\pm\sqrt{-2c_2-1}$, and $\oint db_2$ denotes the integral taken once
along the segment between them, with the radicand replaced by its
modulus.
This is one half of the integral around the closed cycle enclosing the
segment, for which the constant $8$ found below would be $16$.
The $b_2$ integral is a complete elliptic integral, and we evaluate the $c_2$ integral by tanh--sinh quadrature.
At five rational points the ratio
$\mathrm{LS}_{\rm shape}(x)/\bigl(x\,K(x^2)^2\bigr)$ takes the same value, which we identify as $8$ by an integer-relation search, hence
\begin{equation}
  \mathrm{LS}_{\rm shape}(x) = 8\,x\,K(x^2)^2 = 2\pi^2\,x\,\varpi_0(x^2)\,,
  \label{eq:k3gate}
\end{equation}
where $\varpi_0(z)=(2/\pi)^2K(z)^2$ and $\sqrt z\,\varpi_0(z)$ is the
holomorphic Frobenius solution of
$L_{\rm K3}$ \eqref{eq:PFlegendre}.

Dlapa et al.\ treated the 4PM K3 sector with differential
equations~\cite{Dlapa:2022wdu,Dlapa:2023hsl}.
Reducing it instead by IBP on the maximal cut, we find five master integrals, two
of which are odd under the velocity reflection $(u_1,u_2)\to(-u_1,-u_2)$
and vanish.
The three even ones (the scalar integral and its $(k_1^2)$ and
$(2k_1\cdot u_2)^2$ numerator partners), together with the one even
subsector master integral, constitute the $3+1$ elliptic block
of Ref.~\cite{Dlapa:2022wdu}.
The first-type Landau locus~\eqref{eq:alphabet} of the one- and two-loop
topologies coincides with the points $x=0,\pm1$ at which the kernels of the $\epsilon$-form block of the sector are singular.
At order $\epsilon^0$ the maximal cut of the scalar master integral is a combination of the three functions
\begin{equation}
\bigl\{\, x\,K(x^2)^2,\;\; x\,K(x^2)\,K'(x^2),\;\; x\,K'(x^2)^2 \,\bigr\},
\label{eq:sym2span}
\end{equation}
which are the solutions of its Picard--Fuchs equation given in Ref.~\cite{Dlapa:2023hsl}, Eq.~(5.31), with $K'(z)\equiv K(1-z)$ the complementary elliptic integral.

The two master integrals with a doubled propagator in the basis of Dlapa et al.\ are, at order $\epsilon^0$ on the maximal cut, combinations of the scalar one and its first two derivatives, and they also contain the complete elliptic integral of the second kind.
The leading singularity~\eqref{eq:k3gate} is, for $\Psi_1\propto K(x^2)$,
the factor $x\,\Psi_1^2$ by which the scalar integral is divided in
Ref.~\cite{Dlapa:2022wdu}, Eq.~(3.23), to define the first canonical
integral of the elliptic block, while Ref.~\cite{Dlapa:2023hsl} divides by $x\,K'(x^2)^2/\pi^2$.

The hypergeometric threefold CY$_3$ of topology~3 (5PM-1SF) appears only
in its odd-parity sector, which has a dotted linearized
propagator~\cite{Klemm:2024wtd,Brammer:2025rqo}, and its loop-by-loop Baikov
polynomial is $P_8$, Eq.~(3.5) of Brammer et al.
Integrating $d^3t/\sqrt{P_8}$ over the vanishing torus $T^3$ at the MUM
point, with the branch of $\sqrt{P_8}$ followed by homotopy in $x$ from
$x=0$, we find
\begin{equation}
\frac{1}{(2\pi i)^{3}}\oint_{T^3} \frac{d^3t}{\sqrt{P_8}}
\;=\; -\frac{i}{2\sqrt{1-x^2}}\;
{}_4F_3\!\bigl(\tfrac12,\tfrac12,\tfrac12,\tfrac12;1,1,1;x^4\bigr)\,,
\label{eq:cy3-maxcut}
\end{equation}
where ${}_4F_3$ is the holomorphic period $\varpi_0$~\eqref{eq:period4F3} of $L_{{\rm CY}_3}$.
In contrast, the even-parity scalar integral of the same
topology is polylogarithmic (Table~2 of Brammer et al.), and its
maximal cut is not proportional to $\varpi_0$ for any kinematic
prefactor.

The Hadamard threefold CY$_3'$ of topology~37 (5PM-2SF) has the
even-parity Baikov polynomial $Q_8$~\cite{Frellesvig:2023bbf}, Eq.~(3.8) of Brammer et al.
We evaluate its maximal-cut integral over $T^3$ by a residue series and
recover $x\,\varpi_{\rm maxcut}(x) = \tfrac18\,\varpi_0(z)$, Eq.~(A6) of Klemm et al., with $z$ of Eq.~\eqref{eq:PFhadamard}.
From the $(t_1,t_2)$ constant-term reduction of $Q_8$ we obtain a
closed formula for the Frobenius coefficients
$a_n = 1,\,112,\,47376,\,\dots$ of $\varpi_0$,
\begin{equation}
\begin{aligned}
a_n &= 16^n\binom{2n}{n}^{\!2} J_n,\\
J_n&=\sum_{j,k=0}^{n}\binom{2n}{n-j}\binom{2n}{n-k}
\frac{(-1)^{j+k}\,(n+\tfrac12)_{j+k}}{j!\,k!}\\
&=\sum_{j=0}^{n}(-1)^j\binom{2n}{n-j}\binom{n-\tfrac12-j}{n}\frac{(n+\tfrac12)_j}{j!}\,,
\end{aligned}
\label{eq:cy3p-identity}
\end{equation}
where the second form follows from the Chu--Vandermonde identity.
We prove Eq.~\eqref{eq:cy3p-identity} for all $n$: the single sum satisfies an order-two recurrence from
Zeilberger's algorithm, the coefficient
recurrence of $L_{{\rm CY}_3'}$ is a left multiple of it, and induction from three initial terms completes the proof.

A similar formula follows from the Hadamard representation of Klemm et al., who relate $\varpi_0$ to the period of the Hadamard product $C\ast i$, case~2.33 of Almkvist and van Straten~\cite{Almkvist:2023yja}.
The double sum that follows from this relation agrees with Eq.~\eqref{eq:cy3p-identity} for $n\le40$ when the coefficients of the factor $i$ are taken positive.
Almkvist, van Straten and Zudilin~\cite{Almkvist2011} relate case~2.33 by a pullback to the Hadamard square that is entry~115 of Ref.~\cite{Almkvist:2005qoo} and note its second MUM point at $\tilde z=1/256$.
With $\tilde z=x^2/256$ the two MUM points are the high-energy point $x=0$ and the low-velocity point $x=1$.
For the conservative integral containing this threefold, Frellesvig et al.~\cite{Frellesvig:2024rea} and, on its maximal cut, Duhr et al.~\cite{Duhr:2025lbz} have constructed $\epsilon$-factorized differential equations.

At four loops the nested Beta-function integration of the Gram determinant removes eight of the twelve irreducible scalar products for topology~3 and six of the eleven for topology~37, but does not yield the three-dimensional cut surfaces.
For the 5PM rows of Table~\ref{tab:four-geometry} we therefore take as input the loop-by-loop Baikov polynomials $Q_6$, $P_8$ and $Q_8$ of Brammer et
al., which they derived
without reference to the Picard--Fuchs operators of
Klemm et al.
Hence for $Q_6$ and $P_8$ we compare a cut integral and a Frobenius series that were constructed independently.

For each constant of the table we evaluate the contour integral at
three to five rational points inside the Frobenius disk around the MUM point and divide it by the Frobenius series of the
Picard--Fuchs operator.
The ratio is constant across the points to at least $34$ significant figures. It is unity by construction for the Ap\'ery row and is otherwise found by an integer-relation search.
Outside the disks we made the comparison only for topology~40, since for topologies~3 and~37 it would require continuation of the period or a deformed contour, which we have not attempted.

\section{The dissipative function space}\label{sec:functionspace}

In the four dissipative families of 5PM-2SF we keep both routings of the uncut worldline propagators, and the function space of these families has not been determined. We count their master integrals at first stage, find linear relations among them from the response matrices of the auxiliary-mass flow, and compute boundary constants of the smallest family, PX2, at one kinematic point.

\subsection{Master-integral counts and parity}
\label{sec:S1conv}

We reduce the four dissipative families with \textsc{Kira}~\cite{Maierhofer:2017gsa,Klappert:2020nbg,Lange:2025fba} using its automatic sector symmetries (Section~\ref{sec:families}). Table~\ref{tab:families} lists the first-stage master-integral counts we obtain, the parity split for the crossed family PX2 and the
conservative counts of Ref.~\cite{Driesse:2026qiz}.
The four dissipative counts sum to 731, the conservative total.
Since the dissipative counts are taken at a fixed seeding depth and the conservative counts after imposing relations that hold only under time symmetry, we regard this equality as a coincidence and infer no correspondence between the bases.

For the reductions we allow derivatives of the cut delta functions and normalize the integrals of Eq.~\eqref{eq:px2props}, in the kinematics of Section~\ref{sec:kinematics}, as
\begin{equation}
I_{n_1\cdots n_{22}}
 \;=\; \int\prod_{i=1}^{4}\frac{d^dk_i}{i\pi^{d/2}}\;
 \frac{\prod_{c=1}^{4}\,2\pi\,\delta^{(n_c-1)}(D_c)}{\prod_{i=5}^{22} D_i^{\,n_i}}\,,
 \qquad d=4-2\epsilon\,,
\label{eq:S1family}
\end{equation}
where the first four propagators are cut: $n_c=1$ denotes the delta function and $n_c=2$ its first derivative. A negative
exponent $n_i<0$ denotes the numerator $D_i^{|n_i|}$.
We call an integral even or odd according to its sign under $u_i\to-u_i$, which reverses every linearized
propagator: each power of an uncut $D_5,\dots,D_8$ and each derivative of a cut delta function contributes a
factor $-1$.
With this definition the 37
master integrals of PX2 split into 12 even and 25 odd. Even master integrals contribute
to the scattering angle and odd ones to the radiated energy
and momentum~\cite{Duhr:2025lbz}.

\begin{table}[!htb]
\caption{First-stage master-integral counts of the four dissipative 5PM-2SF families
and the conservative counts of Ref.~\cite{Driesse:2026qiz}.
The first-stage counts are taken at the fixed seeding of Appendix~\ref{sec:S5counts} and are not converged basis
dimensions: the PX2 count is unchanged under the deeper seeding described there, and the counts of P, PX1 and NP have not been tested. The 36 sectors in parentheses are the further top sectors of PX2 that contain
the K3 and Calabi--Yau subtopologies. The
parity split is determined for PX2 only.
\label{tab:families}}
\centering
\begin{tabular}{lcccc}
\toprule
Family & Top sectors & Conservative & First stage & Parity split \\
\midrule
P   & 56       & 321 & 195 & --- \\
PX1 & 44       & 144 & 256 & --- \\
PX2 & 4 (+36)  & 46  & 37  & $12 + 25$ \\
NP  & 9        & 220 & 243 & --- \\
\bottomrule
\end{tabular}
\end{table}

\subsection{Response matrices \label{sec:flowreach}}

In the auxiliary-mass flow~\cite{Liu:2017jxz,Liu:2021wks,Liu:2022chg} a subset of the propagators
of a top sector is deformed,
$D_i\to D_i-\eta$, and the master integrals are evolved from $\eta\to\infty$, where they reduce to boundary
integrals of the flow, to $\eta\to0$. On cut worldline families the boundary integrals are not vacuum integrals (cf.~\cite{Liu:2020kpc,Liu:2022tji}). Since the evolution is linear in the boundary data, we can write
\begin{equation}
M(\epsilon) \;=\; A(\epsilon)\, b(\epsilon)\,,
\label{eq:response}
\end{equation}
with $M$ the vector of master integrals of the sector's $\eta$-deformed system at $\eta\to0$,
$b$ the vector of boundary integrals at $\eta\to\infty$ and $A(\epsilon)$ the response matrix.
We call a left null
vector $v$ of $A$, $v^{T}\!A=0$, a flow-free direction.

Table~\ref{tab:flowrank} lists the propagators of the PX2 top sectors $S_1$, $S_2$
and $S_3$, which contain 35 of the 37 master integrals of the family, and the rank of
$A(\epsilon)$. We determine the rank numerically from a gap of more than forty orders of magnitude in its singular-value spectrum at each of twelve values of $\epsilon$.
From the rank deficits we obtain ten flow-free directions, supported on 31 rows of the three response matrices.
Of the ten null vectors we extract, two
have rational, $\epsilon$-independent weights,
\begin{equation}
v^{(S_1)} = \bigl(1,\ \tfrac{16}{3},\ -6\bigr)
\ \text{on}\ (I_0, I_2, I_3)\,,
\qquad
v^{(S_2)}_0 = \bigl(-\tfrac16,\ 1,\ -\tfrac89\bigr)
\ \text{on}\ (I_{r0}, I_{r2}, I_{r3})\,,
\label{eq:rationalnulls}
\end{equation}
with $I_{rk}$ the master integral in row $k$ of the response matrix of $S_2$, whereas the other eight depend on
$\epsilon$.
The three master integrals $I_0$, $I_2$, $I_3$ that support the null vector of $S_1$, given in Eq.~\eqref{eq:S1masters}, contain none of the sector's uncut linearized propagators
$D_5,\dots,D_8$, and we call them the corner integrals.

\label{sec:doublecover}
For the NP family we solve the $\eta$-deformed system on the double cover of the
flow variable, because one coupled block of its system has the local exponents of Eq.~\eqref{eq:cosets}. The last row of Table~\ref{tab:flowrank} lists the first of its nine top
sectors.

\begin{table}[t]
\caption{Response matrices of the three analyzed top sectors of PX2 and
of the first NP top sector: the rank deficits give ten flow-free directions on PX2 and two on NP. $n_J$ and $n_b$ are the numbers of master
integrals of the $\eta$-deformed system and
of boundary integrals, $n_v=n_J-{\rm rank}\,A$ the number of flow-free
directions and $n_{\rm supp}$ the number of
master integrals supporting them.
The matrices and their left null vectors are in the ancillary directory \texttt{response\_matrices/}.
\label{tab:flowrank}}
\centering
\begin{tabular}{llccccc}
\toprule
Sector & Propagators & $n_{J}$ & $n_{b}$ & rank\,$A$ & $n_{v}$ & $n_{\rm supp}$ \\
\midrule
$S_1$ & $D_1,\dots,D_9,\,D_{13},D_{14},D_{15},D_{21}$ & 14 & 27  & 13 & 1 & 3  \\
$S_2$ & $D_1,\dots,D_9,\,D_{11},D_{14},D_{16},D_{21}$ & 27 & 68  & 23 & 4 & 12 \\
$S_3$ & $D_1,\dots,D_9,\,D_{11},D_{14},D_{17},D_{20}$ & 44 & 101 & 39 & 5 & 16 \\
\midrule
NP, first top sector & (double cover)                   & 31 & 98  & 29 & 2 & 6  \\
\bottomrule
\end{tabular}
\end{table}

\subsection{The supporting master integrals}
\label{sec:S1rows}

The left null vectors of Section~\ref{sec:flowreach} are supported on 31 rows: $I_0$, $I_2$, $I_3$ on
$S_1$, twelve rows $I_{rk}$ of $S_2$ and sixteen rows $\tilde I_k$ of $S_3$. Their exponent vectors are in the ancillary file \texttt{tables/\allowbreak PX2\_\allowbreak supporting\_\allowbreak 31.txt}.
The three rows of $S_1$ are the corner integrals
\begin{equation}
\begin{aligned}
I_0 &= \int\prod_{i=1}^{4}\frac{d^dk_i}{i\pi^{d/2}}\;
\frac{(2\pi)^4\,\delta(2k_1\!\cdot\! u_1)\,\delta(2k_2\!\cdot\! u_1)\,
\delta(2k_3\!\cdot\! u_2)\,\delta(2k_4\!\cdot\! u_2)}
{(k_1-k_2)^2\,(k_2-k_4)^2\,(k_3-k_4)^2\,(k_1+q)^2\,k_3^2}\,,\\
I_2 &= I_0\bigl[(k_1-k_3)^2\bigr]\,,\qquad
I_3 = I_0\bigl[(k_1-k_4)^2\bigr]\,,
\end{aligned}
\label{eq:S1masters}
\end{equation}
where $I_0[N]$ denotes $I_0$ with numerator $N$ inserted.
We denote by $I_{r0}$ the corner integral of $S_2$, with denominators $D_9D_{11}D_{14}D_{16}D_{21}$. In the same notation for numerators,
$I_{r1}=I_{r0}[D_5]$, $I_{r2}=I_{r0}[D_{10}]$, $I_{r3}=I_{r0}[D_{12}]$, $I_{r4}$ has $D_{14}$ squared,
$I_{r5}$ has the uncut denominator $D_5$, $I_{r6}=I_{r5}[D_{12}]$ and $I_{r7}=I_{r5}[D_{13}]$. On $S_3$ we denote the corresponding integrals by $\tilde I_0,\dots,\tilde I_7$.

We relate these integrals by three changes of variables.
(i) The relabeling $k_1\leftrightarrow k_2$ maps the
corner sector of $S_1$ (denominators $D_9D_{13}D_{14}D_{15}D_{21}$) onto that
of $S_2$, and gives $I_{r0}=I_0$ and $I_{r3}=I_2$.

(ii) The reflection $k_i\to-k_i-q$ for all $i$ exchanges $D_{14+j}\leftrightarrow D_{18+j}$, fixes
$D_9,\dots,D_{14}$, and reverses every linearized propagator, since $u_i\!\cdot\! q=0$. It maps
$S_2$ onto $S_3$ row by row, with the sign $(-1)^{m}$ for $m$
linearized numerator powers and with the routing sign $\sigma$ of an uncut linearized denominator reversed. This gives $\tilde I_0=I_{r0}$, $\tilde I_2=I_{r2}$, $\tilde I_3=I_{r3}$, $\tilde I_4=I_{r4}$ and $\tilde I_1=-I_{r1}$.
Since $I_{r5}$, $I_{r6}$, $I_{r7}$ are odd in $\sigma$ for Feynman graviton propagators (Section~\ref{sec:ipifive}), it also gives
$\tilde I_{5,6,7}=I_{r5,6,7}$ at equal $\sigma$.
The map (ii) also reverses the momentum of every graviton line. The identities it gives therefore hold for graviton propagators that are even in their momenta, such as Feynman propagators, and the map turns a retarded propagator into an advanced one.
(iii) Within the corner sector of $S_1$ the map $k_1\to-k_3-q$, $k_2\to-k_4-q$, $k_3\to-k_1-q$,
$k_4\to-k_2-q$ together with $u_1\leftrightarrow u_2$ exchanges
$D_{15}\leftrightarrow D_{21}$, $D_9\leftrightarrow D_{14}$ and the
numerators $D_{11}\leftrightarrow D_{12}$. Combined with (i)
it gives $I_{r2}=I_0[D_{12}]=I_0[D_{11}]=I_3$.

Thus, with the identifications of the three corner integrals alone, the 31 rows involve at most 25 distinct integrals, and fewer with the reflection (ii).
Reducing the corner sectors with numerator integrals as seeds, we find $I_{r4}$ equal to
$I_0$ with $D_9$ squared, the image of $D_{14}$ under (iii).

\subsection{Boundary constants of PX2 \label{sec:ipifive}}
\label{sec:closedforms}

At one reference kinematic point,
\begin{equation}
\gamma \equiv u_1\cdot u_2 = \tfrac{5}{4}\,,\qquad q^2 = -1\,,\qquad
x = \gamma - \sqrt{\gamma^2-1} = \tfrac{1}{2}\,,
\label{eq:refpoint}
\end{equation}
we compute twelve of the at most 25 distinct master integrals on which the ten flow-free directions are supported. With the six relabeled corner rows and $\tilde I_4=I_{r4}$, the twelve cover 19 of the 31 supporting rows:
all three of $S_1$, eight of the twelve on $S_2$ ($I_{r0}$ through $I_{r7}$), and eight of the sixteen on $S_3$ (their counterparts $\tilde I_0,\dots,\tilde I_7$).
The others, $I_{r9}, I_{r10}, I_{r18}, I_{r19}$ and eight integrals of $S_3$, contain the uncut propagator $D_7$ or $D_8$, alone or together with
$D_5$, and we have not yet computed them.

The measure is that of Eq.~\eqref{eq:S1masters}, with
$d = 4-2\epsilon$ and no $e^{\gamma_E\epsilon}$ prefactor. Throughout this section the graviton propagators, written without an $i0$ in Eq.~\eqref{eq:px2props}, are Feynman propagators.
The finite parts of the corner integrals and the order-$\epsilon$ term of $I_{r4}$ contain $\ln 3$, $\pi^2$ and $\mathrm{Li}_2(\tfrac{1}{4})$. The other computed values (Table~\ref{tab:ipifive}) are exact and lie in the ring $i\pi^5\,\mathbb{Q}[\ln 2,\gamma_E]$, plus real $\pi^4$ terms where there is an ultraviolet pole.

In the two worldline frequencies
$a$ and $b$ that remain after the cuts, the one graviton propagator that joins the two worldlines, $(k_2-k_4)^2$, contains the indefinite form
$a^2+b^2-2ab\gamma$, which vanishes on two rays for $\gamma>1$ (the
radiation cone). Therefore the corner integrals are radiative,
two-frequency Bessel-$K$ integrals of the same type as the memory-region
integrals of Ref.~\cite{Driesse:2026qiz} that we evaluate in
Section~\ref{sec:memory}, with an ultraviolet pole from a factor
$\Gamma(4\epsilon-1)$. Having no uncut linearized propagator, they are integrals of the family reduced in that reference.
We reduce them to frequency integrals in Appendix~\ref{app:boundary}, together with the dotted and static-slice integrals below, and find for their pole coefficients and finite parts
\begingroup\allowdisplaybreaks
\begin{align}
I_0\big|_{\epsilon^{-1}} &= \pi^4\,\frac{80\ln 2 - 48}{27} + \frac{i\pi^5}{27}\,,\nonumber\\
I_2\big|_{\epsilon^{-1}} &= \pi^4\,\frac{150 - 220\ln 2}{27} - \frac{7\,i\pi^5}{864}\,,\nonumber\\
I_3\big|_{\epsilon^{-1}} &= \pi^4\,\frac{1128 - 1640\ln 2}{243} - \frac{i\pi^5}{972}\,,\nonumber\\
I_0\big|_{\epsilon^{0}} + 4\gamma_E\,I_0\big|_{\epsilon^{-1}} &= \pi^4\Big(- \frac{80}{3} + \frac{22}{9}\pi^2 + \frac{400}{27}\ln 2 - \frac{796}{27}\ln^2 2 - \frac{238}{27}\Lambda\Big)\nonumber\\
&\quad + i\pi^5\Big(\frac{41}{27} + \frac{976}{27}\ln 2 - \frac{638}{27}\ln 3\Big)\,,\nonumber\\
I_2\big|_{\epsilon^{0}} + 4\gamma_E\,I_2\big|_{\epsilon^{-1}} &= \pi^4\Big(\frac{3995}{54} - \frac{977}{144}\pi^2 - \frac{3985}{162}\ln 2 + \frac{17593}{216}\ln^2 2 + \frac{10553}{432}\Lambda\Big)\nonumber\\
&\quad + i\pi^5\Big(- \frac{20035}{5184} - \frac{5287}{54}\ln 2 + \frac{28153}{432}\ln 3\Big)\,,\nonumber\\
I_3\big|_{\epsilon^{0}} + 4\gamma_E\,I_3\big|_{\epsilon^{-1}} &= \pi^4\Big(\frac{14900}{243} - \frac{911}{162}\pi^2 - \frac{14140}{729}\ln 2 + \frac{16399}{243}\ln^2 2 + \frac{9839}{486}\Lambda\Big)\nonumber\\
&\quad + i\pi^5\Big(- \frac{18559}{5832} - \frac{19684}{243}\ln 2 + \frac{26239}{486}\ln 3\Big)\,,
\label{cornerpoles}
\end{align}
\endgroup
with $\Lambda=\mathrm{Li}_2(\tfrac{1}{4})-2\ln 2\,\ln 3$.
The pole coefficients, real parts included, are elementary integrals that we evaluate in exact arithmetic.

The dotted integral $I_{r4}=\tilde I_4$, the corner integral of $S_2$ with the propagator $(k_3-k_4)^2$ squared,
is the only computed master integral with a purely real ultraviolet
pole. We find
\begin{equation}
\begin{aligned}
I_{r4} &= -\frac{\pi^4}{\epsilon}
- \pi^4\Big(4 + \frac{20}{3}\ln 2 - 4\gamma_E\Big)
+ \frac{2}{3}\,i\pi^5\\
&\quad - \epsilon\,\pi^4\Big(28 + \frac{7}{3}\pi^2 + \frac{80}{3}\ln 2 - \frac{232}{3}\ln^2 2\\
&\qquad\qquad - \frac{64}{3}\Lambda - 16 \gamma_E - \frac{80}{3}\gamma_E\ln 2 + 8 \gamma_E^2\Big)\\
&\quad - \epsilon\,i\pi^5\Big(- \frac{8}{3} + \frac{232}{3}\ln 2 - \frac{164}{3}\ln 3 + \frac{8}{3}\gamma_E\Big) + O(\epsilon^2)\,.
\end{aligned}
\label{eq:ir4}
\end{equation}
Its pole arises at a static endpoint of the radial integration, off
the radiation cone, which first contributes at the finite order
through the $+\tfrac{2}{3}i\pi^5$ term.
We identify the finite parts of $I_0$, $I_2$, $I_3$ and the order-$\epsilon$ coefficient of $I_{r4}$ by an integer-relation search, as in Refs.~\cite{Driesse:2026qiz,Dlapa:2026oyq}, on evaluations at 330 and 220 digits. The relations hold for more than 100 digits beyond those used in the fit.
The overall sign of $I_{r4}$ relative to $I_0$ follows from the integration-by-parts differential equations of the corner integrals in $\gamma$.

The odd-parity master integrals $I_{r5}, I_{r6}, I_{r7}$ of $S_2$ and
their counterparts $\tilde I_5,\tilde I_6,\tilde I_7$ in $S_3$ contain
one uncut linearized worldline propagator $1/(2k\cdot u+i\sigma 0)$~\cite{Jakobsen:2022psy,Dlapa:2022lmu}. We split it into a
principal value and a delta function,
\begin{equation}
\frac{1}{b + i\sigma 0} \;=\; \mathrm{PV}\,\frac{1}{b} \;-\; i\pi\sigma\,\delta(b)\,.
\label{eq:pvdelta}
\end{equation}
A reflection of both frequency axes reverses $b$ and the frequency components of the momentum $k_1-k_4$ of the graviton line that joins the two worldlines, and leaves the rest of the integrand unchanged. With Feynman propagators, which depend only on the square of that momentum, the principal-value part therefore integrates to zero and only the $\delta(b)$ term survives. On its support, the static slice, the radiation cone is not crossed, so this term does not depend on the graviton $i0$ prescription. Thus the integrals are purely imaginary and linear in $\sigma$. We obtain them in closed $\Gamma$-function form to all
orders in $\epsilon$, without the $\Gamma(4\epsilon-1)$ pole of the corner
integrals.

The master integrals $I_{r1}$ of $S_2$ and $\tilde I_1$ of $S_3$
have a single odd worldline frequency as numerator and no linearized
denominator. Hence, for graviton propagators that are even in their momenta, they vanish under the same
reflection to all orders in $\epsilon$,
\begin{equation}
I_{r1} \;=\; \tilde I_1 \;=\; 0\,.
\label{eq:vanishing}
\end{equation}

\begin{table}[t]
\caption{Boundary constants of PX2 at the
reference point~\eqref{eq:refpoint}, with
Feynman propagators on the graviton lines. The finite parts of the corner integrals,
given in closed form in Eq.~\eqref{cornerpoles}, are quoted to ten figures, truncated. $I_{rk}$ and $\tilde I_k$
denote the master integrals in row $k$ of the response matrices of $S_2$ and $S_3$. The static-slice
integrals are quoted at $\sigma=+1$ (Eq.~\eqref{eq:pvdelta}).
\label{tab:ipifive}}
\centering
\begin{tabular}{lll}
\toprule
Master & $\epsilon^{-1}$ & $\epsilon^{0}$ \\
\midrule
\multicolumn{3}{l}{\emph{Radiative corner ($S_1$; also realized as $I_{r0},I_{r2},I_{r3}$ in $S_2$ and $\tilde I_0,\tilde I_2,\tilde I_3$ in $S_3$)}} \\
$I_0$ & $\pi^4(80\ln 2 - 48)/27 + i\pi^5/27$        & $\phantom{-}388.8416383\ldots + 161.9427314\ldots\,i$ \\
$I_2$ & $\pi^4(150 - 220\ln 2)/27 - 7\,i\pi^5/864$  & $-131.6922906\ldots - 35.15238209\ldots\,i$ \\
$I_3$ & $\pi^4(1128 - 1640\ln 2)/243 - i\pi^5/972$  & $-52.25287419\ldots - 4.256106616\ldots\,i$ \\
\midrule
\multicolumn{3}{l}{\emph{Static slice ($S_2$), no pole; an entry $c_0\,[\,c_1\,]$ denotes $(c_0+c_1\,\epsilon)\,i\pi^5+O(\epsilon^2)$}} \\[5pt]
$I_{r5}$ & $0$ & $-\dfrac{16}{3}\;\Big[-128 + \dfrac{160}{3}\ln 2 + \dfrac{64}{3}\gamma_E\Big]$ \\[9pt]
$I_{r6}$ & $0$ & $+\dfrac{128}{81}\;\Big[\dfrac{1024}{27} - \dfrac{1280}{81}\ln 2 - \dfrac{512}{81}\gamma_E\Big]$ \\[9pt]
$I_{r7}$ & $0$ & $+\dfrac{32}{81}\;\Big[\dfrac{736}{81} - \dfrac{320}{81}\ln 2 - \dfrac{128}{81}\gamma_E\Big]$ \\[4pt]
\midrule
\multicolumn{3}{l}{\emph{Vanishing master integrals (all orders in $\epsilon$, momentum-even graviton propagators)}} \\
$I_{r1}$        & $0$ & $0$ \\
$\tilde I_1$    & $0$ & $0$ \\
\bottomrule
\end{tabular}
\end{table}

We collect the boundary constants in Table~\ref{tab:ipifive}, having derived each of them and the pole and finite part in Eq.~\eqref{eq:ir4} at least twice. We know the values for
both signs of $\sigma$, but which sign is retarded depends on the worldline frequency
routing~\cite{Jakobsen:2022psy}, and we leave this choice open.
Retarded graviton propagators are Feynman propagators plus cut terms~\cite{Kalin:2022hph,Jakobsen:2022psy}. A cut term is not even in the graviton momentum, so Eq.~\eqref{eq:vanishing} and the vanishing of the principal-value parts do not extend to retarded propagators. We have not evaluated the cut integrals and give no in-in boundary constants.

\subsection{Relations among boundary constants \label{sec:nulllaw}}

The three corner integrals satisfy
\begin{equation}
I_0 + \tfrac{16}{3}\, I_2 - 6\, I_3 \,=\, 0
\label{eq:deltarelation}
\end{equation}
order by order in $\epsilon$: each component of the poles and of the finite parts in Eq.~\eqref{cornerpoles} cancels separately. The weights are the components of
$v^{(S_1)}$ in Eq.~\eqref{eq:rationalnulls}, fixed by the differential
system independently of the values. Equation~\eqref{eq:deltarelation} is
one instance of a general relation. Under three assumptions stated in Appendix~\ref{app:flow}, of which the physical one is the completeness of the enumerated boundary integrals (assumption~(ii), not proved here), every
flow-free direction $v$ satisfies
\begin{equation}
v^T A = 0 \;\;\Longrightarrow\;\; v^T M = 0
\label{eq:nulllaw}
\end{equation}
identically in $\epsilon$ where the rank of $A(\epsilon)$ is constant.

The ten relations $v^TM=0$ on $S_1$, $S_2$ and $S_3$ are reduction identities of the undeformed PX2 family, which we obtain from its integration-by-parts and Lorentz-invariance identities and its sector symmetries.
Each involves integrals with a numerator or a dot, which were not among the targets of our first-stage reduction. Relations found only in an extended reduction are known~\cite{Maierhofer:2018gpa}, including for these families~\cite{Driesse:2026qiz}. With these integrals as targets we obtain all ten relations with coefficients rational in $d$ and $\gamma$, and Eq.~\eqref{eq:deltarelation} for all $d$ and $\gamma$.
Therefore we check our boundary constants below against these identities and do not probe assumption~(ii).

\label{sec:nl-tests}
Seven of the ten flow-free directions are supported on computed master integrals: the direction of Eq.~\eqref{eq:deltarelation} in each of the three sectors (on $S_3$ a combination of two of the $\epsilon$-dependent vectors we extracted), and two directions of the parity-odd block on each of $S_2$ and $S_3$.
For these seven we find that Eq.~\eqref{eq:nulllaw} holds exactly: in rational arithmetic at the pole on $S_1$ and $S_2$, and with the closed forms of Eq.~\eqref{cornerpoles} at the finite order on all three sectors.
At $\epsilon = 0$ all left null
vectors we extracted have rational weights, the $\epsilon$-dependent
ones within the uncertainty of their extrapolation in $\epsilon$. Indeed, the weights are coefficients of the reduction identities at $d=4$ and $\gamma=5/4$, which we give for all ten directions in the ancillary file \texttt{boundary\_\allowbreak constants\_\allowbreak PX2.json}.

The relation~\eqref{eq:nulllaw} applies separately to each assignment of graviton propagators that is even in the graviton momenta
(Feynman, anti-Feynman or their average),
since these share the differential system and differ only in boundary data.

On $S_2$ the two directions of the parity-odd block are
supported on the vanishing $I_{r1}$ and on the pairs
$\{I_{r5}, I_{r6}\}$ and $\{I_{r5}, I_{r7}\}$. Given $I_{r5}$, the
relation~\eqref{eq:nulllaw} fixes the other two,
\begin{equation}
\begin{aligned}
I_{r6}\big|_{\epsilon^0} &= -\tfrac{8}{27}\, I_{r5}\big|_{\epsilon^0}
  = \tfrac{128}{81}\, i\pi^5\,,\\
I_{r7}\big|_{\epsilon^0} &= -\tfrac{2}{27}\, I_{r5}\big|_{\epsilon^0}
  = \tfrac{32}{81}\, i\pi^5\,,
\end{aligned}
\label{eq:oddpred}
\end{equation}
and, applied at fixed $\epsilon$, their order-$\epsilon$ terms, in
agreement with the static-slice values of Table~\ref{tab:ipifive}.
The two directions
occur on $S_3$ with identical weights, and there $\tilde I_5$, $\tilde I_6$, $\tilde I_7$ agree through order $\epsilon$ with their $S_2$ counterparts. The contractions of these two directions
with $M$ change only by an overall sign when $\sigma$ is
flipped and vanish term by term for a principal-value worldline propagator, so they do not distinguish the retarded from the advanced assignment of
$\sigma$.

In the remaining $\epsilon$-dependent direction of $S_2$ the relation~\eqref{eq:nulllaw} fixes, at the
pole, a combination of the four $S_2$ integrals we have not computed,
\begin{equation}
\begin{aligned}
\tfrac{1}{2} I_{r9}^{(-1)} &+ \tfrac{5}{12} I_{r10}^{(-1)}
 + \tfrac{1}{8} I_{r18}^{(-1)} + I_{r19}^{(-1)}\\
 &\,=\, \tfrac{5}{12}\, I_{r0}^{(-1)}
 \,=\, \pi^4\, \frac{100\ln 2 - 60}{81} + \frac{5}{324}\, i\pi^5\,.
\end{aligned}
\label{eq:pvpred}
\end{equation}
Similarly, the two remaining directions of $S_3$, both in its even block,
fix combinations of that sector's eight uncomputed master integrals.

\subsection{K3 and Calabi--Yau sectors \label{sec:geometry}}

The K3 and Calabi--Yau sectors at
5PM-2SF~\cite{Frellesvig:2023bbf,Klemm:2024wtd,Frellesvig:2024rea,Duhr:2025lbz,Brammer:2025rqo} are
subtopologies with six or seven graviton propagators, whereas the
leading top sectors of dissipative PX2 have five
(Table~\ref{tab:flowrank}). Since a period can enter a master integral only
through a sector contained in its top sector, no K3 or
Calabi--Yau period appears in any of the 37 master integrals of the
leading top sectors. Hence any such period of
PX2 must enter through one of its 36 further top
sectors (Table~\ref{tab:families}), and we have not determined which of them couple to the dissipative master integrals.
Brammer et al.\ predict that the 4PM K3 reappears in the dissipative sector~\cite{Brammer:2025rqo}.

The planar family P contains the Ap\'ery K3 sector, topology~40 of Brammer et al.
On its maximal cut we reduce the even-parity block with integration-by-parts and Lorentz-invariance identities~\cite{Tkachov:1981wb,Chetyrkin:1981qh,Gehrmann:1999as} to eleven master integrals. When we also impose the sector's worldline-exchange symmetry we obtain nine, the number found in Ref.~\cite{Duhr:2025lbz}.
At one kinematic point and two numerical values of $d$, all nine elements of that reference's starting basis reduce to linearly independent combinations of our nine master integrals (Eqs.~\eqref{eq:S5duhr8} and~\eqref{eq:S5duhr9}). However, we have not completed the symbolic reduction that would give the sector's differential equation.

\section{Continuation to bound kinematics}
\label{sec:boundorbits}
\label{app:boundperiods}
\label{sec:S3}

As solutions of the Picard--Fuchs equations \eqref{eq:PFlegendre}--\eqref{eq:PFhadamard}, the K3 and Calabi--Yau periods of the 4PM and 5PM
integrals~\cite{Brammer:2025rqo,Driesse:2024feo,Klemm:2024wtd} extend analytically from $\gamma>1$ to bound kinematics, $\gamma<1$.
Driesse et al.\ give these periods at threshold as power series~\cite{Driesse:2024feo,Driesse:2026qiz}, which converge on part of the bound domain, and leave the transition to the bound case pending.
Here we continue the full Frobenius bases of the four operators at $z=0$ to $\gamma<1$, evaluate them with rigorous
enclosures at twelve bound-orbit energies, and compute the monodromy matrices that relate continuations along
inequivalent paths. We also determine the local exponents of the Ap\'ery sector's differential system at threshold.
To our knowledge the full Frobenius bases of the Ap\'ery K3$'$ and ${}_4F_3$ geometries have not been evaluated at bound kinematics before.

\subsection{Bound kinematics and Frobenius frame}
\label{sec:bound-kinematics}
\label{sec:S3ops}
\label{sec:S3frame}

For bound motion we write
\begin{equation}
\gamma = \cos\theta \in (0,1)\,,\qquad x = e^{\mp i\theta}\,,
\label{eq:boundmap}
\end{equation}
and bound orbits map to the unit circle $|x|=1$. We fix the branch by the Feynman prescription $s+i0$ on the
Mandelstam invariant $s$, a prescription distinct from the $i0^+$ routing of the
memory gravitons. Since $s+i0$ implies $\gamma + i0$, and $d\gamma/dx < 0$ for $0<x<1$, we continue
along the lower arc $x = e^{-i\theta}$, $\theta \in (0, \pi/2]$, which we call the physical branch.
Because the four operators have real coefficients, the periods on the upper arc take the
Schwarz-conjugate values.

On the arc we use a separate variable for each geometry,
\begin{equation}
z_{\rm K3} = z_{{\rm K3}'} = x^2\,,\qquad
z_{{\rm CY}_3} = \frac{x^4}{2^{8}}\,,\qquad
z_{{\rm CY}_3'} = \frac{1-\gamma^2}{2^{10}}\,,
\label{eq:zmaps}
\end{equation}
the last from $\gamma=(x+1/x)/2$.
The 4PM operator $L_{\rm K3}$ of Eq.~\eqref{eq:PFlegendre} has indicial exponent $\tfrac12$ at $z=0$, so we work with the conjugated operator
$\tilde L_{\rm K3}=\tfrac18\,z^{-1/2}L_{\rm K3}\,z^{1/2}=\theta^3-z\big(2\theta^3+3\theta^2+2\theta+\tfrac12\big)+z^2(\theta+1)^3$,
whose holomorphic solution at $z=0$ is $\varpi_0={}_2F_1(\tfrac12,\tfrac12;1;z)^2=(2/\pi)^2K(z)^2$.
With $\tilde L_{\rm K3}$ in place of $L_{\rm K3}$, all four operators have a MUM point at $z=0$ with indicial polynomial $\theta^{r}$, where $r$ is the order. We fix the Frobenius basis there
as
\begin{equation}
\varpi_k(z)=\sum_{j=0}^{k}\frac{\ln^{j}\!z}{j!}\;h_{k-j}(z)\,,\qquad k=0,\dots,r-1,\qquad
h_0(0)=1\,,\quad h_j(0)=0\ \ (j\ge1)\,,
\label{eq:S3frob}
\end{equation}
with power series $h_j$. The logarithm is real on the
positive real axis inside the disk of convergence, and we continue it from there along each integration
path. The local monodromy about $z=0$ acts by $\ln z\to\ln z+2\pi i$, and its inverse is the matrix $M_0^{-1}$ of Eq.~\eqref{eq:M0inv}.

\subsection{Singular points on the bound domains}
\label{sec:bound-singularities}
\label{sec:bound-otherarcs}

\begin{figure}[t]
\centering
\includegraphics[width=\linewidth]{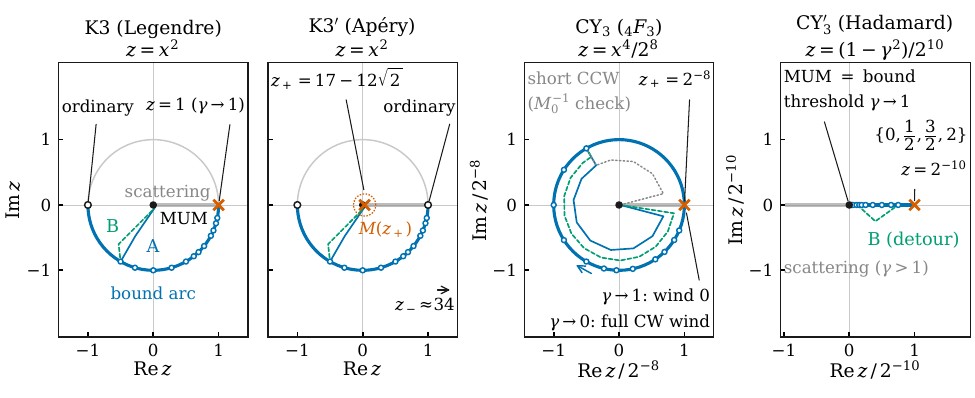}
\caption{Of the four bound domains, only the Ap\'ery K3$'$ arc has no singular point
on it or at its endpoints. Each panel shows one geometry of Table~\ref{tab:four-geometry} in its variable $z$ of
Eq.~\eqref{eq:zmaps}: singular points (crosses), the MUM point $z=0$ (dot), the scattering
image $\gamma>1$ (real segment), the bound arc on the physical branch
with the grid points of Eq.~\eqref{eq:boundgrid} (open circles), and the two
continuation paths from the MUM point (A, solid, and B, dashed). The short
counterclockwise path in the CY$_3$ panel is the second
winding used for Eq.~\eqref{eq:M0inv}, and the dotted loop in the K3$'$ panel
has monodromy $M(z_+)$ of
Eq.~\eqref{eq:Mzplus}.}
\label{fig:bound-zplane}
\end{figure}

The finite singular points of each operator are the zeros of its leading symbol $s_r(z)$,
where $p_r(z) = z^r s_r(z)$ is the leading coefficient of the order-$r$ operator,
\begin{equation}
s_3^{\rm K3}(z) = (1-z)^2,\quad
s_3^{{\rm K3}'}(z) = 1 - 34z + z^2,\quad
s_4^{{\rm CY}_3}(z) = 1 - 256 z,\quad
s_4^{{\rm CY}_3'}(z) = (1 - 1024 z)^3,
\label{eq:symbols}
\end{equation}
and we write $z_+$ for the zero of each symbol nearest the MUM point $z = 0$ (Fig.~\ref{fig:bound-zplane}).
The Ap\'ery symbol has roots $z_\pm = 17 \mp 12\sqrt{2}$, with
$z_+ \approx 0.0294$, the image of $\gamma=3$, inside the unit circle and
$z_-$ outside. Hence the K3$'$ arc and both its endpoints are ordinary points.
The point $\gamma=3$ lies outside the bound domain
$\gamma<1$~\cite{Dlapa:2026oyq} and enters the continuation only through
the monodromy around $z_+$.
In contrast to the K3$'$ arc, each of the other three bound domains has a singular point at an endpoint.

For the 4PM K3 sector~\cite{Dlapa:2022wdu,Dlapa:2023hsl,Klemm:2024wtd} the arc $z = e^{-2i\theta}$ is
the lower half of the unit circle $|z|=1$. It meets the singular point $z = 1$ only at the weak-binding endpoint
$\gamma \to 1$, with the logarithmic threshold behavior
\begin{equation}
\varpi_0 = \Bigl(\frac{2}{\pi}\Bigr)^{2} K(z)^2
\;\sim\; \frac{1}{\pi^{2}}\, \ln^2\!\frac{16}{1-z}
\qquad (\gamma \to 1^-)\,.
\label{eq:legendre-threshold}
\end{equation}
The 4PM results~\cite{Dlapa:2021npj}, applied at bound-orbit energies~\cite{Khalil:2022ylj,Dlapa:2024cje,Buonanno:2024byg,Bini:2024tft}, contain $K\bigl(\tfrac{\gamma-1}{\gamma+1}\bigr)=\tfrac{1+x}{2}\,K(1-z)$, regular at threshold and real for $\gamma<1$.
The open arc and the $\gamma = 0$ endpoint $z = -1$ are ordinary.

The ${}_4F_3$ arc $z = 2^{-8} e^{-4i\theta}$ is the circle $|z| = 2^{-8}$ through the
singular point $z_+$, traversed once clockwise as $\theta$ runs over $(0,\pi/2]$. Only its endpoints are
singular, both at $z_+=2^{-8}$ but on different windings: $\gamma = 1$ at winding zero and
$\gamma = 0$ after one full clockwise turn, on the sheet shifted by the inverse MUM monodromy
$M_0^{-1}$.

The bound domain $z \in (0, 2^{-10})$ of the Hadamard threefold CY$_3'$ is a real segment of
ordinary points on which the Frobenius series at the MUM point
converge and are real. Its endpoints are the MUM point ($\gamma = 1$) and the triple
root $z = 2^{-10}$ of the symbol ($\gamma = 0$), where the local exponents are
$\{0, \tfrac12, \tfrac32, 2\}$ instead of the conifold pattern
$\{0,1,1,2\}$ and the monodromy squares to the identity~\cite{Klemm:2024wtd}.

\subsection{Periods on the bound arc}
\label{sec:bound-atlas}
\label{sec:bound-monodromy}
\label{sec:S3paths}
\label{sec:S3files}
\label{sec:S3mono}

We evaluate the Frobenius basis of each operator at the bound-orbit energies
\begin{equation}
\gamma \in \Bigl\{\tfrac{199}{200},\ \tfrac{99}{100},\ \tfrac{49}{50},\
\tfrac{97}{100},\ \sqrt{\tfrac{8}{9}},\ \tfrac{23}{25},\ \tfrac{9}{10},\
\tfrac{\sqrt3}{2},\ \tfrac45,\ \tfrac{\sqrt2}{2},\ \tfrac35,\ \tfrac12\Bigr\}.
\label{eq:boundgrid}
\end{equation}
The grid is densest at the weak-binding end and includes $\gamma = \sqrt{8/9} \approx 0.9428$, the
specific energy of the Schwarzschild innermost stable circular orbit.
We sum each Frobenius basis at a base point inside the disk of
convergence and continue it to each grid point by integrating the Picard--Fuchs equation in ball arithmetic~\cite{Johansson:2017arb} along two homotopic polygonal paths, A and B in Fig.~\ref{fig:bound-zplane}.
Each integration step has a rigorous remainder bound, and every value has an enclosure radius below $10^{-167}$.

The ancillary files \texttt{bound\_frobenius/values\_*.txt} contain, for each geometry and each energy (48 points in all), the full Frobenius matrix
\begin{equation}
W_{ik}(z)=\frac{d^{\,i-1}}{dz^{\,i-1}}\,\varpi_{k-1}(z)\,,\qquad i,k=1,\dots,r\,,
\label{eq:S3W}
\end{equation}
on the physical branch, in the frame~\eqref{eq:S3frob}
(for K3 the matrix of $\tilde L_{\rm K3}$, whose $W_{11}=(2/\pi)^2K(z)^2$).
We give the paths, the operators in $\theta$ and in $d/dz$ form, and the enclosure radius at each point in \texttt{operators.txt} and in \texttt{bound\_frobenius/}.
For example, the holomorphic period of the Ap\'ery K3$'$ at $\gamma=\sqrt{8/9}$ is $\varpi_0=0.3805554073\,-\,0.3468311375\,i$ to ten decimal places.

The values of $\varpi_0$ for CY$_3'$ are real, as is the CY$_3$ value at
$\gamma=\sqrt2/2$ ($z=-2^{-8}$). The CY$_3$ values at $\gamma$ and $\sqrt{1-\gamma^2}$,
energies that map to $z$ and $\bar z$ on the circle $|z|=2^{-8}$ one turn apart, are complex conjugates because $\varpi_0$ is invariant under the monodromy $M_0$ about the MUM point.
However, the Ap\'ery K3$'$ values are nowhere real on the arc and form no complex-conjugate pairs.

The K3$'$ period $\hat\varpi$ of Ref.~\cite{Driesse:2026qiz} is normalized at threshold by the velocity expansion $(1+x^2)\,\hat\varpi=1+\tfrac{15}{32}v^2+\tfrac{351}{1024}v^4+O(v^6)$ with $v=\sqrt{\gamma^2-1}/\gamma$, and the threefold periods of Ref.~\cite{Driesse:2024feo} are given at $x=1$ as power series in $1-x$.
Since the nearest singular points are $v^2=\tfrac89$ and $x=0$, these series converge at least for $|v^2|<\tfrac89$ and for $|1-x|<1$, which on the bound side is $\gamma>3/\sqrt{17}\approx0.73$ and $\gamma>\tfrac12$.
On the physical branch we express $\hat\varpi$, for which $(1+x^2)\hat\varpi$ is real for $\gamma<1$, in the Frobenius frame~\eqref{eq:S3frob} of $L_A$,
\begin{equation}
\begin{gathered}
\hat\varpi = \hat c_0\,\varpi_0 + \hat c_1\,\varpi_1 + \hat c_2\,\varpi_2\,,\qquad \hat c_0 = 0.3888737022 - 0.5499504638\,i\,,\\
\hat c_1 = -0.5251640086\,,\qquad \hat c_2 = 0.1182034314 + 0.1671648958\,i\,.
\end{gathered}
\label{eq:thresholdframe}
\end{equation}
The ancillary file \texttt{tables/threshold\_frame\_K3prime.txt} gives the three constants to thirty decimal places.

The physical K3$'$ path runs from the scattering segment $z \in (z_+, 1)$ through the lower
half plane ($z - i0$) onto the arc and encloses no singular point. Any other path inside
$|z|<z_-$ differs from it by a word in the unipotent MUM monodromy
$M_0$ and the monodromy around $z_+$,
\begin{equation}
M(z_+) \;=\;
\begin{pmatrix} 0 & 0 & \pi^2/3 \\ 0 & 1 & 0 \\ 3/\pi^2 & 0 & 0 \end{pmatrix},
\qquad M(z_+)^2 = \openone\,.
\label{eq:Mzplus}
\end{equation}
We continue the K3$'$ basis to $z = e^{-i\pi/3}$ ($\gamma = \sqrt3/2$) once through each half plane.
The upper path crosses the real axis between $z_+$ and $z_-$, and the two paths therefore differ by a loop around $z_+$ only. The quotient of the two bases equals the
Picard--Lefschetz reflection~\eqref{eq:Mzplus} entrywise (local exponents $0,\tfrac12,1$ at $z_+$).

On the ${}_4F_3$ arc we integrate to $\gamma = 1/2$ once with the physical clockwise winding
and once along the short counterclockwise path of Fig.~\ref{fig:bound-zplane}, which reaches the same $z$ with one winding fewer around $z=0$. The quotient of the two bases is
the inverse MUM monodromy, with Frobenius-frame entries
\begin{equation}
\bigl(M_0^{-1}\bigr)_{jk} \;=\; \frac{(-2\pi i)^{\,k-j}}{(k-j)!}\,,
\qquad k \ge j\,,
\label{eq:M0inv}
\end{equation}
and zero below the diagonal.

\subsection{Local exponents at threshold}
\label{sec:S3threshold}

At the threshold $x=1$ the periods of the Ap\'ery K3$'$ are regular.
However, for $\epsilon\neq0$ the differential system of Ref.~\cite{Duhr:2025lbz} for the nine-master Ap\'ery sector~\cite{Klemm:2024wtd} on its maximal cut, with the subsector rows not included, is singular there, and we determine its local exponents.
The nine-element basis $I_1,\dots,I_9$ of that reference obeys
$\partial_x I=\Omega(x,\epsilon)\,I$, with the matrix $\Omega$ given in its ancillary files. In this basis the integrals $I_2$ and $I_3$, which differ from the corner integrals of the same names in Section~\ref{sec:ipifive},
are the first and second derivatives of $I_1$.

For $\epsilon\neq0$ the entries of $\Omega$ have poles of order up to three at $x=1$.
With $t=x-1$ and $J=\mathrm{diag}(1,t,t^2,1,t,t,1,1,1)\,I$ we bring the system to the form $t\,\partial_tJ=B(t)\,J$ with $B$
regular at $t=0$, and $x=1$ is therefore a regular singular point.
We compute the eigenvalues of $B(0)$, which are the local exponents in this frame, and find
\begin{equation}
1\,,\qquad -6\epsilon\,,\qquad 2+10\epsilon\,,\qquad 1-4\epsilon\ \ (\text{twice})\,,\qquad
-4\epsilon\ \ (\text{twice})\,,\qquad 10\epsilon\ \ (\text{twice})\,,
\label{eq:S3thresholdexp}
\end{equation}
of which one $-4\epsilon$ and one $10\epsilon$ belong to the block of $I_8$ and $I_9$.

Since a rational change of basis normalization shifts the exponents by integers, we give them modulo integers: they are $0$,
$-6\epsilon$, $10\epsilon$ and $-4\epsilon$, with multiplicities 1, 1, 3 and 4.
At three rational values of $\epsilon$ we find for each class of exponents as many independent solutions of the form $t^{\rho}$ times a
power series as its multiplicity, and we expect no logarithm of $t$ for generic $\epsilon$.
At $x=-1$ we find the same local exponents.

\subsection{Bound-orbit observables}
\label{sec:bound-scope}

A bound-orbit observable requires, beyond the continued periods, a map from scattering data to bound dynamics.
For the local-in-time conservative sector such a map is the boundary-to-bound (B2B)
dictionary of K\"alin and Porto~\cite{Kalin:2019rwq,Kalin:2019inp}. It holds for all
local-in-time contributions and generic orbits~\cite{Cho:2021arx} and has been applied at
4PM~\cite{Dlapa:2021npj,Dlapa:2021vgp,Dlapa:2024cje} and 5PM-1SF~\cite{Dlapa:2025biy}.
Of the periods that we continue, only the Ap\'ery K3$'$ remains in the conservative 5PM-2SF result~\cite{Driesse:2026qiz,Bern:2026oqp}.

Nonlocal-in-time (tail)
terms can be mapped only at large eccentricity and, logarithms aside, do not continue smoothly between
hyperbolic and elliptic motion~\cite{Cho:2021arx,Dlapa:2021vgp}. Moreover, Kankani and McWilliams find with numerical relativity that the B2B relations
do not hold non-perturbatively~\cite{Kankani:2024may}.

At 4PM the dissipative observables contain no tail logarithms and, as Jakobsen et al.\ note,
it may then be possible to map the fluxes directly to bound orbits~\cite{Jakobsen:2023hig}. The flux, reconstructed from scattering data, is the natural object for generic bound orbits,
since the per-period losses of the adiabatic map vanish at even PM orders~\cite{Bini:2020hmy,Cho:2021arx,Dlapa:2023hsl}.
The $\gamma$-dependent coefficient functions of a flux are unchanged by the adiabatic
map and need only be evaluated at $\gamma < 1$, where we give the
Frobenius bases on the branch fixed by $s+i0$.

\section{Evaluation of the memory-region integrals \label{sec:memory}}

We evaluate the memory-region boundary integrals
$I_1^{(M)}$ and $I_2^{(M)}$, Eq.~\eqref{memmasters}, at the static point
and extract the constant $c_M$, which is not fixed by the cancellation conditions
underlying Eq.~\eqref{cMdef}. The two memory
gravitons are routed as in the $\gamma$-3 prescription of
Ref.~\cite{Driesse:2026qiz}, which we take as input, together
with the frequency representations and the integration-by-parts
identity of its supplement.
We derive the Laurent coefficients of $I_1^{(M)}$ and $I_3^{(M)}$ (Eq.~\eqref{I3rep})
in closed form through order $\epsilon^1$, Eq.~\eqref{i1i3closed}.

\subsection{Frequency representations
\label{sec:memlayer}}

At the static point $x=1$, $I_1^{(M)}$ and an auxiliary integral $I_3^{(M)}$ admit
two-fold frequency representations~\cite{Driesse:2026qiz}.
Since $c_M$ is defined at $x=1$ and the Ap\'ery K3 periods enter only the $x$ dependence of
the master integrals, $c_M$ contains no K3 period.
These boundary integrals, and the coefficients of the
identity~\eqref{ibp}, depend on $\gamma$ only through overall
powers of $\sqrt{\gamma^2-1}$, which we remove by
evaluating at $\gamma = \sqrt{2}$, where these powers equal one.

We write
$\omega_3 = \omega_1 + \omega_2$ and
$\int_\omega = \int_{-\infty}^{\infty} d\omega/(2\pi)$.
With $K_\epsilon$ the modified Bessel function of the second kind, the
first integral reads\footnote{The supplement of Ref.~\cite{Driesse:2026qiz} (arXiv version) prints the Bessel factor in its Eqs.~(27)--(28)
as $|\omega_i|^{+\epsilon} K_\epsilon(|\omega_i|)$. The exponent
$-\epsilon$ follows from their Eq.~(26) at unit impact parameter, and only
with it does the $\epsilon^{-3}$ pole of $I_2^{(M)}$ cancel.}
\begin{align}
I_1^{(M)} &= N_1(\epsilon)
\int_{\omega_1\omega_2}
\prod_{j=1,2}(0^+\!-i\omega_j)^{1-2\epsilon}
\prod_{i=1}^{3}|\omega_i|^{-\epsilon} K_\epsilon(|\omega_i|)\,,
\nonumber\\
N_1(\epsilon) &= -\,2^{7\epsilon-1}\,
\frac{\Gamma^2\!\big(\tfrac{2\epsilon-1}{2}\big)\,\Gamma(4\epsilon-1)}
     {\Gamma(2-5\epsilon)\,(4\pi)^{5-4\epsilon}}\,,
\label{I1rep}
\end{align}
and $I_3^{(M)}$, obtained from $I_2^{(M)}$ by
inserting the numerator $-(\ell_3\cdot v_1)^3(\ell_4\cdot v_1)^3$, is given by
\begin{align}
I_3^{(M)} &= N_3(\epsilon)
\int_{\omega_1\omega_2}
\omega_1^3\,\omega_2^3
\prod_{j=1,2}(0^+\!-i\omega_j)^{-1-2\epsilon}
\prod_{i=1}^{3}|\omega_i|^{-\epsilon} K_\epsilon(|\omega_i|)\,,
\nonumber\\
N_3(\epsilon) &= 2^{7\epsilon-3}\,
\frac{\Gamma^2\!\big(\tfrac{2\epsilon+1}{2}\big)\,\Gamma(4\epsilon-2)}
     {\Gamma(3-5\epsilon)\,(4\pi)^{5-4\epsilon}}\,.
\label{I3rep}
\end{align}

The integration-by-parts identity of the supplement, whose
derivation uses propagator symmetries respected by both the $\gamma$-3
prescription and Feynman propagators, expresses
$I_2^{(M)}$ in terms of $I_1^{(M)}$ and $I_3^{(M)}$,
\begin{equation}
I_2^{(M)} = r_1(\epsilon)\, I_3^{(M)} + r_2(\epsilon)\, I_1^{(M)}\,,
\label{ibp}
\end{equation}
with
\begin{align}
r_1(\epsilon) &=
\frac{28\,(5\epsilon-2)(14\epsilon-9)\,
      \big(1088\epsilon^4-898\epsilon^3+275\epsilon^2-38\epsilon+2\big)}
     {3\,\epsilon^3\,\big(96\epsilon^3-116\epsilon^2+43\epsilon-5\big)}\,,
\nonumber\\
r_2(\epsilon) &=
\frac{(1-2\epsilon)^2\,
      \big(121024\epsilon^6-237340\epsilon^5+181552\epsilon^4
           -70363\epsilon^3+14767\epsilon^2-1612\epsilon+72\big)}
     {6\,\epsilon^3\,\big(96\epsilon^3-116\epsilon^2+43\epsilon-5\big)}\,.
\label{ibpcoeffs}
\end{align}
Both coefficients scale as $\epsilon^{-3}$, while $I_1^{(M)}$ and
$I_3^{(M)}$ start at order $\epsilon^{-1}$.
Hence the $\epsilon^{-4}$ and $\epsilon^{-3}$ poles must cancel in
$r_1I_3^{(M)}+r_2I_1^{(M)}$ to leave the $\epsilon^{-2}$ pole of
Eq.~\eqref{cMdef}, and to extract $c_M$ we need $I_1^{(M)}$
and $I_3^{(M)}$ through order $\epsilon^{1}$.

The $i0^+$ prescription enters through each frequency factor
$(0^+ - i s\,\omega)^{a - q\epsilon}$, with integers $a$ and $q$ and a sign $s=\pm1$, and it affects the integer
power $a$ as well as the power $-q\epsilon$. These factors arise when the spatial graviton momentum is integrated out at fixed
worldline frequency. For a retarded propagator raised to the power $n$ the integration gives, in $3-2\epsilon$ spatial
dimensions, a factor proportional to $[-(\omega+i0)^2]^{3/2-n-\epsilon}=(0^+\!-i\omega)^{3-2n-2\epsilon}$,
in which the whole power has a single prescription~\cite{Driesse:2026qiz}. For a retarded ($s=+1$) or advanced ($s=-1$) propagator
\begin{equation}
\ln(0^+ - i s\,\omega) \;=\; \ln|\omega| - \frac{i\pi}{2}\, s\,
\mathrm{sign}(\omega)\,,
\label{phases}
\end{equation}
whereas for a Feynman propagator
$1/(\omega^2-\mathbf{k}^2+i0)^n$ the same integral has $-\omega^2-i0=(0^+\!-i|\omega|)^2$ in place of
$-(\omega+i0)^2$ and gives $(0^+ - i|\omega|)^{a - q\epsilon}$,
with logarithm $\ln|\omega| - i\pi/2$ for either sign of
$\omega$.
In Eqs.~\eqref{I1rep} and~\eqref{I3rep} both memory gravitons have
retarded propagators, $s_1=s_2=+1$, and the $\gamma$-3 prescription averages
this routing with the all-advanced one.
Since the substitution $(\omega_1,\omega_2)\to(-\omega_1,-\omega_2)$ maps each
retarded factor to the advanced one and leaves the rest of the integrand
unchanged, the two routings and their average give the same values.

\subsection{\texorpdfstring{Closed forms of $I_1^{(M)}$ and $I_3^{(M)}$}{Closed forms of I1(M) and I3(M)}}
\label{sec:S4}
\label{sec:S4analytic}

The three Bessel factors of Eqs.~\eqref{I1rep} and~\eqref{I3rep} are coupled through
$\omega_3 = \omega_1+\omega_2$. Following Refs.~\cite{Driesse:2026qiz,Driesse:2024feo}, we Fourier transform to the time domain, where the frequency integrals reduce to one-dimensional moments.
For two functions $A$ and $C$ of one frequency, with Fourier transforms
$a(k)=\int_\omega e^{ik\omega}A(\omega)$ and $c(k)=\int_\omega e^{ik\omega}C(\omega)$, we use
\begin{equation}
\int_{\omega_1\omega_2}A(\omega_1)\,A(\omega_2)\,C(\omega_1+\omega_2)=\int_{-\infty}^{\infty}\!dk\;c(k)\,a(-k)^2\,,
\label{eq:S4parseval}
\end{equation}
which replaces the two coupled frequency integrals by one integral over the conjugate variable $k$, a worldline time. The line
$\omega_3$ has the factor $C=|\omega|^{-\epsilon}K_\epsilon(|\omega|)$, with transform
$g(k)=G\,(1+k^2)^{\epsilon-1/2}$ and $G=2^{-\epsilon-1}\Gamma(\tfrac12-\epsilon)/\sqrt{\pi}$ by Basset's integral~\cite[Eq.~10.32.11]{NIST:DLMF}.

Under the retarded routing both lines $\omega_1$, $\omega_2$ have the factor
$|\omega|^{-3\epsilon}K_\epsilon(|\omega|)\,e^{i\pi\epsilon\,{\rm sign}\,\omega}$, multiplied by $-i\omega$ in
$I_1^{(M)}$ and by $i\omega^2$ in $I_3^{(M)}$. The transform of this factor is
$f_R=\cos(\pi\epsilon)\,C_\epsilon-\sin(\pi\epsilon)\,S_\epsilon$, with
\begin{equation}
C_\epsilon(k)=\frac1\pi\int_0^\infty\!\omega^{-3\epsilon}K_\epsilon(\omega)\cos(k\omega)\,d\omega\,,\qquad
S_\epsilon(k)=\frac1\pi\int_0^\infty\!\omega^{-3\epsilon}K_\epsilon(\omega)\sin(k\omega)\,d\omega\,,
\label{eq:S4CS}
\end{equation}
and the powers of $\omega$ act as derivatives in $k$. Since the products of $C_\epsilon$ and $S_\epsilon$ are odd in
$k$ and integrate to zero, we obtain
\begin{equation}
\begin{aligned}
I_1^{(M)}&=N_1\!\int\!g\,(f_R')^2\,dk=N_1\big[\cos^2(\pi\epsilon)\,E_1+\sin^2(\pi\epsilon)\,O_1\big]\,,\\
I_3^{(M)}&=-N_3\!\int\!g\,(f_R'')^2\,dk=-N_3\big[\cos^2(\pi\epsilon)\,E_3+\sin^2(\pi\epsilon)\,O_3\big]\,,
\end{aligned}
\label{eq:S4EO}
\end{equation}
where $E_1=\int g\,(C_\epsilon')^2dk$, $O_1=\int g\,(S_\epsilon')^2dk$, and $E_3$, $O_3$ are the same integrals
with second derivatives. For Feynman propagators each frequency factor is $(0^+\!-i|\omega|)^{a-2\epsilon}=(-i)^a e^{i\pi\epsilon}|\omega|^{a-2\epsilon}$. Hence
only the odd transform $S_\epsilon$ enters, which gives
$I_{1,F}^{(M)}=-e^{2i\pi\epsilon}N_1O_1$ and $I_{3,F}^{(M)}=e^{2i\pi\epsilon}N_3O_3$.

Bessel's equation for $\omega^{-3\epsilon}K_\epsilon(\omega)$ implies that $C_\epsilon$ and $S_\epsilon$ both solve
\begin{equation}
(1+k^2)\,f''+(3-6\epsilon)\,k\,f'+(1-2\epsilon)(1-4\epsilon)\,f=0\,,
\label{eq:S4ode}
\end{equation}
with $C_\epsilon'(0)=0$, $S_\epsilon(0)=0$ and, from the Mellin transform of $K_\epsilon$~\cite[Eq.~10.43.19]{NIST:DLMF},
$C_\epsilon(0)=2^{-3\epsilon-1}\Gamma(\tfrac12-\epsilon)\Gamma(\tfrac12-2\epsilon)/\pi$ and
$S_\epsilon'(0)=2^{-3\epsilon}\Gamma(1-\epsilon)\Gamma(1-2\epsilon)/\pi$.
We set $k=\sinh t$ and $f=f_0\cosh^{-1+3\epsilon}t\;y(t)$, with $f_0=C_\epsilon(0)$ for the even solution $y_E$ and
$f_0=S_\epsilon'(0)$ for the odd one, $y_O$. Then Eq.~\eqref{eq:S4ode} becomes
$y''=[\epsilon^2+(3\epsilon-9\epsilon^2)\operatorname{sech}^2t]\,y$,
with $y_E(0)=1$, $y_E'(0)=0$ and $y_O(0)=0$, $y_O'(0)=1$, which we solve order by order in $\epsilon$. With $L=\ln\cosh t$ we find
\begin{equation}
y_E=1+3\epsilon L+\epsilon^2\big(\tfrac92L^2-4t^2\big)+O(\epsilon^3)\,,\qquad
y_O=t+\epsilon\,y_1+\epsilon^2y_2+O(\epsilon^3)\,,
\label{eq:S4pert}
\end{equation}
where $y_1'=3\,(t\tanh t-L)$, $y_2''=3\operatorname{sech}^2t\;y_1+(1-9\operatorname{sech}^2t)\,t$ and
$y_1(0)=y_2(0)=y_2'(0)=0$.

Integrating by parts with Eq.~\eqref{eq:S4ode}, we remove the derivatives of $f$ and, with
$z=\operatorname{sech}^2t$, arrive at
\begin{equation}
\int g\,(f')^2dk=G f_0^2\!\int_{-\infty}^{\infty}\!e^{8\epsilon L}B_1\,y^2\,dt\,,\qquad
\int g\,(f'')^2dk=G f_0^2\!\int_{-\infty}^{\infty}\!e^{8\epsilon L}B_2\,y^2\,dt\,,
\label{eq:S4tint}
\end{equation}
where $B_1=\big(5-18\epsilon+16\epsilon^2\big)\,z^2-\big(6-16\epsilon+8\epsilon^2\big)\,z^3$ and
$B_2=\big(130-834\epsilon+1960\epsilon^2-1992\epsilon^3+736\epsilon^4\big)\,z^3-\big(375-2142\epsilon+4308\epsilon^2-3528\epsilon^3+960\epsilon^4\big)\,z^4+\big(252-1332\epsilon+2376\epsilon^2-1584\epsilon^3+288\epsilon^4\big)\,z^5$.
We evaluate all integrals over $t$ with Euler's Beta integral and its derivatives in $n$ and $s$,
\begin{equation}
\int_{-\infty}^{\infty}\!\operatorname{sech}^nt\;e^{st}\,dt
=\frac{2^{n-1}\,\Gamma\big(\tfrac{n+s}{2}\big)\Gamma\big(\tfrac{n-s}{2}\big)}{\Gamma(n)}\,.
\label{eq:S4beta}
\end{equation}
A factor $t^2$ is generated by a second derivative in $s$ at $s=0$ and a factor $L$ by $-\partial_n$, so that
$S_n=\int\operatorname{sech}^nt\,dt=\sqrt{\pi}\,\Gamma(\tfrac n2)/\Gamma(\tfrac{n+1}{2})$,
$L_n=\int\operatorname{sech}^nt\;L\,dt=-\partial_nS_n$ and
$T_n=\int\operatorname{sech}^nt\;t^2dt=\tfrac12S_n\,\psi_1(\tfrac n2)$. The retarded coefficients through order
$\epsilon$ require no other moments, because
$e^{8\epsilon L}y_E^2=1+14\epsilon L+\epsilon^2(98L^2-8t^2)+O(\epsilon^3)$ and $\sin^2(\pi\epsilon)$ is of order
$\epsilon^2$.

At $\epsilon=0$ we find $E_1=\tfrac{1}{30}$, $E_3=\tfrac{8}{105}$, $O_1=\tfrac{1}{90}+\tfrac{1}{3\pi^2}$ and $O_3=\tfrac{8}{315}+\tfrac{4}{9\pi^2}$.
Up to sign and the normalization of $\int_\omega$, $E_1$ and $E_3$ are the leading frequency-space moments, which also follow directly from Basset's integral and Euler's Beta integral,
\begin{align}
\int_{\mathbb{R}^2} \omega_1\omega_2 \prod_{i=1}^3 K_0(|\omega_i|)\,
 d\omega_1 d\omega_2
 &= -\frac{2\pi^2}{15}\,,
\nonumber\\
\int_{\mathbb{R}^2} \omega_1^2\omega_2^2 \prod_{i=1}^3 K_0(|\omega_i|)\,
 d\omega_1 d\omega_2
 &= \frac{32\pi^2}{105}\,.
\label{k0moments}
\end{align}
The leading Laurent coefficients of the prefactors and of the coefficients~\eqref{ibpcoeffs} are
$N_1\big|_{\epsilon^{-1}}=1/(2048\pi^4)$, $N_3\big|_{\epsilon^{-1}}=1/(131072\pi^4)$,
$r_2\big|_{\epsilon^{-3}}=-\tfrac{12}{5}$ and $r_1\big|_{\epsilon^{-3}}=-\tfrac{336}{5}$.
The $\epsilon^{-1}$
coefficient of $I_1^{(M)}$ is therefore $1/(30\cdot2048\,\pi^4)=1/(15\,(8\pi)^4)$ and that of $I_3^{(M)}$ is
$-8/(105\cdot131072\,\pi^4)=-1/(1720320\,\pi^4)$.

Expanding Eq.~\eqref{eq:S4EO} with the series of $N_1$, $N_3$, $G$, $C_\epsilon(0)$ and $S_\epsilon'(0)$, we obtain the subleading Laurent coefficients of $I_1^{(M)}$ and $I_3^{(M)}$ needed for $I_2^{(M)}$ and not listed in Ref.~\cite{Driesse:2026qiz}.
We write $\lambda=\ln 2\pi-\gamma_E$ for the constant from the $(4\pi)^{-4\epsilon}$ and $\Gamma$-function
prefactors of $N_{1,3}$, which contain no factor $e^{\gamma_E\epsilon}$, and find
\begin{align}
I_1^{(M)} &= \frac{1}{15\,(8\pi)^4}
 \Big[\frac{1}{\epsilon}+4\lambda+\frac{232}{15}
 +\epsilon\Big(8\lambda^2+\frac{928}{15}\,\lambda+\frac{50944}{225}-7\pi^2\Big)\Big]+\mathcal{O}(\epsilon^2)\,,
\nonumber\\
I_3^{(M)} &= -\frac{1}{420\,(8\pi)^4}
 \Big[\frac{1}{\epsilon}+4\lambda+\frac{182}{15}
 +\epsilon\Big(8\lambda^2+\frac{728}{15}\,\lambda+\frac{116782}{675}-7\pi^2\Big)\Big]+\mathcal{O}(\epsilon^2)\,.
\label{i1i3closed}
\end{align}
Here $\lambda$ enters only through an overall factor
$e^{4\lambda\epsilon}$.
A two-dimensional quadrature of the frequency integrals reproduces these expansions to more than 110 figures.

\subsection{\texorpdfstring{The value of $c_M$}{The value of cM}
\label{sec:cmresult}}

For $I_2^{(M)}=r_1I_3^{(M)}+r_2I_1^{(M)}$ we combine the two integrands of Eq.~\eqref{eq:S4tint} into
\begin{equation}
Q(z,\epsilon)=\epsilon^3\Big[r_1\,\frac{N_3}{N_1}\,B_2-r_2\,B_1\Big]=q_0+\epsilon\,q_1+\dots\,,\qquad
\frac{N_3}{N_1}=\frac{1-2\epsilon}{32\,(2-5\epsilon)}\,,
\label{eq:S4Q}
\end{equation}
whose leading term is a total derivative, $q_0=d^2R/dt^2$ with $R=\tfrac{3}{4}\,z^2-\tfrac{151}{40}\,z^3+\tfrac{147}{40}\,z^4$. We write $a_k$ and $c_k$ for the
coefficients of $\epsilon^k$ in $\int e^{8\epsilon L}Q\,y_E^2\,dt$ and in $\int e^{8\epsilon L}Q\,y_O^2\,dt$. Then
\begin{equation}
\begin{aligned}
I_2^{(M)}&=-\frac{N_1G}{\epsilon^{3}}\Big[\cos^2(\pi\epsilon)\,C_\epsilon(0)^2\sum_ka_k\epsilon^k
+\sin^2(\pi\epsilon)\,S_\epsilon'(0)^2\sum_kc_k\epsilon^k\Big]\,,\\
I_{2,F}^{(M)}&=\frac{e^{2i\pi\epsilon}N_1G\,S_\epsilon'(0)^2}{\epsilon^{3}}\sum_kc_k\epsilon^k\,,
\end{aligned}
\label{eq:S4I2comb}
\end{equation}
with
\begin{equation}
\begin{aligned}
a_0&=a_1=0\,,&\quad a_2&=\tfrac{2}{3}\,,&\quad a_3&=-\tfrac{26}{3}-\tfrac{28}{3}\,\ln 2\,,\\
c_0&=\tfrac{2}{3}\,,&\quad c_1&=-\tfrac{26}{3}-\tfrac{4}{3}\,\ln 2\,,&\quad c_2&=40+\tfrac{52}{3}\,\ln 2+\tfrac{4}{3}\,\ln^2 2-\tfrac{1}{2}\,\pi^2\,.
\end{aligned}
\label{eq:S4ac}
\end{equation}
Here $c_2$ depends on $y_2$ through the moments
$\int\operatorname{sech}^nt\;y_2'\,dt$, which we reduce by one integration by parts to $S_n$, $L_n$, $T_n$ and
$2\int_0^\infty\operatorname{sech}^2t\;y_1\,dt=6\ln 2+3\ln^2 2-\tfrac{1}{2}\,\pi^2$. Since $N_1$ starts at order $\epsilon^{-1}$ and $a_0=a_1=0$,
the $\epsilon^{-4}$ and $\epsilon^{-3}$ poles of $I_2^{(M)}$ cancel
identically under the retarded routing. The
$\epsilon^{-1}$ pole of $I_1^{(M)}$ agrees with Eq.~\eqref{cMdef}, and for $I_2^{(M)}$ we find
\begin{equation}
I_2^{(M)} = -\frac{5}{6\,(8\pi)^4\,\epsilon^{2}}
            + \mathcal{O}(\epsilon^{-1})\,,
\qquad c_M = 1\,,
\label{cmone}
\end{equation}
exactly. Of this value, the $\cos^2$ term of Eq.~\eqref{eq:S4I2comb} contributes $\tfrac15$ and the $\sin^2$
term, which is due to the retarded phases, contributes $\tfrac45$.
Driesse et al.~\cite{Driesse:2026qiz} generally evaluate the
frequency integrals numerically and reconstruct analytic results from numerical data. With the closed forms above we confirm their value of $c_M$
analytically.

With Feynman propagators on $D_{31}$ and $D_{24}$ the frequency
factors of Eqs.~\eqref{I1rep} and~\eqref{I3rep} become
$(0^+-i|\omega_j|)^{1-2\epsilon}$ and $(0^+-i|\omega_j|)^{-1-2\epsilon}$.
The integrand then changes at order $\epsilon^0$: the leading
factor $-\omega_1\omega_2$ of $I_1^{(M)}$ becomes
$-|\omega_1\omega_2|$.
As a result, the $\epsilon^{-1}$ coefficient of $I_1^{(M)}$ is multiplied by
$-O_1/E_1=-\tfrac13-\tfrac{10}{\pi^2}=-1.3465\ldots$ and no longer satisfies
Eq.~\eqref{cMdef}. Because the $\epsilon^{-1}$ coefficient of $I_3^{(M)}$ is multiplied by a different factor,
$-O_3/E_3=-\tfrac13-\tfrac{35}{6\pi^2}=-0.9243\ldots$,
the $\epsilon^{-4}$ pole of $I_{2,F}^{(M)}$ in Eq.~\eqref{ibp} survives, with coefficient
$1/(6144\,\pi^6)$ exactly (Appendix~\ref{app:boundary}). The second line of Eq.~\eqref{eq:S4I2comb} gives the three leading poles,
\begin{equation}
I_{2,F}^{(M)} = \frac{1}{6144\,\pi^6}\Big[\frac{1}{\epsilon^{4}}
 +\frac{4\lambda+2i\pi}{\epsilon^{3}}
 +\frac{8\lambda^2+8i\pi\lambda-\tfrac{23}{12}\,\pi^2}{\epsilon^{2}}\Big]
 +\mathcal{O}(\epsilon^{-1})\,,
\label{i2feyn}
\end{equation}
and no
constant $c_M$ in the sense of Eq.~\eqref{cMdef} is defined.
Reference~\cite{Driesse:2026qiz} reports an uncanceled $\epsilon^{-4}$ pole for
Feynman $i0^+$ on the gravitons, without its coefficient, together
with a $(\gamma-3)^{-15/2}$ divergence of the resulting
observable.

\begin{sloppypar}
Porto and Riva~\cite{Porto:2026fsd} instead define a local-in-time
conservative sector through a Poincar\'e--Bertrand decomposition of the
principal-value integrals that Feynman's $i0^+$ prescription produces~\cite{Porto:2024cwd}. In
their fifth-post-Newtonian computation the $\gamma$-3 prescription reproduces
the scattering angle of Ref.~\cite{Driesse:2026qiz} at $c_M=1$ over the
common regime of validity, while the
Poincar\'e--Bertrand prescription yields the local memory-like
contribution at $O(G^5\nu^2)$ with opposite sign. They find that this prescription would
imply values of $I_1^{(M)}$ and $I_2^{(M)}$ opposite in sign to those of
Eq.~\eqref{cMdef} at $c_M=1$. With the sign-reversed values the
conservative result retains the singularity at $\gamma=3$ and, pending a
treatment of the infrared divergences of the scattering problem under
Feynman's prescription, would be compatible with a finite $\epsilon\to0$
limit~\cite{Porto:2026fsd}. These values are not evaluations of
Eqs.~\eqref{I1rep}--\eqref{I3rep} with Feynman propagators on $D_{31}$
and $D_{24}$, and the uncanceled pole found here and in
Ref.~\cite{Driesse:2026qiz} does not bear on them.
\end{sloppypar}

\subsection{The singular point and graviton routing}
\label{sec:gamma3-landau}

Driesse et al.~\cite{Driesse:2026qiz} find that the potential-region contribution
diverges at $\gamma=3$ through the integrals of the K3 sector, and that
the memory-region contribution cancels this divergence once its boundary integrals
have the pole structure of Eq.~\eqref{cMdef}, for any
$c_M$.
In Sections~\ref{sec:apery-k3-bootstrap} and~\ref{sec:S9second} we showed that $\gamma=3$
is a second-type Landau singularity of the maximal
cut of the K3 sector~\cite{Klemm:2024wtd} (topology~40~\cite{Brammer:2025rqo}).
The ten on-shell conditions of topology~40 have no simultaneous
solution at real loop momenta, because a real null momentum orthogonal
to a timelike velocity vanishes. By the Coleman--Norton
theorem~\cite{Coleman:1965xm}, this singularity corresponds to no real classical process
(cf.~\cite{Porto:2026fsd}).

The divergence arises in the potential-region integrals, which are by construction insensitive to the
graviton $i0^+$ prescription~\cite{Driesse:2026qiz}, whereas $c_M$ is fixed at $x=1$ by the
routing of $D_{31}$ and $D_{24}$ in the boundary integrals $I_{1,2}^{(M)}$.
The position of the singularity follows from the Landau
locus~\eqref{eq:disc-q6}, but
the value of $c_M$ requires the evaluation above.

Under the in-in worldline rules~\cite{Jakobsen:2022psy,Kalin:2022hph}, every propagator is retarded toward the outgoing vertex of its diagram.
In topology~40, when that vertex is the middle one of three on its worldline, both radiative gravitons are retarded into their common vertex, and otherwise one is advanced relative to it (cf.~\cite{Porto:2026fsd}).
We do not identify these gravitons with $D_{31}$ and $D_{24}$ of $I_{1,2}^{(M)}$, which have five graviton propagators.
We show that, with cut worldline propagators, the four-dimensional scalar integral of topology~40 in impact-parameter space is regular at $\gamma=3$ for either routing of the radiative gravitons (Appendix~\ref{sec:S9position}). However, the complete observable requires this integral in $d=4-2\epsilon$ and the reduction of the in-in integrand.

\section{Conclusions}
\label{sec:outlook}

Next-generation gravitational-wave detectors need more accurate waveform models, and the post-Minkowskian expansion of the two-body problem that enters them has reached fifth order, four loops.
At this order its integrals involve K3 surfaces and Calabi--Yau threefolds, as do Feynman integrals in particle physics, and one can ask, as for scattering amplitudes, which singularities they have and which functions they evaluate to. Many Feynman-integral methods carry over, such as the Landau equations and maximal cuts, and this paper has explored some applications of those methods.

For example, in 5th order in the PM expansion at second order in the mass ratio, the potential-region contribution diverges at relative Lorentz factor $\gamma=3$~\cite{Driesse:2026qiz}.
Nothing is special at that speed.
We traced the divergence to the maximal cut of the planar K3 sector, where it comes from a pinch at infinite loop momentum that corresponds to no real classical process~\cite{Coleman:1965xm}.
It must therefore cancel, and the only region that can cancel it is the memory region, where two gravitons are radiative~\cite{Driesse:2026qiz}.

How the memory region is treated also affects the conservative part of the result, which is the part that carries over to bound orbits~\cite{Kalin:2019rwq,Kalin:2019inp} and effective-one-body models~\cite{Damour:2016gwp,Buonanno:2024byg}.
Separating the conservative from the dissipative part requires an $i0$ prescription for the radiative gravitons, and different prescriptions give different conservative results~\cite{Driesse:2026qiz,Porto:2026fsd}.
The choice should affect only how the full observable is divided, but no rule for it has yet been derived from first principles.
The $\gamma=3$ singularity does not fix the choice either, because with cut worldline propagators the four-dimensional scalar integral of the K3 sector is regular there for either routing of the radiative gravitons.
We computed how the memory region depends on the prescription.
With the prescription of Ref.~\cite{Driesse:2026qiz} we derived the first memory-region integral in closed form through order $\epsilon$ and confirmed analytically that the constant entering the conservative scattering angle is $c_M=1$.
With Feynman propagators the $\epsilon^{-4}$ pole of the second memory-region integral does not cancel~\cite{Driesse:2026qiz}, and we derived its coefficient and the next two poles.

A full 5PM-2SF calculation, with radiation reaction, needs the remaining sectors of the four-loop families, and these families contain no geometries beyond two K3 surfaces and two Calabi--Yau threefolds~\cite{Brammer:2025rqo}.
We fixed the maximal-cut normalizations of the periods of all four geometries, proved a closed formula for the Frobenius coefficients of the period of the threefold at second order in the mass ratio, and continued the four Frobenius bases to bound-orbit energies.
For the dissipative families we made a first count of the master integrals, and in the PX2 family we found ten exact relations with the auxiliary-mass flow~\cite{Liu:2017jxz,Liu:2022chg} and evaluated twelve master integrals at one velocity with Feynman graviton propagators, working from maximal cuts and boundary limits without the integration-by-parts reduction of the four-loop integrand.
With that reduction and the remaining sectors, the differential equations would continue these boundary values to general kinematics and show whether K3 and Calabi--Yau periods enter the radiated energy, as a threefold period does at first order in the mass ratio~\cite{Driesse:2024feo}.

\acknowledgments
We thank Rafael A.\ Porto for comments on a draft of this manuscript. MDS was supported in part by the U.S. Department of Energy, Office of Science, under grant DE-SC0013607. Support and funding for computations in this project were provided by Anthropic. Calculations in this paper were largely performed by AI (Claude Fable 5.1) and AI drafted the manuscript, which the author then edited. The author takes full responsibility for the validity of all the calculations in the paper. 

\appendix
\section{Auxiliary-mass flow on worldline families}\label{app:flow}\label{sec:methods}

As in phase-space integrals~\cite{Liu:2020kpc}, on worldline families the cut propagators
$2\pi\,\delta(2k\cdot u)$ cannot be deformed. The large-$\eta$
regions of a cut sector therefore contain boundary integrals with cut worldline
propagators in place of the vacuum integrals of a Feynman family~\cite{Liu:2017jxz,Liu:2022chg}.
We do not
evaluate these boundary integrals for the 2SF sectors: the boundary constants of
Section~\ref{sec:functionspace} are master integrals that we compute
directly at the reference point~\eqref{eq:refpoint}, and the response matrices are
independent of the boundary-integral values.

\subsection{Assumptions for the boundary relation \label{sec:nl-identity}}

The relation~\eqref{eq:nulllaw} holds under three assumptions.
(i) Closure: the $\eta$-deformed master integrals $J(\eta,\epsilon)$,
with physical values $M = \lim_{\eta\to 0^+} J$,
satisfy a first-order homogeneous linear system
$\partial_\eta J = C(\eta,\epsilon)\, J$ with rational coefficients.
(ii) Boundary completeness: as $\eta \to \infty$ the physical solution matches onto the enumerated boundary
integrals, $J(\eta) = \Phi(\eta,\epsilon)\, b(\epsilon) + o(1)$, with $\Phi$ the power-by-power matching matrix. All powers of $\eta$ are included in the matching, in particular the odd or half-integer-shifted powers that
linearized worldline propagators can generate, and no recessive
exponentials occur on the cut cycles.
(iii) Genericity: the rank of $A(\epsilon)$ is constant
near the sampled $\epsilon$.
Equation~\eqref{eq:nulllaw} then follows from linearity in the boundary data.
We make no claim where the rank jumps, and the relation does not extend to the fixed-point family of Appendix~\ref{sec:eta-fixed-point}, where the boundary vector coincides with $M$.

\subsection{The solver \label{app:solver}}

We carry out the $\eta$-evolution with \textsc{PMflow}~\cite{Schwartz:2026afx,BootLoops:v1}, our generalization of
\textsc{AMFlow}~\cite{Liu:2022chg} to the boundary integrals of cut worldline sectors.
The auxiliary-mass flow with linear and cut propagators~\cite{Liu:2022tji,Liu:2020kpc} is due to Liu, Ma and collaborators.
\textsc{AMFlow} has been applied to one-loop post-Minkowskian integrals and their cuts~\cite{Bohnenblust:2023qmy} and to one-worldline integrals in gravitational Compton scattering at three and four loops~\cite{Bautista:2026fcp,Brunello:2026rdk,Brunello:2026lzf}, including retarded and cut configurations~\cite{Brunello:2026rdk,Brunello:2026lzf}. The sectors that we treat here have two worldlines.
\textsc{PMflow} consists of
a driver, in the \textsc{BootLoops} repository, and a fork of
\textsc{AMFlow.cpp}~\cite{AMFlowCpp}. We do not distribute the reduction tables.

In the fork we evaluate several classes of
boundary integrals of cut sectors in closed form, using the static-source and cusp integrals of heavy-quark effective theory~\cite{Broadhurst:1991fz,Korchemsky:1987wg}.
Along a linearized worldline
propagator, where the standard enumeration $\ell\sim\eta^{0}$ or
$\ell\sim\eta^{1/2}$ is not systematic~\cite{Liu:2022tji}, we include the region $\ell\sim\eta$.
Uncut linearized worldline
propagators introduce half-integer powers of $\eta$, and one coupled block of the dissipative NP system has
local exponents
\begin{equation}
\bigl\{-\tfrac{1}{2},\ -\tfrac{3}{2},\ -1,\ +\tfrac{1}{2}\bigr\} + 4\epsilon
\label{eq:cosets}
\end{equation}
at $\eta\to\infty$. A fundamental
system of power-series solutions therefore exists only in the double-cover variable
$t$, with $t^2$ the flow variable.

We have derived three further classes of two-velocity boundary integrals, which arise in the coupled sectors of the $O(G^3)$ radiative family, but we have not implemented all of them, so we have not solved those sectors through the flow.
Moreover, we have not shown that the implemented classes exhaust the boundary
integrals of the coupled 2SF sectors.

As a test of the solver, we derive graviton-cut boundary data from
the closed forms of Ref.~\cite{Herrmann:2021tct} for two master integrals of its two-loop soft family. With these data the flow reproduces order by order in $\epsilon$ twice the imaginary part
of the standard flow solution~\cite{Liu:2022chg}. The $O(G^3)$ radiated energy at $\gamma=\tfrac{5}{4}$, which we assemble by reverse unitarity~\cite{Anastasiou:2002yz} from the
cut integrand of Ref.~\cite{DiVecchia:2021bdo} and the master values of
Ref.~\cite{Herrmann:2021tct}, matches the closed form of Ref.~\cite{Herrmann:2021lqe}.
On the 4PM cut-eikonal K3 family~\cite{Dlapa:2022wdu,Dlapa:2023hsl} one set of values of the 43 master
integrals satisfies the constraints of two flows with different $\eta$
insertions at each sampled $\epsilon$ (Appendix~\ref{sec:eta-fixed-point}).

\section{Master integrals and identities}\label{app:masters}
\label{sec:S1}\label{sec:S5}

This appendix gives the seeding of the first-stage reductions of Section~\ref{sec:functionspace} and our master integrals for the Ap\'ery sector, which we relate to the nine-element basis of Duhr et al.~\cite{Duhr:2025lbz}.

\subsection{First-stage reductions of the four families}
\label{sec:S5counts}\label{sec:S5px2}

The counts of Table~\ref{tab:families} come from one Laporta reduction per family. We seeded the integrals of the top
sectors with at most $r$ propagator powers in total, $s$ numerator powers and
$n_{\rm d}$ dots, with $(r,s,n_{\rm d})=(14,0,1)$ for P and PX1 and $(14,1,\text{--})$, dots unrestricted, for PX2 and NP. We required the reduction of the integrals with every propagator at unit power ($r=13$, $s=n_{\rm d}=0$) and of their subsector
images (ancillary file \texttt{tables/\allowbreak first\_\allowbreak stage\_\allowbreak seeding.txt}; the four master lists are in \texttt{masters/\allowbreak}).
Repeating the PX2 reduction with the seeds $(r,s,n_{\rm d})=(15,2,2)$ added on the four leading top sectors, we obtained the same 37 master integrals.

The PX2 count holds for the family as defined, in which the uncut linearized propagators have a common
routing sign $\sigma$ and the graviton propagators are even in their momenta.
With the automatic sector relations and symmetries switched off, the leading top sectors share no nonvanishing subsector and reduce independently. At the same seeding, $S_1$ with its subsectors reduces to 30 master
integrals (8 even, 22 odd) and $S_2$ to 28 (8 even, 20 odd).
Since their images under
$k_i\to k_i+q$, $q\to-q$ are $S_3$ and the fourth top sector, the total is
$2\times(30+28)=116$ master integrals, against 37 with the
identifications.

The master integrals of the response-matrix rows are
not in general among the 37, since the auxiliary-mass flow uses the master integrals of the
$\eta$-deformed system.

\subsection{The Ap\'ery-sector master integrals}
\label{sec:S5t40}

The Ap\'ery sector (K3$'$) is topology~40 of Ref.~\cite{Brammer:2025rqo}. Our family for it (ancillary file \texttt{masters/\allowbreak t40\_\allowbreak even\_\allowbreak 44\_\allowbreak eleven.txt}) has the ten
denominators $P_1,\dots,P_{10}$, of which $P_7,\dots,P_{10}$ are linearized worldline propagators, and the numerators $P_{11},\dots,P_{22}$.
On the maximal cut we generated the
integration-by-parts and Lorentz-invariance identities of the even-parity block from all seeds with at most \SItfortyDots{} dots and at most
\SItfortySisp{} powers of the numerators $P_{13},\dots,\allowbreak P_{16}$, imposing no momentum-relabeling symmetry.
At $\gamma=\SItfortyYzero$ and at a second value of
$\gamma$ we find that the system reduces to eleven master integrals $M_1,\dots,M_{11}$, numbered as in the ancillary file.

The worldline exchange
$u_1\leftrightarrow u_2$ with $k_1\to k_2+k_4$, $k_3\to k_2-k_3-q$, $k_4\to k_1-k_2$ is a symmetry of the sector,
permuting its propagators as $(P_1P_4)(P_2P_6)(P_3P_5)(P_7P_{10})(P_8P_9)$, and exchanges $M_2$ and $M_5$.
Imposing it on every even seed integral and reducing again, we obtain nine master integrals: $M_1$, $M_2$, $M_3$, $M_4$, $M_6$, $M_7$,
$M_8$, $M_9$ and $M_{11}$.

\subsection{Relation to the nine-element basis}
\label{sec:S5duhr}

Klemm et al.~\cite{Klemm:2024wtd} report nine master integrals in the Ap\'ery
sector. Duhr et al.\ give in their ancillary files an explicit nine-element basis on the maximal cut, obtained with Feynman propagators and all symmetry relations, together with its differential equation and the rotation to $\epsilon$-factorized form.
The basis is built from \SIduhrNint{} integrals $\hat M_1,\dots,\hat M_8$ (ancillary file \texttt{tables/\allowbreak t40\_\allowbreak duhr\_\allowbreak propagators\_\allowbreak integrals.txt}) and the first and
second $x$-derivatives of $\hat M_1$, with prefactors rational in $x$ and $\epsilon$. The eighth and ninth elements are $\hat M_7$ and $\hat M_8$ times such prefactors.

The loop momenta of Duhr et al.\ are related to ours by
$K_1=k_2-k_3-q$, $K_2=k_1-k_3-q$, $K_3=-k_3-k_4-q$, $K_4=-k_3-q$. Then $\hat M_1$ and $\hat M_3$ are $2^4$ times $M_1$ and $M_3$, and $\hat M_2$ is $2^4$ times the integral with $P_1$ squared, which the worldline exchange maps onto
$M_4$. The factor $2^4$ comes from the normalization of the linearized propagators.

At $\gamma=23/17$, for the spacetime dimensions $d=97/23$ and $d=113/31$,
the first seven basis elements, which close under differentiation in the system of Duhr et al., reduce to linearly independent combinations of our nine integrals
(ancillary file \texttt{masters/\allowbreak t40\_\allowbreak duhr\_\allowbreak map\_\allowbreak point.json}).
The last two elements contain the numerators $P_{11}=k_2^2$ and $P_{18}=(k_3+q)^2$, which do not occur in the seeds used above.
Adding the identities of the seeds with one power of $P_{11}$ or $P_{18}$, we reduce both onto our nine integrals,
\begin{align}
\hat M_7&=-32\,(d-4)\,M_6\,,\label{eq:S5duhr8}\\
\hat M_8&=\Bigl(\tfrac{3358}{15}-\tfrac{1540}{27}\,d\Bigr)M_1+\tfrac{578}{135}\,\tfrac{d-5}{d-4}\,M_3-\tfrac{16}{3}\,M_4\nonumber\\
&\quad+\tfrac{1564}{135}\,M_6+\tfrac{578}{45}\,M_7+\tfrac{578}{45}\,\tfrac{d-3}{d-4}\,M_8-\tfrac{782}{135}\,M_9\,.\label{eq:S5duhr9}
\end{align}

Equation~\eqref{eq:S5duhr8} is a combination of two integration-by-parts identities and one exchange relation, none of which contains
$\gamma$, and hence it holds for all $\gamma$ and $d$.
Equation~\eqref{eq:S5duhr9} holds at $\gamma=23/17$, with its dependence on $d$ interpolated (ancillary file \texttt{masters/\allowbreak t40\_\allowbreak duhr\_\allowbreak map\_\allowbreak u8\_\allowbreak u9.json}), and we have not determined its dependence on $\gamma$.
It follows that the nine basis elements without the prefactors of Duhr et al.\ are linearly independent combinations of our nine integrals at both values of $d$.

\section{Derivations for the boundary integrals}
\label{app:boundary}
\label{sec:S2}

We reduce the boundary constants of PX2 quoted in Section~\ref{sec:ipifive} to frequency integrals, in the conventions of that section.
We then extend the closed forms of the memory-region integrals of Section~\ref{sec:memory} by one order in $\epsilon$ and give their values for Feynman propagators on the two memory gravitons.

\subsection{The corner integrals \texorpdfstring{$I_0$, $I_2$, $I_3$}{I0, I2, I3}}
\label{sec:S2corner}

In the corner integral $I_0$ of Eq.~\eqref{eq:S1masters} and its numerator partners
$I_2$, $I_3$, each delta function confines its loop momentum to the Euclidean
hyperplane orthogonal to a worldline
velocity, $k_1,k_2\in V_1=u_1^\perp$ and $k_3,k_4\in V_2=u_2^\perp$, of dimension $\tilde d=3-2\epsilon$.
We write $k_2=b\,\hat e_1+k_{2T}$ and $k_4=a\,\hat e_2+k_{4T}$, where $\hat e_1\in V_1$ and $\hat e_2\in V_2$
are the unit vectors in the plane spanned by $u_1$, $u_2$, the transverse parts $k_{iT}$
live in $T=2-2\epsilon$ Euclidean dimensions, and $a$, $b$ are the two worldline frequencies.
We first integrate over $k_1$ and $k_3$, which form massless one-loop bubbles inside $V_1$ and $V_2$,
\begin{equation}
\int\frac{d^{\tilde d}\kappa}{\pi^{\tilde d/2}}\,\frac{1}{\kappa^2\,(\kappa+P)^2}
=G_{11}\,(P^2)^{-\frac12-\epsilon}\,,\qquad
G_{11}=\frac{\Gamma(\tfrac12+\epsilon)\,\Gamma(\tfrac12-\epsilon)^2}{\Gamma(1-2\epsilon)}\,,
\label{eq:S2bubble}
\end{equation}
with $P^2=b^2+|k_{2T}+q|^2$ for the $k_1$ bubble and
$P^2=a^2+|k_{4T}|^2$ for the $k_3$ bubble. The vector bubble is $\tfrac12P^\mu$ times the scalar one, and a numerator equal to one of the bubble's own denominators
gives a scaleless integral. The numerators of $I_2$ and $I_3$ then become
\begin{equation}
\begin{aligned}
(k_1-k_3)^2&\;\to\; \mathcal{N}_2=p_1\!\cdot\! q-1+\tfrac{\gamma}{2}\,s\,\frac{w_1w_2}{r^2}+\tfrac12\,p_1\!\cdot\! p_3-q\!\cdot\! p_3\,,\\
(k_1-k_4)^2&\;\to\; \mathcal{N}_3=p_1\!\cdot\! q-1+\gamma\,s\,\frac{w_1w_2}{r^2}+p_1\!\cdot\! p_3-2\,q\!\cdot\! p_3-\big(a^2+|k_{4T}|^2\big)\,,
\end{aligned}
\label{eq:S2M2M3}
\end{equation}
where $p_1=k_{2T}+q$ and $p_3=k_{4T}$, the dot products
are Euclidean and transverse, and $ab=s\,w_1w_2/r^2$ in the rescaled variables introduced below, with $s=\pm1$ the relative sign of the two frequencies.

The graviton propagator $(k_2-k_4)^2$, which joins the two hyperplanes, contains the indefinite form
\begin{equation}
-(k_2-k_4)^2=F+|k_{2T}-k_{4T}|^2\,,\qquad F=a^2+b^2-2ab\gamma\,,
\label{eq:S2F}
\end{equation}
which vanishes for $\gamma>1$ on the two rays $a/b=x^{\pm1}$ (the radiation cone). The Feynman prescription is
$F\to F-i0$.

After the two bubble integrations, a two-loop chain in the transverse space remains: a line of
mass $b$ and index $\tfrac12+\epsilon$, the exchange of
mass $\sqrt{F}$ and index one, and a line of mass $a$ and index $\tfrac12+\epsilon$.
In transverse position space the integrand of the chain is a product of three Bessel functions.
At fixed transverse distance $r$ we rescale the frequencies, $(b,a)=(w_1,w_2)/r$, and the radial integral is of Weber--Schafheitlin type~\cite[Eq.~10.22.43]{NIST:DLMF}.
With $\nu=\tfrac12-2\epsilon$ the scalar line
factors are $w^{\nu}K_{\nu}(w)$, a line differentiated by a vector insertion gives $-w^{\nu+1}K_{\nu+1}(w)$, and the exchange gives
$\hat F^{-\epsilon/2}K_\epsilon\big(\sqrt{\hat F-i0}\,\big)$ with $\hat F=w_1^2+w_2^2-2s\gamma\,w_1w_2$. We define the
core of a term with insertion $\mathcal{I}(w_1,w_2,s)$ as
\begin{equation}
W(\epsilon)=2\sum_{s=\pm1}\int_0^\infty\!\!\int_0^\infty dw_1\,dw_2\;
\mathrm{sh}_1(w_1)\,\mathrm{sh}_3(w_2)\;\hat F^{-\epsilon/2}K_\epsilon\big(\sqrt{\hat F-i0}\,\big)\;
\mathcal{I}(w_1,w_2,s)\,,
\label{eq:S2core}
\end{equation}
where $\mathrm{sh}_{1,3}$ are the two line factors. For $s=+1$ and $w_1/w_2\in(x,1/x)$ the form $\hat F$ is negative and we continue the Bessel
function through $K_\nu(-iy)=\tfrac{i\pi}{2}e^{i\nu\pi/2}H^{(1)}_\nu(y)$. The cores depend
on the graviton $i0$ only through this continuation, which makes them complex.

Collecting the constants, we obtain
\begin{equation}
I_0(\epsilon)=-\pi\,2^{5\epsilon}\,
\frac{\Gamma(\tfrac12-\epsilon)^4\,\Gamma(4\epsilon-1)}{\Gamma(1-2\epsilon)^2\,\Gamma(2-5\epsilon)}\;W_A(\epsilon)\,,
\label{eq:S2I0closed}
\end{equation}
with $W_A$ the core~\eqref{eq:S2core} with unit insertion. Similarly, the integrals
$I_2$ and $I_3$ are sums of terms of the same form, one for each term of Eq.~\eqref{eq:S2M2M3}, with radial
factors of the type $\Gamma(4\epsilon-2)$. Four cores occur: $A$ with unit insertion, $B$ with the insertion $s\,w_1w_2$, $C$ with one
differentiated line and $E$ with two. At $\epsilon=0$ we find
\begin{equation}
\begin{aligned}
W_A(0)&=\frac{\pi\,(320\ln2-192)}{27}+\frac{4\pi^2}{27}\,i\,,\\
W_B(0)&=\frac{\pi\,(3056640-4409344\ln2)}{2187}+\frac{352\pi^2}{2187}\,i\,,\\
W_C(0)&=\frac{\pi\,(20480\ln2-14592)}{243}-\frac{68\pi^2}{243}\,i\,,\\
W_E(0)&=\frac{\pi\,(4014080\ln2-2777088)}{2187}+\frac{1252\pi^2}{2187}\,i\,.
\end{aligned}
\label{eq:S2W0}
\end{equation}
With the residues of the $\Gamma$-function prefactors these values give the pole coefficients of Eq.~\eqref{cornerpoles}, for example $\tfrac{\pi^3}{4}W_A(0)$ for $I_0$.

\subsection{Feynman-parameter representation and the finite parts}
\label{sec:S2feynman}

For the finite parts we need the cores at first order in $\epsilon$. We extend the two frequency integrals of Eq.~\eqref{eq:S2core}
to the plane $(w_1,w_2)\in\mathbb{R}^2$, where $s$ is the sign of $w_1w_2$ and $F=w_1^2+w_2^2-2\gamma\,w_1w_2$, and
Fourier transform the three factors,
\begin{equation}
\begin{aligned}
\int_{-\infty}^{\infty}\!dw\;e^{-ikw}\,|w|^{\nu}K_\nu(|w|)&=\sqrt{\pi}\,2^{\nu}\,\Gamma(\nu+\tfrac12)\,(1+k^2)^{-\nu-\frac12}\,,\\
\int\! d^2w\;e^{-i\vartheta\cdot w}\,F^{-\epsilon/2}K_\epsilon\big(\sqrt{F-i0}\,\big)&=
\frac{i\,2^{1-\epsilon}\,\pi\,\Gamma(1-\epsilon)}{\sqrt{\gamma^2-1}}\;\big(1+\Pi+i0\big)^{-1+\epsilon}\,,
\end{aligned}
\label{eq:S2fourier}
\end{equation}
with $\Pi=-\kappa\,(\vartheta_1^2+\vartheta_2^2+2\gamma\,\vartheta_1\vartheta_2)$ and $\kappa=1/(\gamma^2-1)$. A differentiated line raises the power of its denominator by one, and the insertion
$s\,w_1w_2$ becomes the numerator $\vartheta_1\vartheta_2$ with both powers raised by one. We then introduce Feynman parameters and perform the Gaussian integral
over $\vartheta$, which gives
\begin{equation}
\begin{aligned}
W_X(\epsilon)&=\frac{i\pi\,c_X\,2^{-5\epsilon}\,\Gamma(n_X-5\epsilon)}{\sqrt{\gamma^2-1}}\;\Phi_X(\epsilon)\,,\\
\Phi_X(\epsilon)&=\int\limits_{x_1+x_2+x_3=1}\!\!\!\!d^2x\;\;x_1^{a_1-2\epsilon}\,x_2^{a_2-2\epsilon}\,x_3^{a_3-\epsilon}\,
\big(x_1x_2-\kappa\,x_3+i0\big)^{-m_X/2}\,,
\end{aligned}
\label{eq:S2feynman}
\end{equation}
where the constants $(c_X,n_X)$, the exponents $(a_1,a_2,a_3)$ and the power $m_X$ are $(1,2)$, $(0,0,0)$, $1$ for
$A$; $(-2,3)$, $(1,0,0)$, $1$ for $C$; $(4,4)$, $(1,1,0)$, $1$ for $E$; and $(-2\kappa\gamma,3)$, $(1,1,1)$, $3$ for
$B$. For $B$ the power $-\tfrac32$ is understood after one integration by parts in $x_2$. Where $x_1x_2<\kappa x_3$ the prescription gives
$(x_1x_2-\kappa x_3+i0)^{-1/2}=-i\,|x_1x_2-\kappa x_3|^{-1/2}$, so that
$\Phi_X=\Phi_X^{+}-i\,\Phi_X^{-}$ with real $\Phi_X^{\pm}$.

In the variables $x_1$ and $x_2$ the combination $x_1x_2-\kappa\,x_3$ is linear in $x_2$, and at each order in $\epsilon$ the $x_2$ integral is elementary. At order zero
and $\gamma=\tfrac54$ we reproduce the four cores of Eq.~\eqref{eq:S2W0} by exact integration, real parts
included.
At first order in $\epsilon$ the $x_2$ integrals produce $\ln x_1$, $\ln(1-x_1)$, $\ln(x_1+\kappa)$ and
$\arctan\sqrt{x_1/\kappa}$. We evaluated the remaining integrals over $x_1$ numerically and searched for integer
relations in the basis $\pi$, $\pi\ln2$, $\pi\ln3$ for $\Phi^{+}$ and
$1$, $\ln2$, $\ln3$, $\ln^22$, $\ln2\ln3$, $\ln^23$, $\pi^2$, $\mathrm{Li}_2(\tfrac14)$ for $\Phi^{-}$.
We list the relations that we found, with the order-zero values, in the ancillary file \texttt{tables/px2\_feynman\_parameter\_integrals.txt}.
Combining these relations with the Laurent expansions of the prefactors, we obtain the finite
parts of $I_0$, $I_2$, $I_3$ in Eq.~\eqref{cornerpoles}.

\subsection{The dotted integral \texorpdfstring{$I_{r4}$}{Ir4}}
\label{sec:S2dotted}

In the dotted integral $I_{r4}$ the
$k_3$ bubble has indices $(2,1)$ and vanishes linearly as $\epsilon\to0$, and the radial integral has
a pole from $\Gamma(4\epsilon)$ that cancels this zero.
Collecting the constants as for Eq.~\eqref{eq:S2I0closed}, we obtain
\begin{equation*}
I_{r4}(\epsilon)=\pi\,2^{1+5\epsilon}\,
\frac{\Gamma(\tfrac12-\epsilon)^3\,\Gamma(-\tfrac12-\epsilon)\,\Gamma(4\epsilon)}
{\Gamma(1-2\epsilon)\,\Gamma(-2\epsilon)\,\Gamma(1-5\epsilon)}\;W'(\epsilon)\,,
\end{equation*}
with the overall sign opposite to that of Eq.~\eqref{eq:S2I0closed}, because $I_{r4}$ has six propagator powers where $I_0$ has
five and each propagator is minus a Euclidean form on the support of the delta functions.

The core $W'$ has the representation~\eqref{eq:S2feynman} with $(c_X,n_X)=(\tfrac12,1)$, exponents $(0,-1,0)$ and $m_X=1$. Here the endpoint $x_2\to0$ gives a pole,
$\Phi_{r4}=\tfrac{3i}{4\epsilon}+\Phi_{r4}^{+}-i\,\Phi_{r4}^{-}$, whose first-order terms we identified as for $\Phi_X^{\pm}$ and list in the same ancillary file.
The prefactor equals $2\pi^3$ at $\epsilon=0$, and $W'$ has the residue $-\tfrac{\pi}{2}$ there, which gives the
pole of Eq.~\eqref{eq:ir4}. The finite part and the order-$\epsilon$ coefficient follow from $\Phi_{r4}^{\pm}$.

\subsection{The static-slice integrals \texorpdfstring{$I_{r5}$, $I_{r6}$, $I_{r7}$}{Ir5, Ir6, Ir7}}
\label{sec:S2static}

The corner sector of $S_2$ has the same structure as that
of $S_1$, with $k_1$ and $k_2$ interchanged. The $k_2$ and $k_3$ integrations are the bubbles, and $k_1\in
V_1$ (frequency $b$) and $k_4\in V_2$ (frequency $a$) are joined by the exchange $(k_1-k_4)^2$. The integrals
$I_{r5}$, $I_{r6}$, $I_{r7}$ contain in addition the uncut propagator $1/(D_5+i\sigma0)$ with
$D_5=2k_1\!\cdot\! u_2=-2\sqrt{\gamma^2-1}\;b$, which we split as
\begin{equation}
\frac{1}{D_5+i\sigma0}=\mathrm{PV}\frac{1}{D_5}-i\pi\sigma\,\delta(D_5)\,,\qquad
\delta(D_5)=\frac{\delta(b)}{2\sqrt{\gamma^2-1}}=\frac{2}{3}\,\delta(b)\quad(\gamma=\tfrac54)\,.
\label{eq:S2pv}
\end{equation}
For Feynman graviton propagators the principal-value part integrates to zero
(Section~\ref{sec:ipifive}). Only the slice $b=0$ then contributes, on which the form $F=a^2$ of Eq.~\eqref{eq:S2F} is positive, the momentum $k_1$ is purely transverse and one frequency integral remains,
\begin{equation}
I_{r5}=\sigma\,\frac{2\pi i}{3}\,\pi^{2\epsilon-1}\,G_{11}^2\,J\,,\qquad
J=\int_{-\infty}^{\infty}\!da\int d^Tk_{1}\,d^Tk_{4}\;
\frac{\big(|k_{1}+q|^2\big)^{-\frac12-\epsilon}\,\big(a^2+|k_{4}|^2\big)^{-\frac12-\epsilon}}{a^2+|k_{1}-k_{4}|^2}\,,
\label{eq:S2J}
\end{equation}
with $k_1$, $k_4$ transverse and Euclidean.

In position space $J$ factorizes into a massless line, the exchange of mass $|a|$ and a
massive line of mass $|a|$. Rescaling $a=w/r$, we are left with a Weber--Schafheitlin integral in
$r$ and a product of two Bessel functions in $w$, which we integrate with
\begin{equation}
\int_0^\infty t^{-\varrho}K_\mu(t)K_\nu(t)\,dt
=\frac{2^{-2-\varrho}}{\Gamma(1-\varrho)}\,
\prod_{\pm,\pm'}\Gamma\Big(\frac{1-\varrho\pm\mu\pm'\nu}{2}\Big)\,,
\label{eq:S2KK}
\end{equation}
where the product runs over the four sign choices. Hence $I_{r5}$ is a
product of $\Gamma$ functions to all orders in $\epsilon$. The integral representation
converges for $\tfrac18<\epsilon<\tfrac14$ and the closed form is its continuation.
For $I_{r6}=I_{r5}[(k_2-k_3)^2]$ and $I_{r7}=I_{r5}[(k_2-k_4)^2]$ the two bubbles reduce the numerators as in Eq.~\eqref{eq:S2M2M3},
and each term is again a Weber--Schafheitlin
integral times a Bessel product~\eqref{eq:S2KK} with shifted indices, which gives the closed forms of Table~\ref{tab:ipifive}.

\subsection{Higher orders of the memory-region integrals}
\label{sec:S4cores}
\label{sec:S4feynman}

Carrying the expansion of Section~\ref{sec:S4analytic} one order further for the retarded routing, we find the next order of Eq.~\eqref{i1i3closed},
\begingroup\small
\begin{equation}
\begin{aligned}
I_1^{(M)}&=\frac{e^{4\lambda\epsilon}}{61440\,\pi^4}\Big[\frac1\epsilon+\tfrac{232}{15}+\epsilon\big(\tfrac{50944}{225}-7\pi^2\big)+\epsilon^2\big(\tfrac{10774648}{3375}-\tfrac{1624}{15}\,\pi^2-\tfrac{2284}{3}\,\zeta(3)\big)\Big]+O(\epsilon^3)\,,\\
I_3^{(M)}&=-\frac{e^{4\lambda\epsilon}}{1720320\,\pi^4}\Big[\frac1\epsilon+\tfrac{182}{15}+\epsilon\big(\tfrac{116782}{675}-7\pi^2\big)+\epsilon^2\big(\tfrac{73302157}{30375}-\tfrac{1274}{15}\,\pi^2-\tfrac{2284}{3}\,\zeta(3)\big)\Big]+O(\epsilon^3)\,.
\end{aligned}
\label{eq:S4closedR2}
\end{equation}
\endgroup
From Eq.~\eqref{eq:S4I2comb} we obtain the expansion of $I_2^{(M)}$ through its finite part,
\begin{equation}
I_2^{(M)}=-\frac{5}{24576\,\pi^4}\Big[\frac{1}{\epsilon^2}+\frac{4\lambda}{\epsilon}+8\lambda^2-\tfrac{3}{5}\,\pi^2\Big]+O(\epsilon)\,.
\label{eq:S4I2R}
\end{equation}

With Feynman propagators on the two memory gravitons we find from Eq.~\eqref{eq:S4I2comb} that the right-hand side of Eq.~\eqref{i2feyn} equals $e^{(4\lambda+2i\pi)\epsilon}\,[1+\tfrac{1}{12}\,\pi^2\,\epsilon^2+O(\epsilon^3)]/(6144\,\pi^6\epsilon^4)$, so the
coefficient $1/(6144\,\pi^6)$ is exact.
Similarly, for the other two integrals we find, through the finite part,
\begin{equation}
\begin{aligned}
I_{1,F}^{(M)}&=-\frac{1}{184320\,\pi^6}\Big[\frac{\pi^2+30}{\epsilon}
+(4\lambda+2i\pi)(\pi^2+30)+\frac{232}{15}\,\pi^2+518-24\zeta(3)\Big]+O(\epsilon)\,,\\
I_{3,F}^{(M)}&=\frac{1}{10321920\,\pi^6}\Big[\frac{2\pi^2+35}{\epsilon}
+(4\lambda+2i\pi)(2\pi^2+35)+\frac{364}{15}\,\pi^2+\frac{8543}{18}-48\zeta(3)\Big]+O(\epsilon)\,.
\end{aligned}
\label{eq:S4I13F}
\end{equation}

The ancillary file \texttt{cM\_laurent.json} contains the polynomial $B_2$ and the coefficients $a_k$ and $c_k$.
It also lists eight further coefficients, which we identified by integer-relation fits to quadrature values and have not derived: the $\epsilon^{3}$ coefficients of $I_1^{(M)}$ and $I_3^{(M)}$ for the retarded routing,
their $\epsilon^{1}$ and $\epsilon^{2}$ coefficients for Feynman propagators, and the $\epsilon^{-1}$ and
$\epsilon^{0}$ coefficients of $I_{2,F}^{(M)}$.
The six two-frequency cores of the integrand expanded in $\epsilon$ are in \texttt{tables/memory\_cores.txt}.

\section{Proofs and further analysis for the Ap\'ery sector}
\label{app:apery}

We prove here that the torus period of topology~40 satisfies the Ap\'ery equation and evaluate the scalar integral of the topology in position space. We also obtain the alphabet of topology~41 from its leading singularities.

\subsection{\texorpdfstring{Two proofs of $L_A\varpi=0$}{Two proofs of the torus-period identity}}
\label{sec:S8gd}\label{sec:S8fermi}

The first proof that the torus period~\eqref{eq:period} satisfies $L_A\varpi=0$, with $L_A$ of Eq.~\eqref{eq:PFapery},
is a Griffiths--Dwork reduction of the integrand at symbolic $x$.
In the variables $\xi_1=t_1/t_2$ and $\xi_2=t_2$ of Section~\ref{sec:apery-baikov} we have $Q_6=t_1^2t_2^2R$ with
$R=(xw-1-x^2)^2-4x^2$, and $\varpi$ is the constant term in $\xi_1$, $\xi_2$ of $R^{-1/2}$ expanded in powers of $x$.
With $\delta_i=\xi_i\partial_{\xi_i}$ and $\theta=(x/2)\,d/dx$ in $L_A$, we find
\begin{equation}
  L_A\,R^{-1/2}=\delta_1\bigl(U_1R^{-5/2}\bigr)+\delta_2\bigl(U_2R^{-5/2}\bigr),
  \qquad
  U_1=\frac{1-x^2}{8}\sum_{\substack{i\ge1,\ j\ge0\\ i+j\le4}}
  x^{i+j}\,v_{ij}\,\bigl(\xi_1^{i}-\xi_1^{-i}\bigr)\,e_j(\xi_2),
  \label{eq:S8gdcert}
\end{equation}
where $U_2(\xi_1,\xi_2)=U_1(\xi_2,\xi_1)$, $e_0=1$, $e_j(\xi)=\xi^j+\xi^{-j}$ for $j\ge1$, and
\begin{equation}
\begin{aligned}
  v_{10}&=x^8-7x^6-280x^4-7x^2+1, & v_{11}&=(x^2+1)(3x^4+128x^2+3),\\
  v_{12}&=-3(3x^4+38x^2+3), & v_{13}&=5(x^2+1),\\
  v_{20}&=-(x^2+1)(x^4-42x^2+1), & v_{21}&=-10(x^4+14x^2+1),\\
  v_{22}&=11(x^2+1), & v_{30}&=-x^4-26x^2-1,\\
  v_{31}&=7(x^2+1), & v_{40}&=x^2+1.
\end{aligned}
  \label{eq:S8gdv}
\end{equation}
Multiplied by $R^{7/2}$, Eq.~\eqref{eq:S8gdcert} is an identity between Laurent polynomials in $\xi_1$, $\xi_2$ with 85 monomials, which fixes the ten polynomials $v_{ij}$ uniquely.
They are given, with the prefactor $(1-x^2)x^{i+j}/8$ included, in the ancillary file \texttt{apery\_proof/gd\_certificate.json}.
Since the constant term of $\delta_iG$ vanishes for every Laurent series $G$ in $\xi_1$, $\xi_2$, the constant term of Eq.~\eqref{eq:S8gdcert} is $L_A\varpi=0$.

For the second proof we use Eq.~\eqref{eq:fermi} and the recurrence of the return numbers $B_N$ of the simple cubic lattice~\cite{PetersStienstra:1989}, which we rederived with Zeilberger's algorithm (ancillary file \texttt{apery\_proof/fermi\_certificate.json}).
We write the return numbers as $B_N=\sum_jF(N,j)$ with $F(N,j)=\binom{2N}{N}\binom{N}{j}^{2}\binom{2j}{j}$.
Zeilberger's algorithm then gives
\begin{equation}
\begin{aligned}
  &(N+2)^3F(N{+}2,j)-2(2N+3)(10N^2+30N+23)F(N{+}1,j)\\
  &\quad+36(N+1)(2N+1)(2N+3)F(N,j)
  =G(N,j{+}1)-G(N,j),\\
  &G(N,j)=4j^3(N+1)(2N+1)(2N+3)(-4N+3j-8)\,H(N,j),\\
  &H(N,j)=\binom{2N}{N}\binom{2j}{j}\frac{N!^2}{j!^2\,(N+2-j)!^2},
\end{aligned}
  \label{eq:S8wz}
\end{equation}
where $1/(N+2-j)!=0$ for $j>N+2$, so that $G(N,j)=0$ for $j\ge N+3$, and $G(N,0)=0$.
Summing the first relation over $0\le j\le N+2$, we therefore reproduce the recurrence of Section~3 of Peters and Stienstra,
\begin{equation}
  N^3B_N-2(2N-1)(10N^2-10N+3)B_{N-1}+36(N-1)(2N-3)(2N-1)B_{N-2}=0.
  \label{eq:S8rec}
\end{equation}
Hence the series $f(\zeta)=\sum_NB_N\zeta^N$ is annihilated by
\begin{equation}
  L_F=\theta_\zeta^3-2\zeta(2\theta_\zeta+1)(10\theta_\zeta^2+10\theta_\zeta+3)+36\zeta^2(\theta_\zeta+1)(2\theta_\zeta+1)(2\theta_\zeta+3),
  \qquad \theta_\zeta=\zeta\,\frac{d}{d\zeta}.
  \label{eq:S8LF}
\end{equation}
For $\zeta=z/(1+z)^2$ the operators $L_A$ and $L_F$ satisfy $L_A\bigl[(1+z)^{-1}f(\zeta)\bigr]=(1-z)\,(L_Ff)(\zeta)$ for every function $f$, which is the pullback relation of Peters and Stienstra in these variables. Thus $L_A\varpi=0$ by Eq.~\eqref{eq:fermi}.

\subsection{The scalar integral in position space}
\label{sec:S9position}

We consider the scalar integral of topology~40 with unit numerator and cut worldline propagators, in four
dimensions and in impact-parameter space, in units $|b|=1$.
With $\tau$ and $\tau'$ the proper times at which the two radiative gravitons attach to the worldlines and $\Delta_2$ the
squared interval between these two vertices, negative for spacelike separation, we define
\begin{equation}
f_{\rm same}(\gamma)=\frac12\int_{\Delta_2<0}\!d\tau\,d\tau'\;\mathcal{G}\,,\qquad
f_{\rm mix}(\gamma)=\frac12\int_{\Delta_2>0}\!d\tau\,d\tau'\;\mathcal{G}
\label{eq:S9f}
\end{equation}
with
\begin{equation}
\begin{gathered}
\mathcal{G}(\tau,\tau')=\frac{2\pi}{\sqrt{1+(\gamma^2-1)\tau^2}\,\sqrt{1+(\gamma^2-1)\tau'^2}}\;
\frac{1}{\sqrt{P^2-\Delta_2^2}}\,\ln\frac{P+\sqrt{P^2-\Delta_2^2}}{|\Delta_2|}\,,\\
\Delta_2=\tau^2+\tau'^2-2\gamma\tau\tau'-1\,,\qquad
P=2(\tau-\gamma\tau')(\gamma\tau-\tau')-\gamma\Delta_2\,,\\
P^2-\Delta_2^2=\big[(\gamma-1)(\tau+\tau')^2+\gamma+1\big]\big[(\gamma+1)(\tau-\tau')^2+\gamma-1\big]\,.
\end{gathered}
\label{eq:S9G}
\end{equation}
The scalar integral is proportional to $f_{\rm same}$ when both radiative gravitons are retarded into their common vertex and to $f_{\rm mix}$ when one of them is advanced relative to it (Section~\ref{sec:gamma3-landau}).
Both $f_{\rm same}$ and $f_{\rm mix}$ converge for $\gamma>1$, and we evaluate them by numerical quadrature.

To compare with the result of Driesse et al.~\cite{Driesse:2026qiz}, let $\hat\varpi(x)$ be their K3 period, with $x=\gamma-\sqrt{\gamma^2-1}$ and
$(1+x^2)\,\hat\varpi=1+\tfrac{15}{32}v^2+O(v^4)$ for $v=\sqrt{\gamma^2-1}/\gamma$. Let $\mathcal{I}[\,\dots;x]$ be the
iterated integrals with base point $x=1$ defined in that reference. With $\phi=x\,\hat\varpi$,
$\mathcal{K}_1=(1+x^2)\,\hat\varpi/x$, $\mathcal{K}_2=1/(x\,r\,\hat\varpi)$ and $r=\sqrt{-1+34x^2-x^4}$, we define
\begin{equation}
F_0^{\rm pot}=\sqrt2\,\pi^2\,\phi\;\mathcal{I}[\mathcal{K}_2;x]\,,\qquad
F_0^{\rm mem}=-16\,\phi\;\mathcal{I}[\mathcal{K}_2,\mathcal{K}_2,\mathcal{K}_1;x]\,.
\label{eq:S9F0}
\end{equation}
The sum of these two functions is $f^{(O)}_{20}/(2\gamma)$ in the basis of Driesse et al. In their region-by-region result $F_0^{\rm pot}$ appears in the part labeled potential and tail and $F_0^{\rm mem}$ in the part labeled memory.
The coefficients of $(3-\gamma)^{1/2}$ in $F_0^{\rm pot}$ and $F_0^{\rm mem}$ are $+\sqrt2\,\pi^2/4$ and $-\sqrt2\,\pi^2/4$. The odd powers of
$(3-\gamma)^{1/2}$ cancel in the sum, as Driesse et al.\ found for the potential and memory regions and as we checked through $(3-\gamma)^{199/2}$.
We find numerically, to at least 41 digits, that
\begin{equation}
(\gamma^2-1)\,f_{\rm same}=-2\pi\,\big(F_0^{\rm pot}+F_0^{\rm mem}\big)\,.
\label{eq:S9identity}
\end{equation}
We also apply to $f_{\rm same}$ and $f_{\rm mix}$ the operator
$\mathcal{M}=(\gamma^2-9)(\gamma^2-1)\,\partial_\gamma^3+6\gamma(\gamma^2-5)\,\partial_\gamma^2+(7\gamma^2-12)\,\partial_\gamma+\gamma$,
which annihilates $\phi$, and find numerically
\begin{equation}
\begin{aligned}
\mathcal{M}\big[(\gamma^2-1)\,f_{\rm same}\big]&=\frac{16\pi\gamma}{\sqrt{\gamma^2-1}}\,,\\
\mathcal{M}\big[(\gamma^2-1)\,f_{\rm mix}\big]&=-\frac{16\pi\gamma}{\sqrt{\gamma^2-1}}+20\pi+\frac{32\pi}{\gamma^2-1}
+\frac{32\pi\gamma\,\ln x}{(\gamma^2-1)^{3/2}}\,,
\end{aligned}
\label{eq:S9ode}
\end{equation}
to at least 27 digits on $1.6\le\gamma\le4.2$.

At $\gamma=3$ we find $f_{\rm same}=2.75532463456$ and $f_{\rm mix}=1.52462878840$.
Both functions are analytic at $\gamma=3$, so neither has a $(3-\gamma)^{1/2}$ term and the two routings cannot be distinguished by regularity at $\gamma=3$.

To show analyticity we write $\gamma=\cosh\eta$ and substitute $\tau=\sinh a/\sinh\eta$ and $\tau'=\sinh a'/\sinh\eta$.
With $u=a-a'$, $s=(a+a')/2$, $C=\cosh2s$ and $c=\cosh u$ we have $\Delta_2\sinh^2\eta=(C+\gamma)(c-\gamma)$, so that $\Delta_2<0$ for $|u|<\eta$. Since $P=\gamma(\tau^2+\tau'^2)-2\tau\tau'+\gamma>0$, the logarithm in Eq.~\eqref{eq:S9G} equals $\operatorname{arcsinh}\bigl(\sqrt{P^2-\Delta_2^2}/|\Delta_2|\bigr)$, and
\begin{equation}
f_{\rm same}=\frac{4\pi}{\sinh\eta}\int_0^\infty\!ds\int_0^{\eta}\!du\;\frac{\operatorname{arcsinh}X}{\sqrt{AB}}\,,\qquad
f_{\rm mix}=\frac{4\pi}{\sinh\eta}\int_0^\infty\!ds\int_{\eta}^{\infty}\!du\;\frac{\operatorname{arcsinh}X}{\sqrt{AB}}\,,
\label{eq:S9rapidity}
\end{equation}
with
\begin{equation}
A=(C-1)(c+1)+(\gamma+1)^2\,,\qquad B=(C+1)(c-1)+(\gamma-1)^2\,,\qquad
X=\frac{\sinh\eta\,\sqrt{AB}}{(C+\gamma)\,|c-\gamma|}\,.
\end{equation}
We set $u=\eta t$ with $0<t<1$ in $f_{\rm same}$ and $u=\eta+t$ with $t>0$ in $f_{\rm mix}$, so that the integration domains do not depend on $\gamma$. We continue $|c-\gamma|$ to complex $\eta$ as $\gamma-c$ and $c-\gamma$, respectively.
For complex $\eta$ close to $\operatorname{arccosh}3$ the quantities $A$, $B$ and $X$ remain in a narrow sector around the positive real axis for all $t$ and $s$, since $C\pm1\ge0$ and the arguments of $\gamma\pm1$, $C+\gamma$, $c+1=2\cosh^2(u/2)$, $c-1=2\sinh^2(u/2)$ and $\pm(c-\gamma)=\pm2\sinh\tfrac{u+\eta}{2}\,\sinh\tfrac{u-\eta}{2}$ are $O(|\operatorname{Im}\eta|)$ uniformly in $t$.
The integrands are therefore holomorphic in $\eta$ and bounded, uniformly in $\eta$, by constant multiples of $e^{-s}(1-t)^{-1/2}$ and $e^{-s-t/4}\,t^{-1/2}$, respectively, which are integrable.
Hence $f_{\rm same}$ and $f_{\rm mix}$ are holomorphic in a neighborhood of $\gamma=3$.
As a check, the mean of the integrals~\eqref{eq:S9rapidity} over 48 equally spaced points on the circle $|\gamma-3|=\tfrac12$ reproduces $f_{\rm same}(3)$ and $f_{\rm mix}(3)$ to 29 digits, which is the accuracy of the 48-point rule.

For a second check we use Eq.~\eqref{eq:S9ode}.
The operator $\mathcal{M}$ has exponents $0$, $\tfrac12$ and $1$ at $\gamma=3$ with no logarithmic solution, and the right-hand sides are analytic there.
For $\gamma<3$ and $f$ either $f_{\rm same}$ or $f_{\rm mix}$, every solution is therefore of the form $(\gamma^2-1)f=\alpha(\gamma)+8\,c_1\,(3-\gamma)^{1/2}\,h(\gamma)$, with $\alpha$ analytic at $\gamma=3$, $c_1$ the coefficient of $(3-\gamma)^{1/2}$ in $f$, and $h=1+\tfrac14(3-\gamma)+\tfrac{11}{160}(3-\gamma)^2+\dots$ fixed by $\mathcal{M}$.
Taking Eq.~\eqref{eq:S9ode} as exact, we matched this form to values of $f_{\rm same}$ and $f_{\rm mix}$ accurate to 75 digits at 48 points in $2.4<\gamma<3$ and found $|c_1|/f(3)<10^{-74}$ for both, which is zero at the accuracy of the values. This held for every interval, subset of points and series order that we tried.
For $f_{\rm same}$ the $(3-\gamma)^{1/2}$ term cancels between $F_0^{\rm pot}$ and $F_0^{\rm mem}$ in Eq.~\eqref{eq:S9identity}.

\subsection{The topology-41 alphabet}
\label{sec:np41}

Topology~41 of Brammer et al.~\cite{Brammer:2025rqo} is the thirteen-propagator top topology of the non-planar family.
From the $d\log$ structure of its six-dimensional leading singularities, given there and in Ref.~\cite{Frellesvig:2024zph}, and from the factorization over
$\mathbb{Q}[x]$ of the algebraic ones we obtain the alphabet
\begin{equation}
\mathcal{A}_{\text{top.41}}=\{x,\;x-1,\;x+1,\;x^2+1\}\,.
\label{eq:np41-alphabet}
\end{equation}
This alphabet coincides with the multiple-polylogarithm alphabet of the 5PM-1SF result~\cite{Driesse:2024feo}
and is a subset of the kernels of the 5PM-2SF observables~\cite{Driesse:2026qiz}.
The letter $x^2+1$ does not arise from a first-type Landau pinch, which on each face we analyzed would require
$u_1\cdot u_2=0$.
It comes from zeros of the leading singularities, such as
$\mathrm{LS}^{(6D)}_{\rm even}=q^2(x^2+1)/(x^2-1)$, which become poles of the canonically normalized master
integrals.
Since we have computed no master integral of this topology, the alphabet~\eqref{eq:np41-alphabet} is a prediction, and the absence of the cyclotomic letters $x^2\pm x+1$ remains to be tested.

\section{Beyond the maximal cut at 4PM}\label{app:fourpm}

For the three-loop K3 sector of the conservative 4PM
dynamics~\cite{Bern:2021dqo,Dlapa:2021npj,Dlapa:2022wdu,Dlapa:2023hsl}, we expand the solutions of the maximal-cut block of the $\epsilon$-form system of Dlapa et al.\ through order $\epsilon^2$. We also constrain the master integrals beyond the maximal cut of
Section~\ref{sec:bootstrap-validation} with two auxiliary-mass flows~\cite{Liu:2017jxz,Liu:2022chg}. However, with these flows we do not determine the individual master integrals as functions of $\epsilon$.

\subsection{Sector labels and maximal-cut values}
\label{sec:S7}\label{sec:checks-4pm}

In the notation of Ref.~\cite{Dlapa:2023hsl} the sector whose maximal cut we reduce to Eq.~\eqref{eq:euler1d}
is $M^{(112;+++)}_{000;010101110}$, equal to
$G_{1,1,1,0,0,0,0,0,1,1,1,0,0,1,1}$ of Ref.~\cite{Dlapa:2022wdu} under
$k_2\leftrightarrow k_3$, $u_1\leftrightarrow u_2$.
The values of $\mathrm{LS}_{\rm shape}(x)$ and of $x\,K(x^2)^2$ at the five rational points used for Eq.~\eqref{eq:k3gate} are in the ancillary file \texttt{tables/fourpm\_maxcut\_values.txt}.

\subsection{Function space on the maximal cut}
\label{sec:collapse}

The Landau locus~\eqref{eq:alphabet}~\cite{Landau:1959fi} gives the singular points of the sector, but not the functions in which its master integrals are expressed.
The $\epsilon$-form system of Dlapa et al., their Eqs.~(5.33)--(5.34) and (B.15), contains besides logarithmic kernels the kernel $\pi^2/[x(1-x^2)K'^2]$ and rational functions multiplied by $K'^2$ and $K'^4$, with $K'\equiv K(1-x^2)$, and they write its solutions as iterated integrals of these kernels.
On the maximal cut, through first order in $\epsilon$ relative to its leading term, the scalar master integral is a combination of the three functions~\eqref{eq:sym2span} with coefficients linear in $\log[x/(1-x^2)]$.
At the next order it contains the iterated integral of the two kernels $\pi^2/[x(1-x^2)K'^2]$ and $(1+110x^2+x^4)K'^2/[x(1-x^2)]$. Comparing exact series expansions at $x=1$, we find that this integral is not a linear combination of products of the three functions with polylogarithms of weight at most two in the letters~\eqref{eq:alphabet}.
Which combination of the three functions appears is fixed by boundary data. Determining it requires values of the master integrals, which we do not have, and to our knowledge none are published for this sector.

\subsection{A fixed point of the flow}
\label{sec:eta-fixed-point}

The
4PM cut-eikonal K3 family, with five graviton propagators and three cut
linearized worldline propagators
$-2k\cdot u$~\cite{Kalin:2020mvi,Mogull:2020sak,Dlapa:2023hsl}, is
a fixed point of the boundary recursion of the auxiliary-mass flow.
Because cut propagators cannot be deformed, we insert the auxiliary mass
in the same five graviton propagators at every level, and the
$\eta\to\infty$ rescaling reproduces the family.
Then the boundary vector $b$ in Eq.~\eqref{eq:response}
coincides with the vector $\mathbf{M}$ of the 43 master integrals, and the
response relation becomes
\begin{equation}
\label{eq:fixedpoint}
\mathbf{M} \;=\; A(\epsilon)\,\mathbf{M}\,.
\end{equation}
Of the rows of $A$, 25 are rows of the identity matrix, and from the 18 coupled rows
we find
\begin{equation}
\label{eq:rank15}
\operatorname{rank}(A-I)=15\,,\qquad \dim\ker(A-I)=28\,.
\end{equation}
The fixed-point condition determines 15 independent linear combinations of the 43 master integrals, and
the 28 kernel directions are the single-scale cut-eikonal vacuum
integrals in $1+\eta$.

We compute the rank at two points, $\gamma=5/3$ ($x=1/3$) and $\gamma=13/5$ ($x=1/5$). At both, 73 of the 76 nonzero entries of $A$ are constant over the sampled values of $\epsilon$ and agree with rational numbers. If these 73 entries take the same rational values at every $\epsilon$, then $A^2=A$ and $\operatorname{tr}A=28$ for any values of the remaining three, and $\operatorname{rank}(A-I)=43-\operatorname{tr}A=15$ for every $\epsilon$.

In the PM literature the master integrals of such cut-eikonal families
are obtained from differential equations in $x$ with static-limit
boundary
values~\cite{Dlapa:2021vgp,Dlapa:2023hsl,Driesse:2024xad},
which we have not implemented for this family.
Instead we constrain the kernel directions with a second flow, in which we insert $\eta$ on all non-cut
propagators,
\label{sec:multiflow-outcome}and whose recursion gives
the linear response $\mathbf{M}=C(\epsilon)\,\mathbf{v}$ of the 43 master integrals to the 68
boundary values $\mathbf{v}$.
We impose the fixed-point condition~\eqref{eq:fixedpoint} on this $\mathbf{M}$, together with the closed-form value of sector 63 (a cube of cut tadpoles),
\begin{equation}
\label{eq:J63}
J_{63} \;=\; -\tfrac{i}{8}\,\pi^{-3/2}\,
\Gamma\!\big(\epsilon-\tfrac12\big)^3\,,
\end{equation}
and the equality $J_{111}=J_{63}$ with sector 111, and we solve for $\mathbf{v}$ at both kinematic points.
All 43 constraint rows of the second flow are then satisfied, and at every sampled $\epsilon$ the first two master
integrals $m_0$, $m_1$ of the cut top sector obey
\begin{equation}
\label{eq:m1m0}
\frac{m_1}{m_0} \;=\; -\frac{2}{5}\,,
\end{equation}
an integration-by-parts identity recovered numerically.

\subsection{\texorpdfstring{Master ratios at leading order in $\epsilon$}{Master ratios at leading order in epsilon}}
\label{sec:eps-family-pathology}
\label{sec:rank1-result}
\label{sec:S6}

At the sampled values of $\epsilon$ each of the eight top-sector master integrals $m_0,\dots,m_7$ is dominated by its leading Laurent term. The ratios $m_r/m_0$,
$r=1,\dots,7$, are then ratios of leading coefficients.
Extrapolating to $\epsilon=0$, we find that each ratio is a rational number at both kinematic points: rational reconstruction returns a fraction $p/q$, and the partial quotient that follows $p/q$ in the continued fraction exceeds $10^{68}$ (ancillary file \texttt{tables/fourpm\_master\_ratios.txt}).

At $\gamma=13/5$ the eight master integrals are rescaled, at three of the eighteen sampled values of $\epsilon$, by a common
complex factor that the subsector master integrals do not share and that we have not derived. At $\gamma=5/3$ the factor at the same three values of $\epsilon$ differs from one by less than $3\times10^{-11}$ but by far more than its scatter over the other sampled values. Since the factor is common to the eight master integrals, it cancels in the ratios.
We have not determined the next order, at which the three K3 periods of
Eq.~\eqref{eq:sym2span} can first enter these ratios, nor derived the
ratios from the differential equation at $\epsilon=0$.

\bibliographystyle{jhep/JHEP}
\bibliography{jhep/refs}

\end{document}